\documentclass[preprint,3p]{elsarticle}

\usepackage{lineno,hyperref}
\usepackage{gensymb}   
\modulolinenumbers[5]

\usepackage{amssymb,amsmath}
\usepackage{multirow}
\usepackage{gensymb}
\usepackage{bm}   
\usepackage{epstopdf}
\usepackage{graphicx}
\usepackage{float}
\usepackage{subfig}
\usepackage{supertabular}
\usepackage{verbatim}
\usepackage{array}
\usepackage{color}
\usepackage{makecell}
\usepackage{framed}
\usepackage{longtable,tabularx}
\usepackage{nomencl}
\usepackage{booktabs}
\usepackage{natbib}
\makenomenclature

\newcommand{\reffig}[1]{figure \ref{#1}}
\newcommand{\reffigs}[1]{figures \ref{#1}}
\newcommand{\refFig}[1]{Figure \ref{#1}}
\newcommand{\refFigs}[1]{Figures \ref{#1}}
\newcommand{\reftab}[1]{table \ref{#1}}
\newcommand{\refTab}[1]{Table \ref{#1}}
\newcommand{\refeq}[1]{Eq. \eqref{#1}}
\newcommand{\refse}[1]{section \ref{#1}}

\journal{Journal of Fluids and Structures}

\begin{document}

\begin{frontmatter}

\title{Aeroelastic mode decomposition framework and mode selection mechanism in fluid-membrane interaction}


\author[mymainaddress]{Guojun Li\corref{mycorrespondingauthor1}}
\cortext[mycorrespondingauthor1]{Corresponding author}
\ead{li.guojun@u.nus.edu}

\author[mysecondaryaddress]{Rajeev Kumar Jaiman}

\author[mymainaddress]{Boo Cheong Khoo}

\address[mymainaddress]{Mechanical Engineering, National University of Singapore, 9 Engineering Drive 1, Singapore 117576}
\address[mysecondaryaddress]{Mechanical Engineering, University of British Columbia, Vancouver, BC Canada V6T 1Z4}

\begin{abstract}
In this study, we present a global Fourier mode decomposition framework for unsteady fluid-structure interaction. We apply the framework to isolate and extract the aeroelastic modes arising from a coupled three-dimensional fluid-membrane system. The proposed framework is employed to decompose the physical variables in the fluid and structural domains into frequency-ranked aeroelastic modes in a unified way. We observe the frequency synchronization between the vortex shedding and the structural vibration via mode decomposition analysis. We examine the role of flexibility in the aeroelastic mode selection and perform a systematic comparison of flow features among a rigid wing, a rigid cambered wing and a flexible membrane. The camber effect can enlarge the pressure suction area on the membrane surface and suppress the turbulent intensity, compared to the rigid flat wing counterpart. With the aid of our mode decomposition technique, we find that the dominant structural mode exhibits a chordwise second and spanwise first mode at different angles of attack. The structural natural frequency corresponding to this mode is estimated using an approximate analytical formula. By examining the dominant frequency of the coupled system, we find that the dominant membrane vibrational mode is selected via the frequency lock-in between the dominant vortex shedding frequency and the structural natural frequency. From the fluid modes and the mode energy spectra at $\alpha=20^\circ$ and $25^\circ$, the aeroelastic modes corresponding to the non-integer frequency components lower than the dominant frequency are found to be associated with the bluff body vortex shedding instability. The non-periodic aeroelastic response observed at higher angles of attack are related to the interaction between aeroelastic modes caused by the frequency lock-in and the bluff-body-like vortex shedding. Using the mode decomposition analysis, we suggest a feedback cycle for flexible membrane wings undergoing synchronized self-sustained vibration. This feedback cycle reveals that the dominant aeroelastic modes are selected through the mode and frequency synchronization during fluid-membrane interaction to exhibit similar modal shapes in the membrane vibration and the pressure pulsation.
\end{abstract}


\end{frontmatter}

\section{Introduction}
During the past decades, morphing wings with flexible membrane components have received substantial attention from the aerospace engineering community in the context of bio-inspired flying vehicles \cite{lian2003membrane,shyy2005membrane,platzer2008flapping,aldheeb2016review}. A flexible membrane can passively deform and vibrate by highly interacting with an unsteady flow, thereby forming a coupled fluid-membrane system. The coupled system exhibits a variety of correlated vibrational and fluid modes with a wide range of spatial and temporal scales. These correlated aeroelastic modes and their scales are closely related to the aerodynamic performance of the membrane wings and play an important role in efficient flight and control strategies. Hence, identifying and isolating the most influential aeroelastic modes from the coupled system is essential to further understand the aeroelastic mode selection mechanism and to promote the design of active or passive control strategies.

Numerous experimental and computational studies on fluid-membrane interaction have been carried out during the past years. Song et al. \cite{song2008aeromechanics} examined the aeromechanics of membrane wings as a function of aspect ratio, flexibility and pre-strain value. It can be observed from the phase map of membrane mode that the dominant mode switched from the first mode to higher modes as the angle of attack and Reynolds number increased. Rojratsirikul et al. \cite{rojratsirikul2009unsteady,rojratsirikul2010effect,rojratsirikul2011flow} performed a series of experiments to study the dynamic behaviors of flexible membrane wings at moderate Reynolds numbers. Different types of dominant vibrational modes and vortical structures have been observed in the coupled fluid-membrane system. Bleischwitz et al. \cite{bleischwitz2015aspect} investigated the effect of aspect ratio on membrane dynamics in wind tunnel experiments. The frequencies corresponding to the dominant structural modes were found to be correlated with the frequencies of the force fluctuations. In view of the limitations of collecting physical data of interest in wind tunnel experiments, high-fidelity numerical simulation methods become effective tools to gain further insight into the coupled mechanism of the flexible membrane. With the aid of advanced numerical simulations, Sun et al. \cite{sun2016nonlinear,sun2017effect,sun2017nonlinear,sun2018bifurcations} systematically studied the nonlinear dynamic behaviors of flexible membrane wings. The vibration of the flexible membrane excited by the unsteady flow gradually transitioned from the periodic state to the non-periodic state as the relevant aeroelastic parameters changed. From the frequency spectra analysis of the membrane responses in the aforementioned literature, it was found that the membrane vibration usually exhibited multiple frequency peaks, which were closely correlated with the vortical structures with a variety of temporal and spatial scales. These multi-modal mixed responses of the coupled fluid-membrane system pose a serious challenge to identify and isolate the correlated aeroelastic modes of interest from the system. The understanding of the underlying mechanism of how a specific aeroelastic mode is selected in the coupled system is limited.

With respect to the dominant mode identification in a coupled fluid-membrane system, some ingenious methods have been adopted to distinguish the most influential modes. Standard deviation analysis has been widely applied to the fluid-membrane interaction problem to determine the dominant structural modes \cite{rojratsirikul2009unsteady,rojratsirikul2011flow,tregidgo2011fluid}. As Bleischwitz et al. \cite{bleischwitz2015aspect} pointed out, the excited structural modal shapes overlapped together, which increased the difficulty of isolating the structural modal shapes of interest from the coupled system. The standard deviation analysis of the membrane deflection could reflect the dominant structural modal shape to some extent. However, occasionally appearing modes or overlapping modes with small energies may be covered by the dominant modes, which makes them hard to be identified from the overall membrane vibration responses. Additionally, the frequency spectra and the vibration state analysis at a single point are not reliable indicators to reflect the dynamic characteristics of the whole membrane structure \cite{sun2016nonlinear}. Therefore, global mode identification methods are naturally desirable to capture the dynamic behaviors of the entire physical field of interest. The relevant dynamic information corresponding to each mode such as mode energy and mode frequency could help us gain further insight into the whole dynamic characteristics when performing the global mode identification.

Data-based modal decomposition techniques have been widely used in the analysis of the flow features and the structural vibrations to identify coherent structures. These mode decomposition techniques separate the temporal-spatial data into energy-ranked or frequency-ranked modes with physical meaning to represent different characteristics of the field. The space-only proper orthogonal decomposition (POD) method \cite{lumley1970stochastic,sirovich1987turbulence,berkooz1993proper} can decompose the collected physical fields of interest into a set of energy-ranked POD modes by diagonalizing the spatial correlation matrix. These decomposed modes are orthogonal in space and multiple frequency components can be observed for each space-only POD mode when the physical fields collected for mode decomposition are complex. The spectral mode decomposition methods such as the Fourier mode decomposition (FMD) \cite{ma2015fourier} and the dynamic mode decomposition (DMD) \cite{schmid2010dynamic} project the spatial-temporal physical data into the spatial-frequency space to obtain the decomposed modes in frequency ranking. The FMD method is based on the discrete Fourier transform, which is a superposition of harmonic modes. The mode energy and the phase information are included in the transformed Fourier coefficients. The DMD method and its variants (e.g., Optimized DMD, Sparsity Promoting DMD) can extract the dynamic modes with growth rates and oscillation frequencies from nonlinear systems by approximating the modes of the Koopman operator. Chen et al. \cite{chen2012variants} mathematically demonstrated that the DMD will be reduced to the FMD when the fluctuations of the physical variables are performed in the mode decomposition process. 

To bridge the gap between the energetically optimal space-only POD with spatial orthogonalization and the spectral mode decomposition technique with temporal orthogonalization, some mode decomposition approaches based on POD, the so-called spectral proper orthogonal decomposition method, were recently studied to obtain decomposed modes evolving coherently both in space and time. Sieber et al. \cite{sieber2016spectral} developed a spectral proper orthogonal decomposition method to build a connection between the energetically optimal POD and the spectrally pure FMD. To achieve this goal, a low-pass filter was applied along the diagonals of the correlation matrix to enforce the diagonal similarity. The filtered correlation matrix was employed to perform the eigendecomposition as the space-only POD to obtain mode energy and temporal coefficients. The proposed spectral POD method converges to the space-only POD when the filter length is set to zero, and it changes to the FMD as the filter length becomes the maximum value. Towne et al. \cite{towne2018spectral} presented another spectral POD method derived from a space-time formulation of POD to identify spatial-temporal coherent structures. The decomposed modes that are energetically ranked within each frequency can be obtained by employing this type of spectral POD method. Based on the proposed spectral POD algorithm, the linear combinations of the FMD modes at each frequency are equivalent to the spectral POD modes.

Aforementioned data-based modal decomposition techniques have been extensively employed to identify the physically meaningful modes from the physical fields, like flow past a cylinder \cite{schmid2010dynamic,chen2012variants,ma2015fourier}, wall-bounded flows \cite{berkooz1993proper} and cavity flows \cite{seena2011dynamic}. Regarding the application of the mode decomposition methods for the fluid-structure interaction problems, the mode decomposition methods were employed to extract modes from the fluid or structure fields and usually treated these modes separately in the data analysis \cite{michelin2008vortex,bozkurttas2009low,liu2016interaction,miyanawala2019decomposition}. Specific to the fluid-membrane interaction problems, the POD method \cite{bleischwitz2017fluid}, the DMD method \cite{schmid2010dynamic} and the FMD method \cite{serrano2018fluid} were employed to identify the dominant modes of interest in the fluid domain. The dominant structural modal shapes were successfully identified from the whole structural responses with multi-modal mixed responses by the POD method \cite{bleischwitz2016aeromechanics} and the DMD method \cite{tiomkin2019membrane}, which avoided the drawbacks of the standard deviation analysis. While these applications of mode decomposition approaches to the fluid-membrane interaction problems have attempted to address some relevant questions, a fully-coupled relationship between the fluid modes and the structure modes can be lost due to individual treatments. Only a handful of literature can be found to build a bridge between the decomposed modes in both fluid and structural fields to explore the coupled mechanism. Recently, Goza et al. \cite{goza2018modal} developed a combined framework based on the POD and DMD methods for the mode decomposition of flapping flags immersed in an unsteady flow. In this combined formulation, the collected data in both fluid and structure fields are decomposed in a unified matrix, which naturally ensured the inherent correlation between the dynamic modes in both fields. To gain further insight into the coupled mechanism during fluid-membrane interaction, a global mode decomposition framework for extracting the aeroelastic modes in a unified manner is highly desirable in the mode analysis of the coupled systems.

In this paper, we propose an effective framework based on the radial basis function (RBF) interpolation method and the FMD method for the aeroelastic mode decomposition of the coupled fluid-membrane system. Of particular interest is to present physical insight into the underlying mechanism of how the unsteady turbulent flow interacts with the extensible 3D membrane to excite particular wake patterns and select specific vibrational modes in a frequency synchronized way. The main purpose of the current study is to present the utility of an RBF-FMD based mode decomposition framework to extract the frequency-ranked aeroelastic modes, rather than compare the advantages against other specific mode decomposition techniques. The proposed procedure can be used to interpret the frequency lock-in phenomenon and the associated mechanisms during fluid-membrane interaction. Using the Fourier-decomposed modes of the coupled fluid-structure system, we attempt to answer the specific questions that are relevant to the membrane aeroelasticity: (i) Which types of membrane vibrations and wake patterns are dominant during fluid-membrane interaction and how do we identify these dominant aeroelastic modes in the coupled system? (ii) How does membrane flexibility affect the membrane aeroelastic characteristics? (iii) What is the aeroelastic mode selection mechanism during fluid-membrane interaction? To address these questions, we extend the original FMD method for fluid-only analysis to the coupled fluid-membrane system. 

To facilitate the mode decomposition, we develop an efficient data projection to a stationary reference grid via RBF which allows handling of physical data at time-varying grids in the fluid domain. The physical variables of interest in both the fluid and structure fields are then stored into a total vector and form a time sequence to compute the global Fourier modes via FMD. The contribution of the dominant aeroelastic mode to the overall membrane dynamics is calculated quantitatively. To obtain reliable dominant mode frequencies, Welch's method \cite{welch1967use} combined with a proper window function is applied to suppress the aliasing effect and to reduce noise in the spectrum analysis. We then present the correlated Fourier modes in the fluid and structure fields. A comparison of the Fourier modes between a rigid wing, a rigid cambered wing and a flexible membrane wing is conducted to investigate the role of flexibility in the membrane aeroelasticity. To explore the connection between the flow-excited vibration and the natural frequency of the flexible membrane immersed in an unsteady flow, an approximate analytical formula of the nonlinear membrane natural frequency is derived to estimate the natural frequency corresponding to a specific structural mode of interest. Based on the mode decomposition analysis for three types of wings, the relationship between the aeroelastic response and the bluff-body-like vortex shedding is investigated to interpret the non-periodic responses observed at higher angles of attack. To our knowledge, this is the first time to utilize such a framework to extract the aeroelastic modes for 3D fluid-membrane interaction problems. Furthermore, there has not been an investigation to build a direct connection between the fluid and structure modes to explore the role of flexibility and explain the underlying mode selection mechanism during fluid-membrane interaction.

The rest of this paper is organized as follows. In section 2, the governing equations for the coupled fluid-structure system and the FMD algorithm are introduced. The description of the fluid-membrane interaction problem and the verification for the proposed fluid-structure interaction simulation framework is provided in section 3. We present the dynamic behavior of the flexible membrane at different angles of attack in section 4. The application of the proposed FMD framework to fluid-membrane interaction and the exploration of the mode selection mechanism are then discussed. In section 5, we provide the major conclusions of this work.

\section{Numerical methodology}
\subsection{Coupled fluid-structure system}
The governing equations for the incompressible unsteady viscous flow with an arbitrary Lagrangian-Eulerian (ALE) reference frame are discretized via stabilized Petrov-Galerkin variational formulation \cite{jaiman2016stable,li2018novel}. The generalized-$\alpha$ method is utilized for the temporal discretization of the ALE flow equations. The governing equations for the multibody flexible structures are discretized via the standard Galerkin finite element method. To capture the separated flow and to reduce the computational cost, a hybrid RANS/LES model based on delayed detached eddy simulation treatment is used for turbulent flow simulation. The fluid equations and the multibody structural equations are integrated in a partitioned manner. The coupled governing equations are solved based on a typical predictor-corrector scheme. The compactly-supported radial basis function method is employed to transfer the interface data between the nonmatching meshes and update the body-fitted fluid meshes in spaces in a unified manner. The fluid forces are corrected based on the recently developed nonlinear interface force correction scheme \cite{jaiman2016stable} at each iterative step to ensure the numerical stability for fluid-structure interaction problems with significant added mass effect. A detailed description of the coupled fluid-multibody structure framework can be referred to Li et al. \cite{li2018novel}. The developed fluid-multibody structure interaction framework has been validated for flexible flapping wings \cite{li2018novel,li2021high} and flow-induced vibration of two-dimensional flexible membrane wings \cite{li2020computational}.

\subsection{Global Fourier mode decomposition of fluid-membrane interaction}
In this section, the algorithm of FMD for fluid-membrane interaction problems is presented. To identify and extract the global spatial modes at specific frequencies of interest, the FMD method projects the physical data of interest in a selected spatial domain from the temporal space to the frequency space based on the discrete Fourier transform.
Before we perform the FMD analysis for the fluid-membrane interaction problem, the physical data of interest in the fluid and structural domains is firstly collected from experiments or numerical simulations in a time-discrete way. To maintain the temporal consistency, the snapshot-based data is equispaced in time with the same sampling frequency $f_{sam}$. Considering the deformation of the flexible membrane under aerodynamic loads, the body-fitted fluid grid is updated at each iterative step by following the motion of the membrane. To avoid the difference of the time-varying fluid grid locations at each snapshot during mode decomposition, the physical variables in the fluid domain $\Omega^f$ are decomposed via the FMD method at a stationary grid in the Euler coordinate as a reference for simplicity. Hence, we project the physical data $\boldsymbol{y}^{f}_s$ at the spatial points $\boldsymbol{x}^f_s$ in the moving grid as sources onto $M^f_p$ projected points $\boldsymbol{X}^f_p$ in the stationary grid to collect the projected data $\boldsymbol{y}^{f}_p$ for the mode decomposition. 

\begin{figure}[H]
	\centering
	\includegraphics[width=0.9\textwidth]{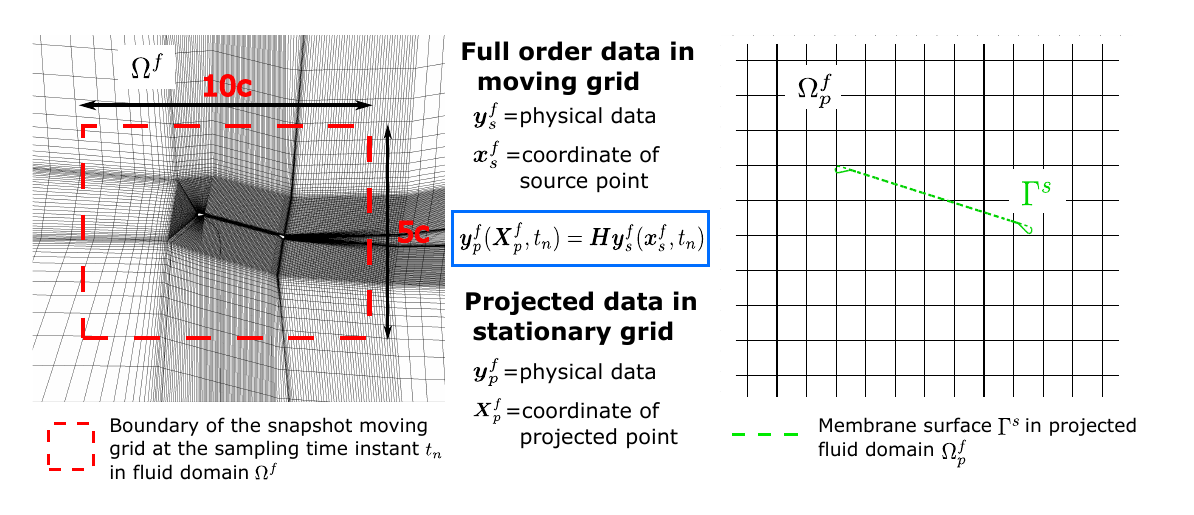}
	\caption{Illustration of projection process from moving grid to stationary reference grid based on radial basis function method for physical variable collection of the mode decomposition procedure for fluid-membrane interaction.}
	\label{FMD_RBF}
\end{figure}

An efficient projection method based on the RBF approach is employed to perform the projection between these two sets of nonmatching grids. \refFig{FMD_RBF} illustrates this projection process based on the RBF method at a specific sampling time instant $t_n=\frac{n}{f_{sam}}$ for a two-dimensional slice of the $n$-th snapshot for demonstration. In the performed Fourier mode decomposition shown in Section \ref{results}, the physical variables in the three-dimensional domains are considered. In this study, we are interested in the flow features in the near field and the wake of the coupled fluid-membrane system. Hence, only the physical data within a box in red dashed line with a size of $10c \times 5c \times 5c$ as shown in \reffig{FMD_RBF} is projected onto the stationary grid with the same size, where $c$ denotes the membrane chord length. This stationary grid consists of $300 \times 150 \times 150$ points. The physical data $\boldsymbol{y}^{s}$ on the deformed membrane surface $\Gamma^s$ is collected at $M^s$ discrete points $\boldsymbol{X}^s$ in a Lagrangian coordinate for the mode decomposition.

In the current study, we collect the projected spanwise vorticity ${\boldsymbol{\omega_y}}_n={\boldsymbol{\omega_y}}_n(\boldsymbol{X}^f_p,t_n) \in \mathbb{R}^{{M^f_p} \times 1}$ and the projected spatial pressure coefficient ${\boldsymbol{C_p}}_n={\boldsymbol{C_p}}_n(\boldsymbol{X}^f_p,t_n) \in \mathbb{R}^{{M^f_p} \times 1}$ at the $M^f_p$ discrete points $\boldsymbol{X}^f_p$ of a stationary reference grid and for the sampling time instant $t_n$. These two physical variables are related to the chordwise vortex structures and the spatial flow perturbations in the fluid domain. The membrane displacement normal to the chord ${\boldsymbol{\delta_n}}_n={\boldsymbol{\delta_n}}_n(\boldsymbol{X}^s,t_n) \in \mathbb{R}^{{M^s} \times 1}$ and the pressure coefficient difference between the upper and lower surfaces ${\boldsymbol{C_p^d}}_n={\boldsymbol{C_p^d}}_n(\boldsymbol{X}^s,t_n) \in \mathbb{R}^{{M^s} \times 1}$ at the $M^s$ discrete points $\boldsymbol{X}^s$ and the same sampling time instant $t_n$ are stored in vector forms. These two variables can reflect the structural modal shapes and the flow perturbations applied on the membrane surface. For simplicity, we store the collected physical data in the fluid and structural domains at $t_n$ into a total state vector written as
\begin{equation}
\boldsymbol{y}_n = \boldsymbol{y}_n(\boldsymbol{X},t_n) =  
\begin{bmatrix}
\boldsymbol{y}_{n1} (\boldsymbol{X}^f_p,t_n) \\
\boldsymbol{y}_{n2} (\boldsymbol{X}^f_p,t_n) \\
\boldsymbol{y}_{n3} (\boldsymbol{X}^s,t_n) \\
\boldsymbol{y}_{n4} (\boldsymbol{X}^s,t_n)
\end{bmatrix}
=
\begin{bmatrix}
{\boldsymbol{\omega_y}}_n \\
{\boldsymbol{C_p}}_n \\
{\boldsymbol{\delta_n}}_n \\
{\boldsymbol{C_p^d}}_n
\end{bmatrix}    \in \mathbb{R}^{M \times 1}
\label{FMD1}
\end{equation}
where $M = 2 M^f_p + 2 M^s$ denotes the total number of the discrete points. The collected snapshot physical data at $N$ time instants is stored in a matrix form
\begin{equation}
\boldsymbol{Y} = [ \boldsymbol{y}_0 \quad  \boldsymbol{y}_1  \ldots \boldsymbol{y}_n  \ldots  \boldsymbol{y}_{N-1} ] \in \mathbb{R}^{M \times N}
\label{FMD2}
\end{equation}

In this matrix, the organized data in each column represents the physical state at a specific sampling time instant and the data in each row provides the time history for a selected physical variable at one spatial point. The fluctuation component $\boldsymbol{y}^{\prime}_n$ can reflect the perturbation based on the mean physical values in the time-varying global physical field. For instance, the flexible membrane is excited to vibrate around its mean membrane shape and the fluctuations are connected with the structural modal shapes. In the current study, we transfer the fluctuations of the collected physical data to the spatial-frequency domain due to their connection with the aeroelastic modes. Following the idea of the Fourier series, the fluctuation function $\boldsymbol{y}^{\prime}_n$ can be written as a discrete Fourier series in an exponential form
\begin{equation}
\boldsymbol{y}^{\prime}_n = \frac{1}{N} \sum_{k=0}^{N-1} \boldsymbol{c}_k e^{i \frac{2 \pi k}{N}n}
\label{FMD3}
\end{equation}
where $\boldsymbol{c}_k$ denotes the Fourier complex coefficient vector at $M$ discrete spatial points $\boldsymbol{X}$ for a specific discrete frequency $f_k$. The Fourier complex coefficient vector consists of four components corresponding to the four selected physical variables, which can be expressed as
\begin{equation}
\boldsymbol{c}_k = \boldsymbol{c}_k (\boldsymbol{X},f_k) = 
\begin{bmatrix}
\boldsymbol{c}_{k1} (\boldsymbol{X}^f_p,f_k) \\
\boldsymbol{c}_{k2} (\boldsymbol{X}^f_p,f_k) \\
\boldsymbol{c}_{k3} (\boldsymbol{X}^s,f_k) \\
\boldsymbol{c}_{k4} (\boldsymbol{X}^s,f_k)
\end{bmatrix}
=
\begin{bmatrix}
{\boldsymbol{\omega_y}}_k \\
{\boldsymbol{C_p}}_k \\
{\boldsymbol{\delta_n}}_k \\
{\boldsymbol{C_p^d}}_k
\end{bmatrix}    \in \mathbb{C}^{M \times 1}
\label{FMD4}
\end{equation}
By employing the discrete Fourier transform, the Fourier complex coefficient vector $\boldsymbol{c}_k $ is expressed as
\begin{equation}
\boldsymbol{c}_k = \mathcal{F}(\boldsymbol{y}^{\prime}_n )  = \sum_{n=0}^{N-1} \boldsymbol{y}^{\prime}_n e^{-i \frac{2 \pi k}{N}n}
\label{FMD5}
\end{equation}

In the global FMD analysis, we perform the discrete Fourier transform to transfer the fluctuation function from the time domain to the frequency space $\boldsymbol{f}=[f_0 \quad f_1 \ldots f_k \ldots f_{N-1}] \in \mathbb{R}^{1 \times N}$ to obtain the global spatial mode matrix $\mathcal{\boldsymbol{C}} = [\boldsymbol{c}_0 \quad \boldsymbol{c}_1  \ldots \boldsymbol{c}_k  \ldots \boldsymbol{c}_{N-1}] \in \mathbb{C}^{M \times N}$. The fast Fourier transform algorithm is used to speed up the Fourier transform process. The global spatial mode matrix $\mathcal{\boldsymbol{C}}$ consists of the decomposed global spatial mode at different frequencies $f_k$. The real and imaginary parts of the spatial mode show an explicit physical significance for the spatial disturbance structures (modal shapes) and their intensity as well as initial phase difference. We define the global amplitude spectrum $\boldsymbol{A}_k$ and the global phase spectrum $\boldsymbol{\theta}_k$ at $M$ discrete points for a specific frequency $f_k$ as
\begin{equation}
\boldsymbol{A}_k  = \boldsymbol{A}_k (\boldsymbol{X},f_k)  = 
\begin{bmatrix}
\boldsymbol{A}_{k1} (\boldsymbol{X}^f_p,f_k) \\
\boldsymbol{A}_{k2} (\boldsymbol{X}^f_p,f_k) \\
\boldsymbol{A}_{k3} (\boldsymbol{X}^s,f_k) \\
\boldsymbol{A}_{k4} (\boldsymbol{X}^s,f_k)
\end{bmatrix}
=
\begin{bmatrix}
| \boldsymbol{c}_{k1} (\boldsymbol{X}^f_p,f_k) | \\
| \boldsymbol{c}_{k2} (\boldsymbol{X}^f_p,f_k) | \\
| \boldsymbol{c}_{k3} (\boldsymbol{X}^s,f_k)   |\\
| \boldsymbol{c}_{k4} (\boldsymbol{X}^s,f_k)   |
\end{bmatrix} \in \mathbb{R}^{M \times 1}
\label{FMD6}
\end{equation}
\begin{equation}
\boldsymbol{\theta}_k  = 
\begin{bmatrix}
{\theta_k}_{ij}
\end{bmatrix} =
\begin{bmatrix}
\text{arg}({c_k}_{ij})
\end{bmatrix} \in \mathbb{R}^{M \times 1}
\label{FMD7}
\end{equation}
where $|\cdot|$ represents the modulus of the Fourier complex coefficient. The argument of the the Fourier complex coefficient represents the phase angle. To identify the dominant modes from the entire decomposed mode matrix, we define the Frobenius norm $\lVert \cdot \rVert$ of the global amplitude spectrum matrix corresponding to each physical variable in the whole spatial domain to calculate the global power spectrum. The global power spectrum matrix $\boldsymbol{s}_k$ corresponding to each physical variable field at a discrete frequency $f_k$ is expressed as
\begin{equation}
\boldsymbol{s}_k = \boldsymbol{s}_k (f_k) =
\begin{bmatrix}
s_{k1} (f_k) \\
s_{k2} (f_k) \\
s_{k3} (f_k) \\
s_{k4} (f_k) 
\end{bmatrix}
=
\begin{bmatrix}
\lVert \boldsymbol{A}_{k1} (\boldsymbol{X}^f_p,f_k)  \rVert\\
\lVert \boldsymbol{A}_{k2} (\boldsymbol{X}^f_p,f_k)  \rVert \\
\lVert \boldsymbol{A}_{k3} (\boldsymbol{X}^s,f_k)    \rVert \\
\lVert \boldsymbol{A}_{k4} (\boldsymbol{X}^s,f_k)    \rVert
\end{bmatrix} \in \mathbb{R}^{4 \times 1}
\label{FMD8}
\end{equation}
where $s_{k1} (f_k)$, $s_{k2} (f_k)$, $s_{k3} (f_k)$ and $s_{k4} (f_k)$ denote the global power spectrum for the spatial modes ${\boldsymbol{\omega_y}}_k$, ${\boldsymbol{C_p}}_k$, ${\boldsymbol{\delta_n}}_k$ and ${\boldsymbol{C_p^d}}_k$, respectively. The contribution of individual spatial mode to the overall dynamic responses at a discrete frequency $f_k$ is defined as the global mode energy $\boldsymbol{e}_k$ written as
\begin{equation}
\boldsymbol{e}_k = \boldsymbol{e}_k (f_k) =
\begin{bmatrix}
e_{k1} (f_k) \\
e_{k2} (f_k) \\
e_{k3} (f_k) \\
e_{k4} (f_k) 
\end{bmatrix}
=
\begin{bmatrix}
\frac{s_{k1} (f_k)}{\sum_{k=0}^{N-1} s_{k1} (f_k)} \\
\frac{s_{k2} (f_k)}{\sum_{k=0}^{N-1} s_{k2} (f_k)} \\
\frac{s_{k3} (f_k)}{\sum_{k=0}^{N-1} s_{k3} (f_k)} \\
\frac{s_{k4} (f_k)}{\sum_{k=0}^{N-1} s_{k4} (f_k)}
\end{bmatrix} \in \mathbb{R}^{4 \times 1}
\label{FMD9}
\end{equation}
After that, the global power spectrum and the global mode energy spectrum in the transformed frequency space $\boldsymbol{f}=[f_0 \quad f_1 \ldots f_k \ldots f_{N-1}] \in \mathbb{R}^{1 \times N}  $ can be stored in the matrix form as follows:
\begin{eqnarray}
\boldsymbol{S}=[\boldsymbol{s}_0 \quad \boldsymbol{s}_1  \ldots \boldsymbol{s}_k \ldots \boldsymbol{s}_{N-1}] \in \mathbb{R}^{4 \times N}
\\
\boldsymbol{E}=[\boldsymbol{e}_0 \quad \boldsymbol{e}_1  \ldots \boldsymbol{e}_k \ldots \boldsymbol{e}_{N-1}]  \in \mathbb{R}^{4 \times N}
\label{FMD10}
\end{eqnarray}

By employing the global FMD analysis for the fluid-membrane interaction problems, the aeroelastic modes are calculated by projecting the simultaneously collected aeroelastic responses from the spatial-temporal space to the spatial-frequency space. The obtained aeroelastic modes in the fluid and structural domains and their global mode energy spectrum are stored in frequency ranking. The dominant modes can be determined as the modes with the most mode energies in the mode energy spectrum. Subsequently, these dominant modes are extracted from the frequency-ranked global mode sequence at the selected frequencies with relatively large mode energies. 
In the current study, we are interested in four physical variables ($\omega_y$, $C_p$, $\delta_n$ and $C^d_p$) in the coupled fluid-membrane system due to their close connections with the aeroelastic modes. Hence, the aforementioned mode decomposition procedure is demonstrated for these four selected physical variables. 
The proposed global FMD approach can be naturally extended to the mode decomposition for other physical variables of interest (e.g., flow velocity and structural velocity) in the fluid and structural domains. The collected physical variables can be stored together into the total state vector form shown in \refeq{FMD1} for further mode decomposition process. 

The global FMD method avoids the limitation of the traditional Fourier transform analysis at a single point. It provides a global view to reflect the dynamics of the entire physical field by containing the information of the decomposed modes at each spatial point. The global FMD analysis can establish direct correspondences of the decomposed fluid and structural Fourier modes by choosing the modes in both domains at the same selected frequency. Hence, it is helpful to build an intrinsic relationship between the flow-induced vibrations and the coherent flow structures to reveal the physical mechanism of fluid-membrane interaction because of the explicit physical interpretation of the FMD results. A summary of the RBF-FMD algorithm is presented in Algorithm 1.

In the mode decomposition analysis, it is difficult to determine the dominant frequency with physical meaning by directly employing the discrete Fourier transform when the aeroelastic response tends to be non-periodic. Some techniques are required to improve the quality of power spectrum estimation to help determine the dominant frequency. To suppress the aliasing effect and to ensure sufficient frequency resolution, a proper sampling frequency $f_{sam}$ and a suitable number of sampling time instants $N$ are selected to satisfy the Nyquist criterion ($f_{sam} > 2 f^{max}$), where $f^{max}$ denotes the maximum frequency component in the collected responses. Due to the finite-length sampling in the discrete Fourier transform, an appropriate window function is usually selected to minimize the influence of the spectral leakage effect. To reduce the noise caused by imperfect and finite data in the power spectrum analysis, the averaged periodogram technique is widely employed in the power spectrum estimation. The nonparametric Welch's method \cite{welch1967use} is one of the popular methods used for estimating power spectrum. In Welch's method, before we proceed to calculate the power spectrum via fast Fourier transform, the collect data is firstly divided into several overlapping segments, and then a proper window function is applied to each split segment. The fast Fourier transform is employed to compute the windowed discrete Fourier transform for each segment and obtain the periodogram value. Finally, the periodogram values for each segment are averaged to obtain Welch's estimation of the power spectrum. In the application of a specific problem, the window function, the number of the split segments and the sample numbers of overlap from segment to segment should be carefully selected to cope with the trade-off between detection performance and frequency resolution \cite{solomon1991psd,jwo2021windowing}.

\underline{ALGORITHM 1: RBF-FMD}
\begin{framed}
	Input: Collection of physical variables of interest in fluid and structural domains with a sampling frequency $f_{sam}$
	
	Output: Selected FMD modes $\mathcal{\boldsymbol{C}} = [\boldsymbol{c}_0 \quad \boldsymbol{c}_1  \ldots \boldsymbol{c}_k  \ldots \boldsymbol{c}_{N-1}] \in \mathbb{C}^{M \times N}$ and mode energies $\boldsymbol{E}=[\boldsymbol{e}_0 \quad \boldsymbol{e}_1  \ldots \boldsymbol{e}_k \ldots \boldsymbol{e}_{N-1}]  \in \mathbb{R}^{4 \times N}$
	
	(i) Project physical variables in fluid domain from each point in moving grids onto $M_p^f$ points in a stationary reference grid via RBF method
	
	(ii) Collect physical variables in structural domain at $M^s$ discrete points in a Lagrangian coordinate
	
	(iii) Store all physical variables in fluid and structural domains in a matrix form $\boldsymbol{Y} = [ \boldsymbol{y}_0 \quad  \boldsymbol{y}_1  \ldots \boldsymbol{y}_n  \ldots  \boldsymbol{y}_{N-1} ] \in \mathbb{R}^{M \times N}$ ($M=a M_p^f + b M^s$, where $M$ is the total number of spatial points, $a$ denotes the number of fluid variables and $b$ represents the number of structural variables; $N$ is the number of total time samples)
	
	(iv) Extract fluctuation physical variable matrix from time-varying global physical field: $\boldsymbol{Y}^{\prime}= \boldsymbol{Y}- \overline{\boldsymbol{Y}}$
	
	(v) Calculate global Fourier coefficients $\boldsymbol{c}_k$ by applying FMD approach
	
	(vi) Find mode energy $\boldsymbol{e}_k$ for its corresponding spatial mode at a discrete frequency $f_k$
	
	(vii) Determine dominant spatial modes by detecting the peaks of mode energy distributions as a function of frequency
\end{framed}

\section{Problem set-up and verification}
\subsection{Problem set-up}
\begin{figure}[H]
	\centering
	\includegraphics[width=0.6\textwidth]{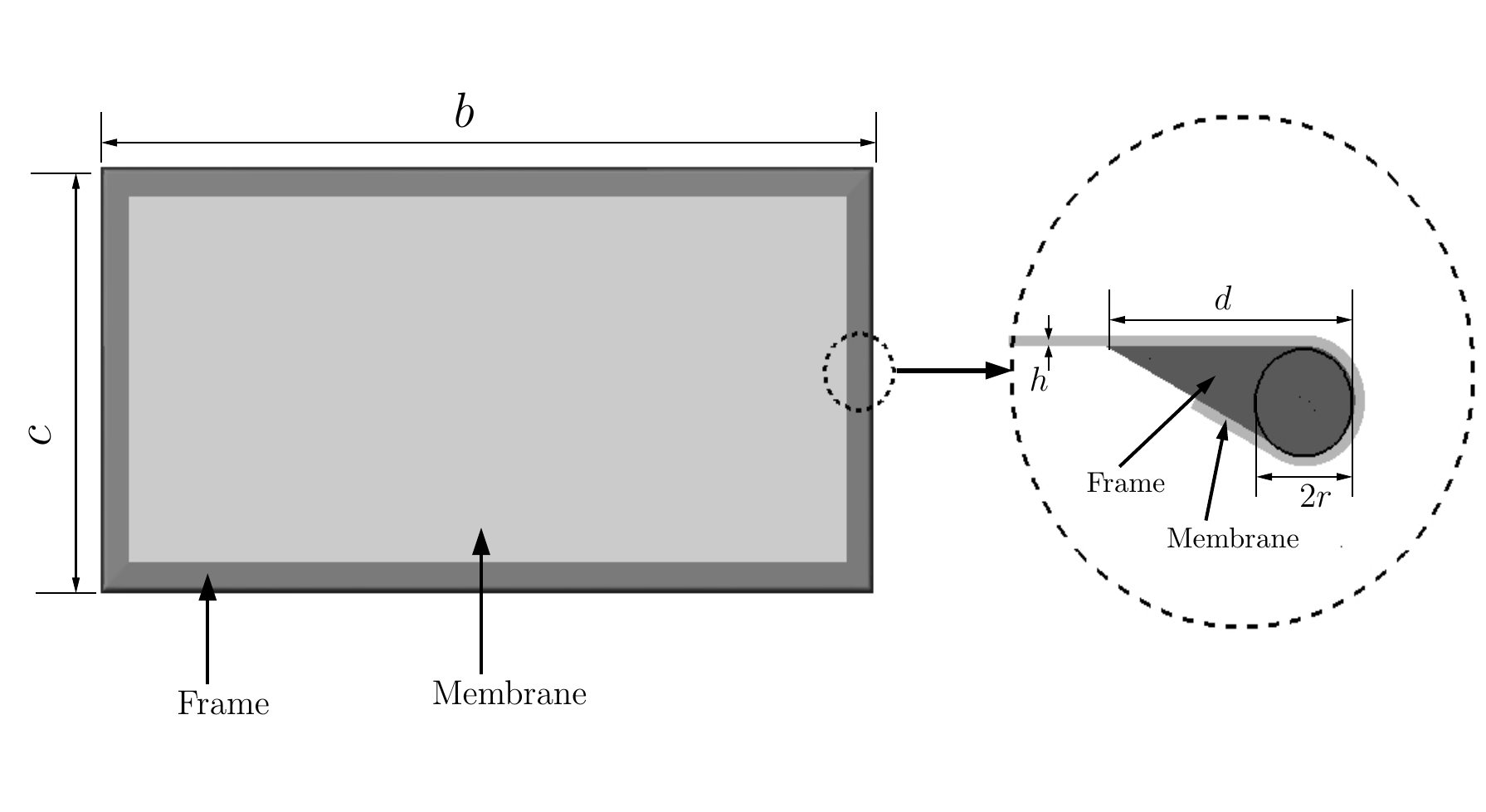}
	\caption{Membrane wing geometry and section of supporting frame.}
	\label{membranegeo}
\end{figure}

In this study, we explore the mode selection mechanism for fluid-membrane interaction problems through aeroelastic mode decomposition with the aid of the FMD technique. For that purpose, we perform a series of numerical simulations for a 3D flexible membrane wing at different angles of attack to gain further insight into the coupled mechanism. This 3D membrane wing with a supporting rigid frame was conducted in the wind tunnel experiments by Rojratsirikul et al. \cite{rojratsirikul2011flow}. The geometry information and the section of the supporting frame is presented in \reffig{membranegeo}. The membrane has a chord length of $c=68.75$ mm and an aspect ratio of $AR=2$. The thickness of the membrane is $h=0.2$ mm. The flexible membrane is made of latex rubber with the material density of $\rho^s = 1000$ kg/m$^3$ and Young's modulus of $E^s=2.2$ MPa. The aerofoil-like section of the supporting frame has a length of $d=5$ mm and the diameter of the rod is $2r =2$ mm. In the current study, the membrane wing is simulated at the same Reynolds number of $Re =24300$ as that in the experiment. The findings of the membrane aeroelasticity at several angles of attack are compared against the results obtained from the experiments to validate the coupled fluid-structure framework. 

\begin{figure}[H]
	\centering
	\subfloat[][]{\includegraphics[width=0.5\textwidth]{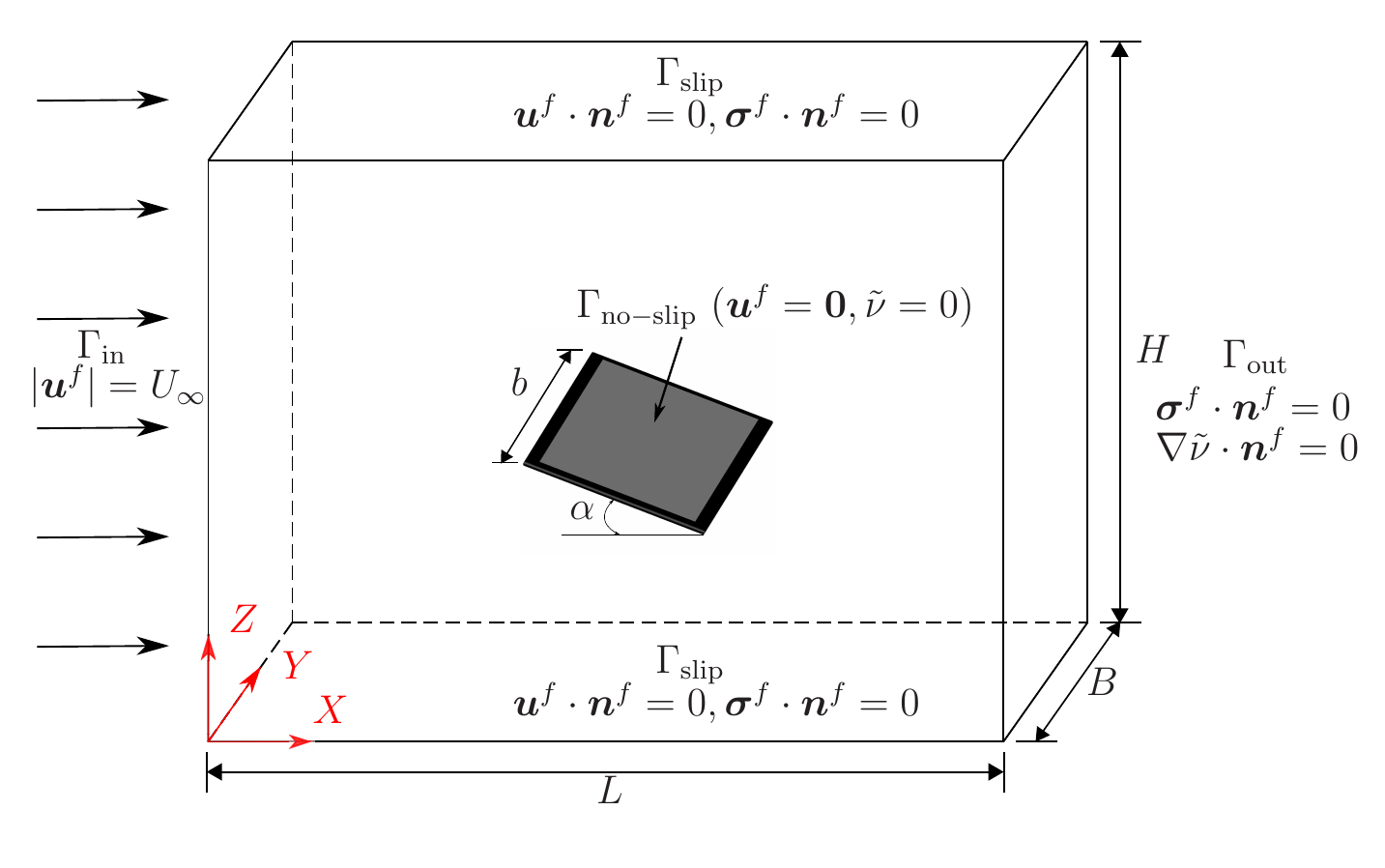}\label{membrane_domaina}}
	\quad
	\subfloat[][]{\includegraphics[width=0.45\textwidth]{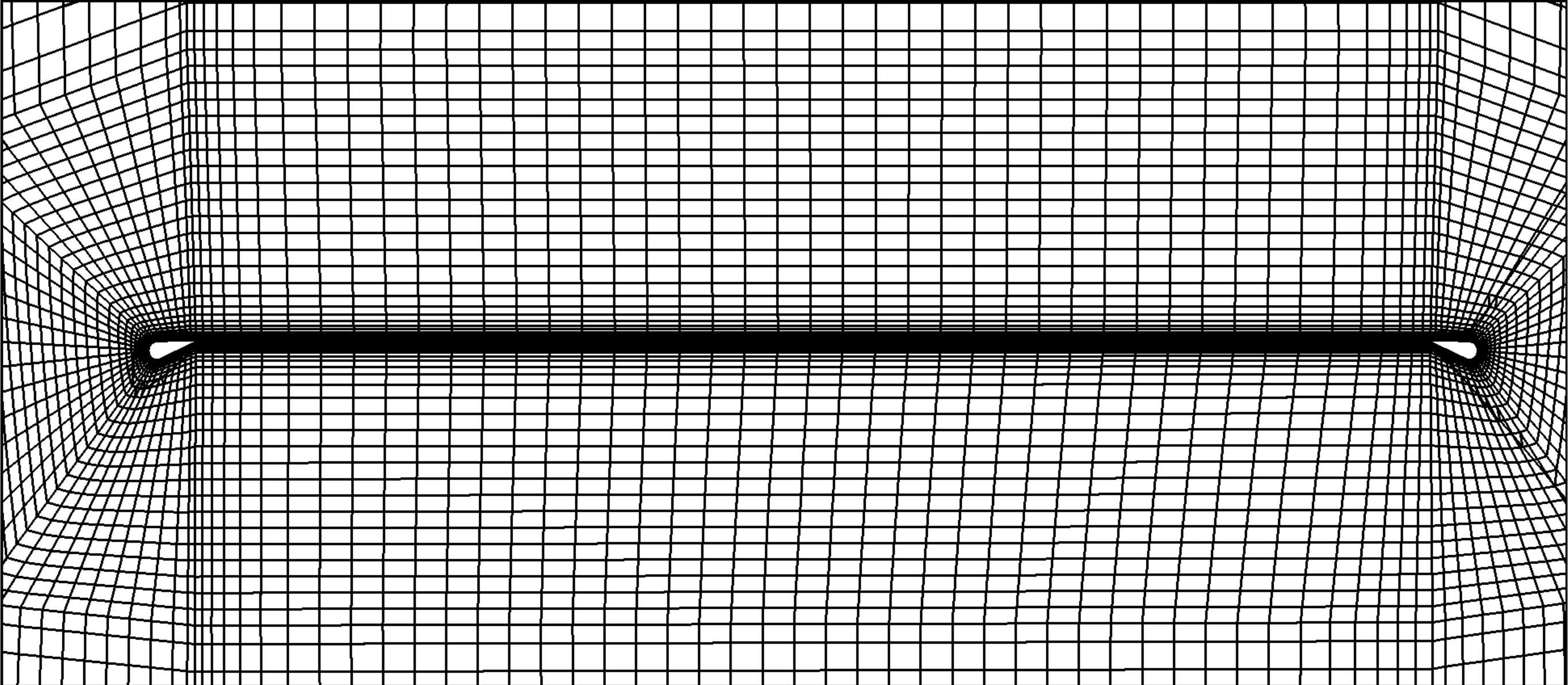}\label{membrane_domainb}}
	\caption{Three-dimensional computational set-up for fluid-membrane interaction: (a) schematic diagram of the computational domain and the boundary conditions and (b) representative mesh distribution in (Y,Z)-plane in the fluid domain.}
	\label{membrane_domain}
\end{figure}

\refFig{membrane_domain} \subref{membrane_domaina} depicts the computational domain and boundary conditions for a 3D flexible membrane immersed in an unsteady flow with a fixed angle of attack $\alpha$. The length, width and height of the computational domain are all set to $50c$. A stream of oncoming flow with uniform velocity of $\boldsymbol{u}^f=(U_{\infty}, 0, 0)$ enters the domain through the inlet boundary $\Gamma_{\rm{in}}$. The slip-wall boundary condition is applied on four side boundaries ($\Gamma_{\rm{slip}}$). The boundary condition on the membrane surface ($\Gamma_{\rm{no-slip}}$) is set to the no-slip boundary condition. The outlet boundary $\Gamma_{\rm{out}}$ has a traction-free boundary condition. In the numerical simulation, all the degrees of freedom of the rigid frame are fixed, and the passive deformation of the flexible membrane is allowed under aerodynamic loads.

To evaluate the aerodynamic characteristics of the membrane wing, we integrate the surface traction for the first layer of elements on the membrane surface to obtain the instantaneous lift, drag and normal force coefficient, which are defined below:
\begin{align}
C_L = \frac{1}{\frac{1}{2} \rho^f U_{\infty}^2 S} \int_{\Gamma} (\boldsymbol{\bar{\sigma}}^f \cdot \boldsymbol{n}) \cdot \boldsymbol{n}_z \rm{d \Gamma}
\label{CL} 
\end{align}
\begin{align}
C_D = \frac{1}{\frac{1}{2} \rho^f U_{\infty}^2 S} \int_{\Gamma} (\boldsymbol{\bar{\sigma}}^f \cdot \boldsymbol{n}) \cdot \boldsymbol{n}_x \rm{d \Gamma}
\label{CD} 
\end{align}
\begin{align}
C_N = \frac{1}{\frac{1}{2} \rho^f U_{\infty}^2 S} \int_{\Gamma} (\boldsymbol{\bar{\sigma}}^f \cdot \boldsymbol{n}) \cdot \boldsymbol{n}_c \rm{d \Gamma}
\label{CN} 
\end{align}
where $U_{\infty}$ is the freestream velocity and $\rho^f$ represents the air density. The area of the membrane surface is denoted as $S=bc$. $\boldsymbol{n}_x$ and $\boldsymbol{n}_z$ are the projection of the unit normal $\boldsymbol{n}$ to the membrane surface on the $X$-axis and $Z$-axis, respectively. The unit normal vector $\boldsymbol{n}_c$ is perpendicular to the membrane chord. $\boldsymbol{\bar{\sigma}}^f$ denotes the fluid stress tensor. The deformation of the flexible membrane is mainly driven by the pressure acting on its surface, and the pressure coefficient is given as:
\begin{align}
C_p = \frac{p-p_\infty}{\frac{1}{2} \rho^f U_{\infty}^2}
\label{Cp} 
\end{align}
where $p$ and $p_{\infty}$ represent the local pressure and the far-field pressure, respectively.

\subsection{Convergence study and verification}
To choose a proper mesh with sufficient resolution for the following numerical simulation, we conduct a mesh convergence study for the 3D flexible membrane by designing three sets of meshes namely M1, M2 and M3. The unstructured finite element is adopted to discretize the 3D fluid domain and the structure domain is discrete by the structured finite element. These three sets of meshes consist of 341821, 823864 and 1304282 eight-node brick elements in the fluid domain and the corresponding element numbers in the structure domain are 160, 228 and 352, respectively. A stretching ratio of $\Delta y_{j+1}/ \Delta y_j = 1.25$ is set within the boundary layer mesh to maintain $y^+ < 1.0$. The representative mesh distribution in the (Y,Z)-plane at the mid-chord position is presented in \reffig{membrane_domain} \subref{membrane_domainb}. The non-dimensional time step is set to $\Delta t U_{\infty}/c = 0.0364$ and the Reynolds number is given as $Re=24300$ with a corresponding freestream velocity of $U_{\infty}=5$ m/s in the numerical simulation. 

The flexible membrane at an angle of attack of $\alpha = 15^\circ$ with obvious vortex shedding phenomenon is considered for the mesh convergence study. \refTab{3dmeshforce} summarizes the aerodynamic forces, the vortex shedding frequency statistics and the structural dynamics for the three sets of meshes. We calculate the percentage discrepancies for M1 and M2 with respect to the finest mesh M3 to evaluate the mesh convergence discrepancy. It can be observed that the discrepancy of the mean lift and mean drag is less than $1 \%$ and the maximum difference of the force fluctuation is $2.8 \%$. The discrepancy of the mean membrane deflection and its root-mean-squared value of the membrane deflection fluctuation for M2 relative to M3 is smaller than $2 \%$. The dominant frequencies of the vortex shedding process and the membrane vibration for M2 are the same as those for M3. Thus, we choose mesh M2 as the reference mesh for further validation study due to its adequate resolution.

A time step convergence study is further conducted to select a proper time step for the numerical simulations. Three non-dimensional time steps of $\Delta t U_{\infty}/c$=0.0728, 0.0364 and 0.0182 are chosen for the time step convergence study. The statistical data related to the aerodynamic forces and structural vibrations for different time steps is summarized in \reftab{3dtimeforce}. The percentage discrepancies are calculated by using the results of $\Delta t U_{\infty}/c = 0.0182$ as the reference. It can be seen that the discrepancies are less than $3\%$ for the non-dimensional time step of  $\Delta t U_{\infty}/c$=0.0364. Thus, a non-dimensional time step of 0.0364 is selected to validate the coupled framework and investigate the aeroelasticity of the flexible membranes at different angles of attack.

\begin{table}[H]
	\centering
	\caption{Mesh convergence of a 3D rectangular flexible membrane wing at $Re$=24300 and $\alpha=15^\circ$ with non-dimensional time step $\Delta t U_{\infty}/c = 0.0364$. The percentage discrepancies are calculated by using M3 results as the reference.}{\label{3dmeshforce}}
	\begin{tabular}{ccccccc}
		\toprule  
		Mesh & M1    &    M2    &     M3 \\[3pt]
		\midrule 
		Fluid elements & 341 821  & 823 864 & 1 304 282  \\
		Structural elements & 160 & 228 & 352 \\
		Mean lift $\overline{C}_L$ & 0.9022 (0.49$\%$) & 0.8902 (0.85$\%$)& 0.8978\\
		Mean drag $\overline{C}_D$ & 0.2289 (2.97$\%$)& 0.2346 (0.55$\%$)& 0.2359\\
		r.m.s. lift fluctuation ${C_L^{\prime}}^{rms}$ & 0.0788 (16.17$\%$) &0.0928 (1.28$\%$)& 0.0940   \\
		r.m.s. drag fluctuation ${C_D^{\prime}}^{rms}$ &0.0289 (10.25$\%$)&0.0313 (2.80$\%$)& 0.0322   \\
		Dominant shedding frequency $f^{vs}c/U_{\infty}$ & 0.9668  (2.70$\%$)& 0.9937 (0$\%$)& 0.9937  \\
		Maximum mean deflection $\overline{\delta}_n^{max}/c$ & 0.03391 (1.14$\%$) & 0.03415 (0.44$\%$) & 0.03430 \\
		Maximum r.m.s. deflection fluctuation ${\delta_n^{\prime}}^{rms}$  & 0.000979 (2.88$\%$) & 0.000994 (1.39$\%$) & 0.001008  \\
		Dominant vibration frequency $f^{s}c/U_{\infty}$ & 0.9668  (2.70$\%$)& 0.9937 (0$\%$)& 0.9937 \\
		\bottomrule 
	\end{tabular}
\end{table}

\begin{table}[H]
	\centering
	\caption{Convergence of the aerodynamic forces and the dynamic responses at different time steps for a 3D rectangular flexible membrane wing at $Re$=24300 and $\alpha=15^\circ$ based on mesh M2. The percentage discrepancies are calculated by using $\Delta t U_{\infty}/c = 0.0182$ results as the reference.}{\label{3dtimeforce}}
	\begin{tabular}{ccccccc}
		\toprule  
		Non-dimensional time step $\Delta t U_{\infty}/c$ & $0.0728$    &    $0.0364$    &     $ 0.0182$ \\[3pt]
		\midrule 
		Mean lift $\overline{C}_L$  & 0.8336 ($7.8386 \%$) & 0.8902 ($1.58 \%$)&  0.9045  \\
		Mean drag $\overline{C}_D$  & 0.2223 ($4.6332 \%$)& 0.2346 ($0.6435 \%$)& 0.2331 \\
		r.m.s. lift fluctuation ${C_L^{\prime}}^{rms}$ & 0.0736  ($ 22.20 \%$) &0.0928 ($1.90 \%$)& 0.0946 \\
		r.m.s. drag fluctuation ${C_D^{\prime}}^{rms}$ & 0.0252 ($20.75 \%$)&0.0313 ($1.57 \%$)& 0.0318  \\
		Dominant shedding frequency $f^{vs}c/U_{\infty}$ & 0.9399 ($5.41 \%$)& 0.9937 ($0 \%$)&  0.9937 \\
		Maximum mean deflection $\overline{\delta}_n^{max}/c$ & 0.0351  ($1.45\%$) & 0.03415 ($1.30 \%$) & 0.0346  \\
		Maximum r.m.s. deflection fluctuation ${\delta_n^{\prime}}^{rms}$  & 0.000830 ($18.79\%$) & 0.000994 ($2.74 \%$) & 0.001022  \\
		Dominant vibration frequency $f^{s}c/U_{\infty}$ & 0.9399 ($5.41 \%$) & 0.9937 ($0 \%$)& 0.9937  \\
		\bottomrule 
	\end{tabular}
\end{table}

\begin{figure}[H]
	\centering
	\subfloat[][]{\includegraphics[width=0.5\textwidth]{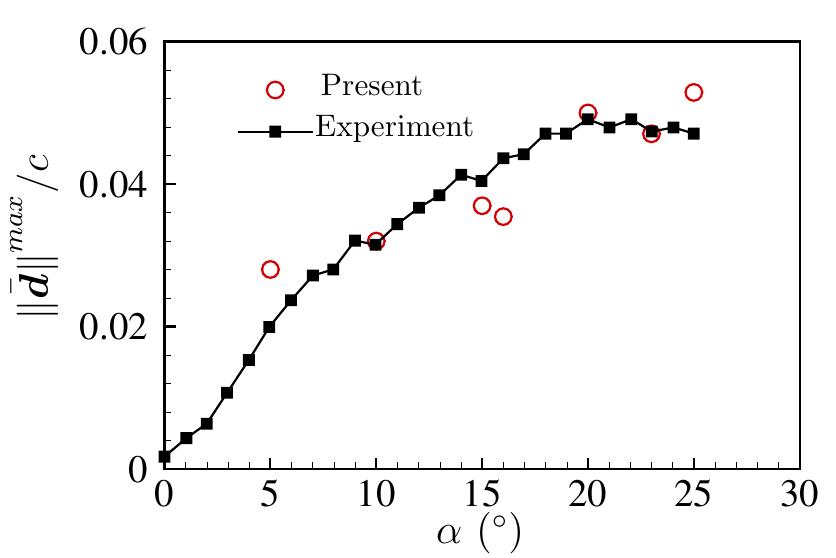}\label{averagediscna}}
	\subfloat[][]{\includegraphics[width=0.5\textwidth]{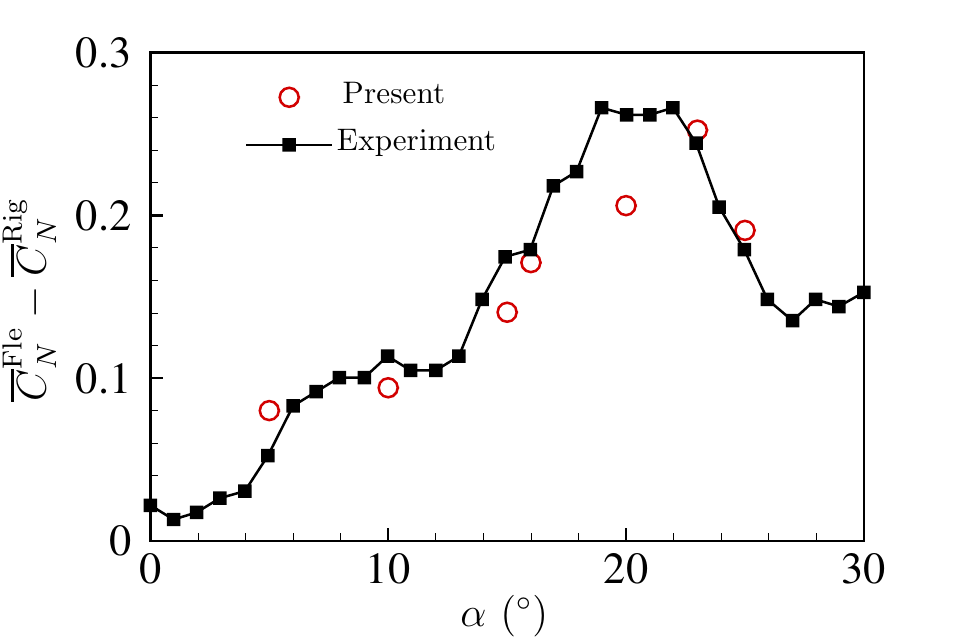}\label{averagediscnb}}
	\\
	\subfloat[][]{\includegraphics[width=0.5\textwidth]{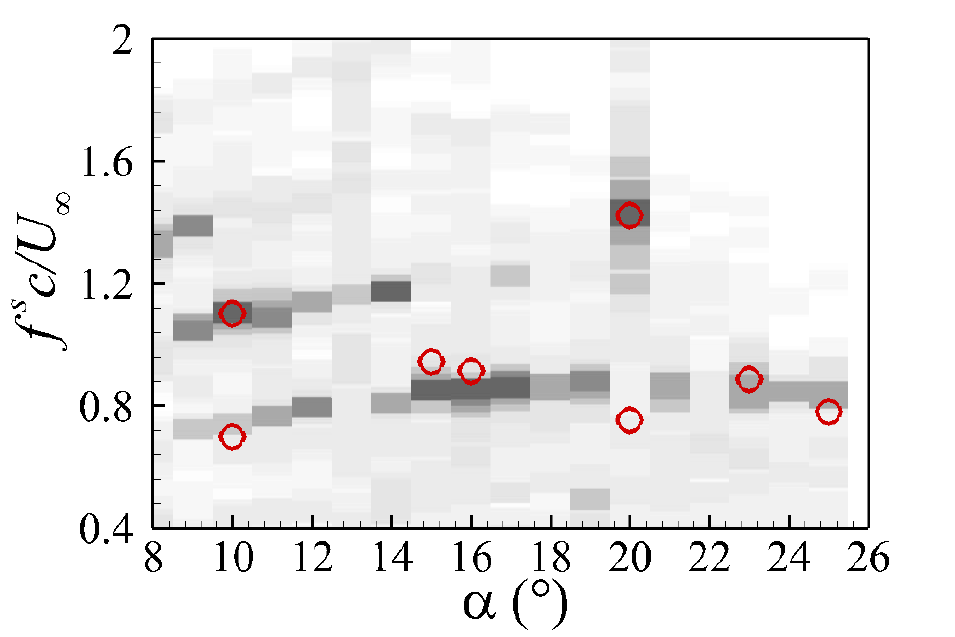}\label{averagediscnc}}
	\subfloat[][]{\includegraphics[width=0.5\textwidth]{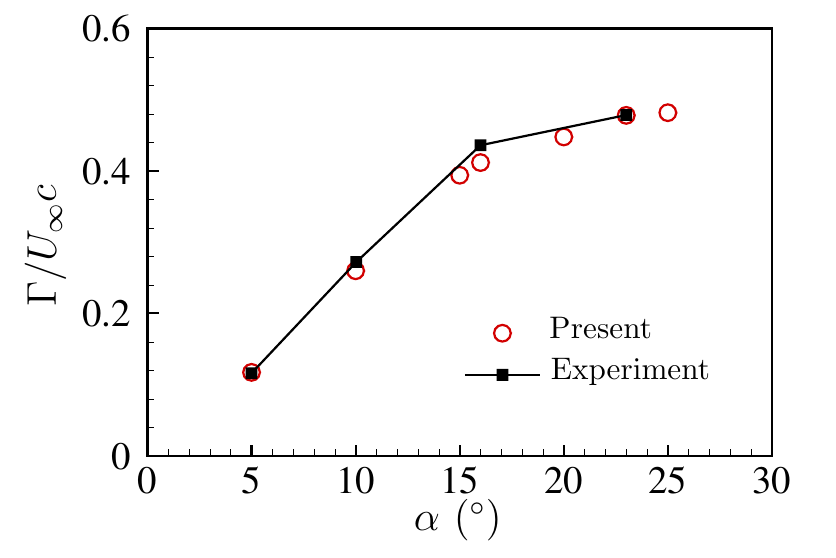}\label{averagediscnd}}
	\caption{Comparison between the present simulations and the data obtained from experiments \cite{rojratsirikul2011flow,rojratsirikul2010unsteady} for: (a) the magnitude of time-averaged normalized maximum membrane deformation ($\bar{\left\|  \bm d \right\|}^{{max}}/c$), (b) the time-averaged normal force coefficient  difference ($\overline{C}_N^{\rm{Fle}}-\overline{C}_N^{\rm{Rig}}$) between the flexible membrane wing and rigid wing, (c) the membrane vibration frequency spectra at the location with maximum standard deviation of the membrane deflection, and (d) normalized circulation ($\Gamma / U_{\infty} c$) of the vortices at the wingtip on a plane normal to the freestream direction.}
	\label{averagediscn}
\end{figure}

\begin{figure}[H]
	\centering
	\subfloat[][]{\includegraphics[width=0.38\textwidth]{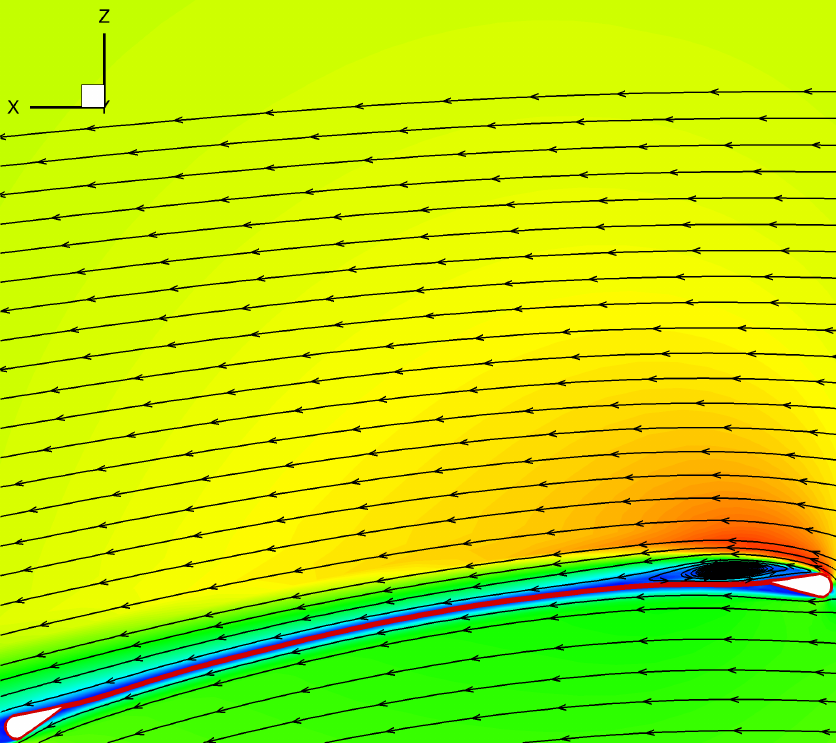}}
	\quad
	\subfloat[][]{\includegraphics[width=0.38\textwidth]{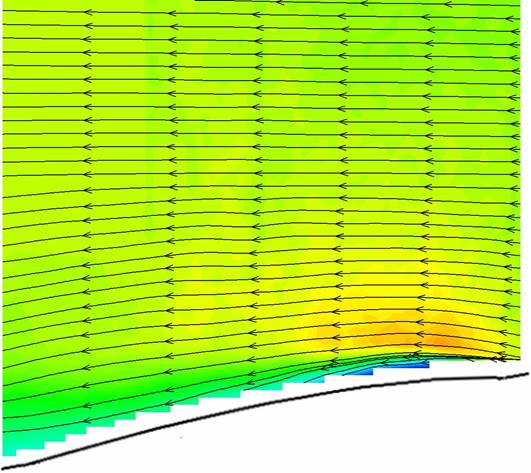}}
	\\
	\subfloat[][]{\includegraphics[width=0.38\textwidth]{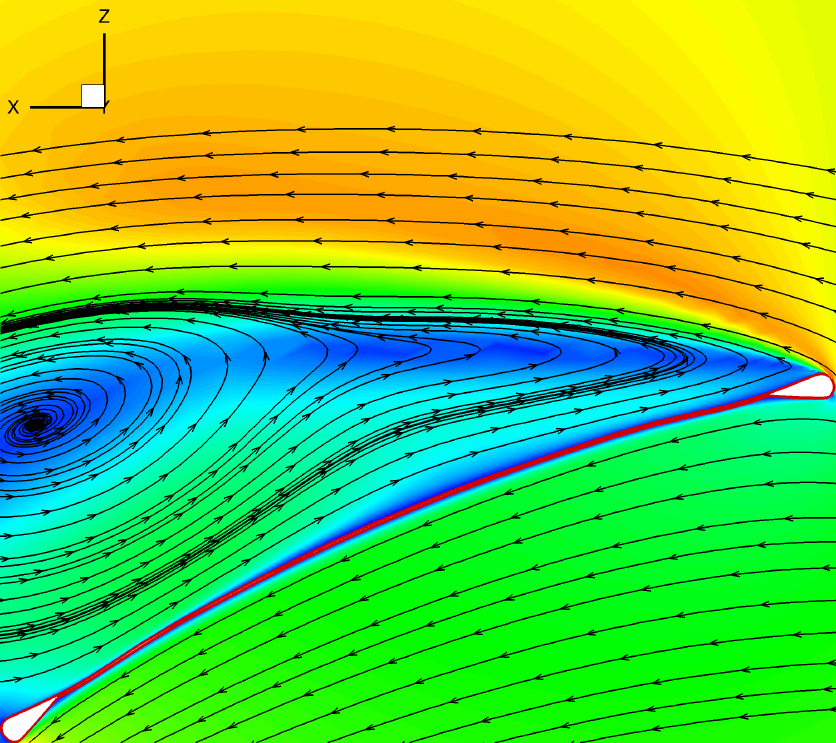}}
	\quad
	\subfloat[][]{\includegraphics[width=0.38\textwidth]{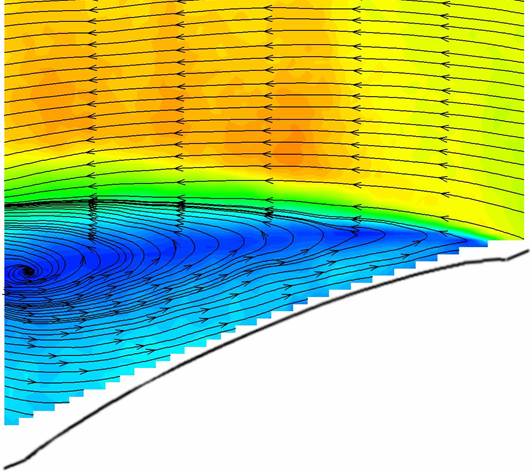}}
	\\
	\includegraphics[width=0.5\textwidth]{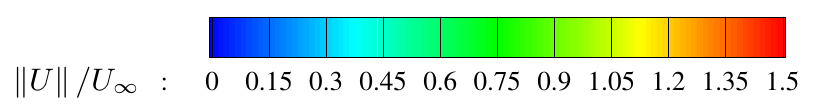}
	\caption{Time-averaged normalized velocity magnitude and streamlines on the mid-span plane obtained from: (a) present simulation at $\alpha=10^\circ$, (b) experiment \cite{rojratsirikul2011flow} at $\alpha=10^\circ$, (c) present simulation at $\alpha=23^\circ$, (d) experiment \cite{rojratsirikul2011flow} at $\alpha=23^\circ$.}
	\label{vmcom}
\end{figure}

The 3D membrane immersed in the unsteady flow at different angles of attack is simulated to compare against the experimental data \cite{rojratsirikul2011flow,rojratsirikul2010unsteady} for validation purpose. \refFig{averagediscn} presents the magnitude of the mean maximum membrane deformation, the mean normal force coefficient difference between the flexible membrane and the rigid wing, the vibration frequency spectra and the normalized circulation of the wingtip vortices, respectively. The frequency spectra presented in \reffig{averagediscn} (c) is obtained from the experimental results \cite{rojratsirikul2011flow}. The color of the frequency spectra changing from white to black corresponds to the increasing intensity of the frequency. The red circle in \reffig{averagediscn} (c) represents the dominant vibration frequency computed from our numerical simulations. It can be observed that the overall trend is well predicted by our numerical simulations. In figure \ref{vmcom}, we also compare the flow field and the streamlines around the membrane to the available results of the experiment at  $\alpha=10^\circ$ and  $\alpha=23^\circ$, respectively. Good agreements can be seen for the distribution of the mean velocity magnitude and the size as well as the location of the separation flow.

\section{Results and discussion} \label{results}
In this section, we present the coupled fluid-membrane dynamics at three different angles of attack with apparent separated flows. To identify the dominant aeroelastic modes, the global mode decomposition method based on the FMD technique is adopted to decompose the physical variables into frequency-ranked modes and extract these modes at the selected energetic frequencies. The flow features and the decomposed modes of rigid wings, rigid cambered wings and flexible membrane wings are compared. Finally, we explore the role of flexibility and reveal the aeroelastic mode selection mechanism.

\subsection{Membrane aeroelasticity}  \label{comfr}
\begin{figure}[H]
	\centering 
	\subfloat[][]{\includegraphics[width=0.45\textwidth]{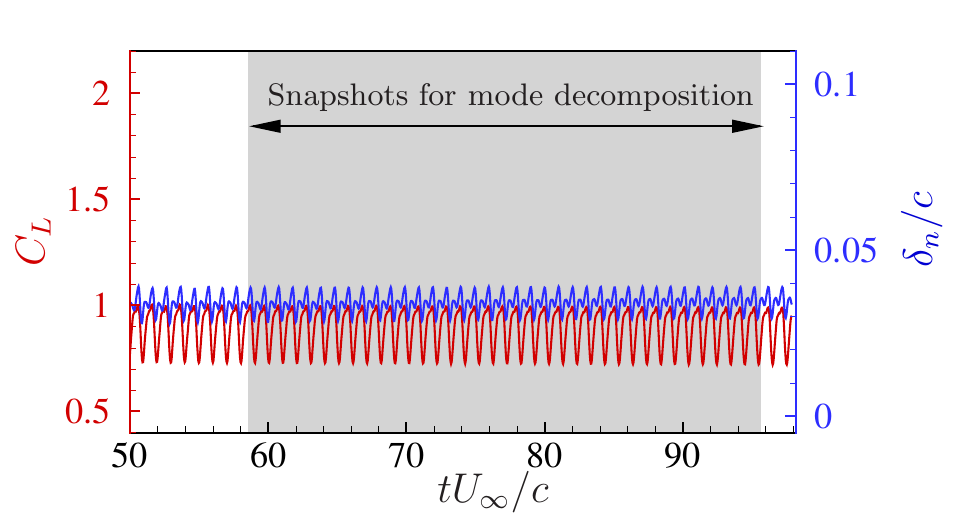}\label{time_historya}}
	\subfloat[][]{\includegraphics[width=0.45\textwidth]{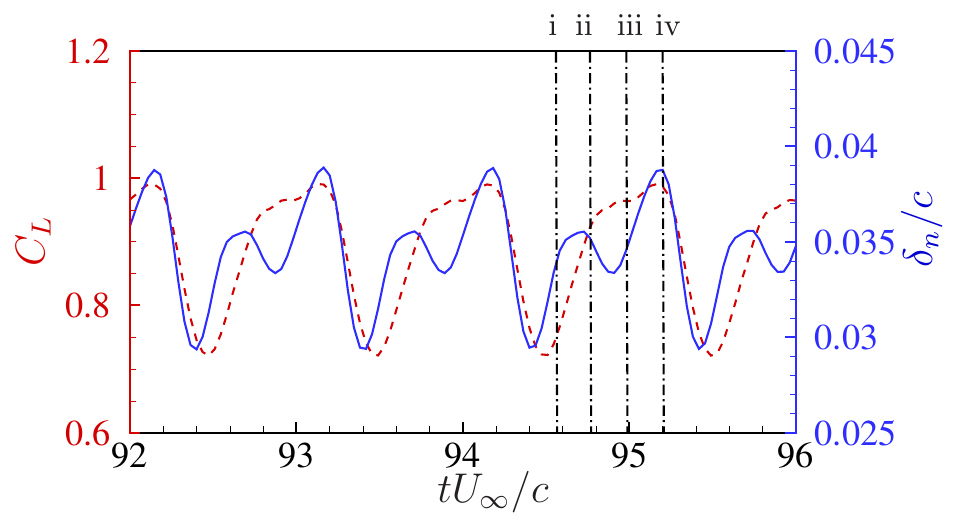}\label{time_historyb}}
	\\
	\subfloat[][]{\includegraphics[width=0.45\textwidth]{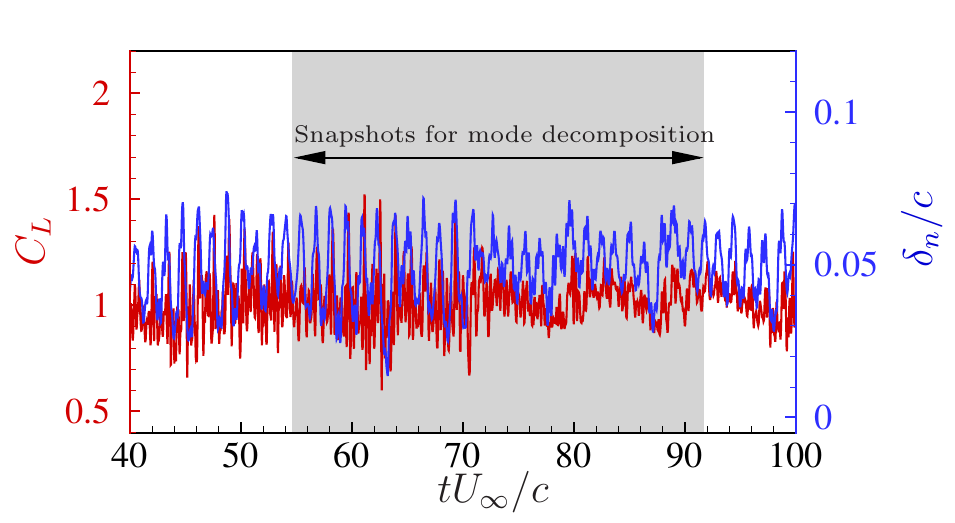}\label{time_historyc}}
	\subfloat[][]{\includegraphics[width=0.45\textwidth]{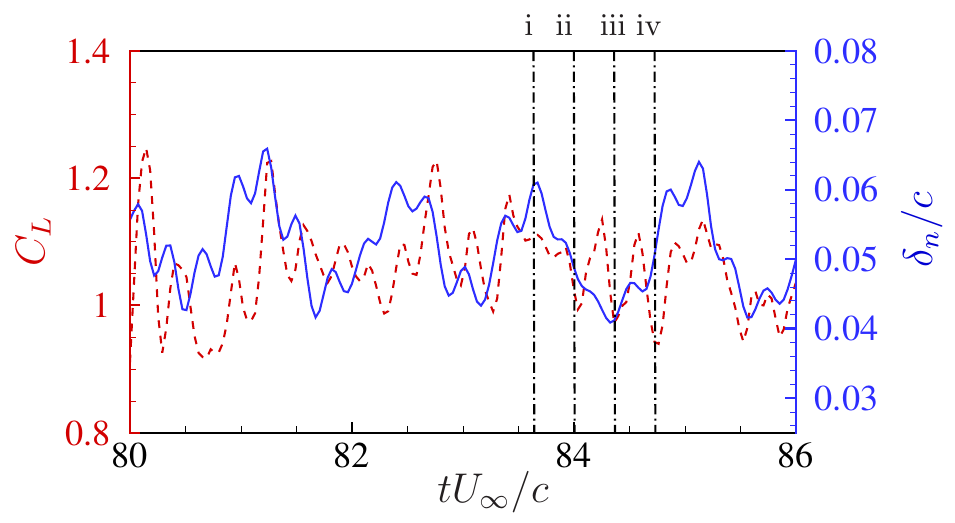}\label{time_historyd}}
	\\
	\subfloat[][]{\includegraphics[width=0.45\textwidth]{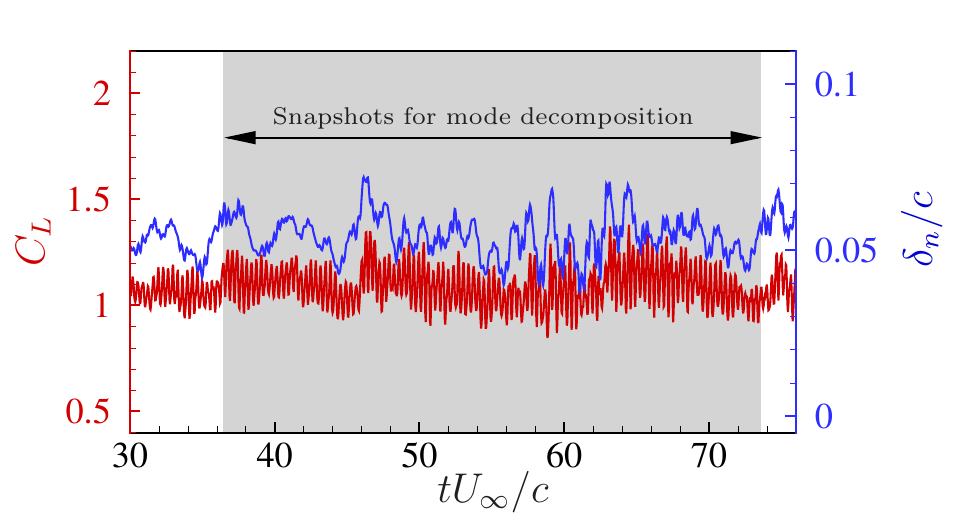}\label{time_historye}}
	\subfloat[][]{\includegraphics[width=0.45\textwidth]{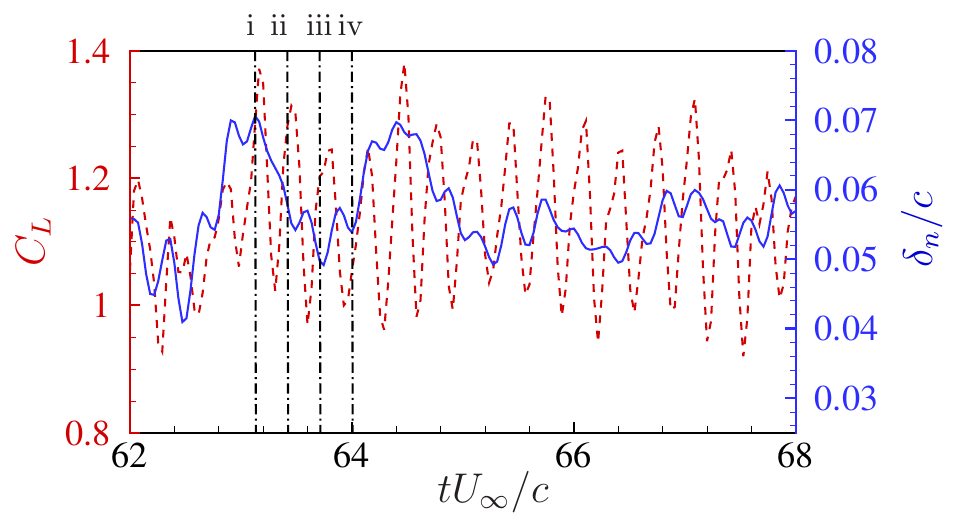}\label{time_historyf}}
	\caption{Time histories of the lift coefficient $C_L$ and the normalized membrane displacement $\delta_n/c$ at the membrane center for $U_\infty$=5m/s at $\alpha$= (a,b) 15$^\circ$, (c,d) 20$^\circ$ and (e,f) 25$^\circ$.} 
	\label{time_history}
\end{figure}

\begin{figure}[H]
	\centering 
	\subfloat[][]{\includegraphics[width=0.3\textwidth]{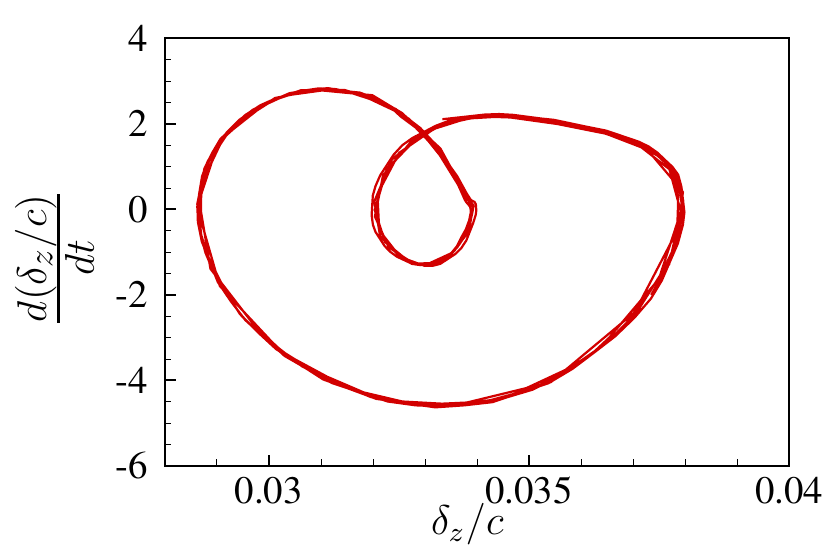}\label{z_va}}
	\subfloat[][]{\includegraphics[width=0.3\textwidth]{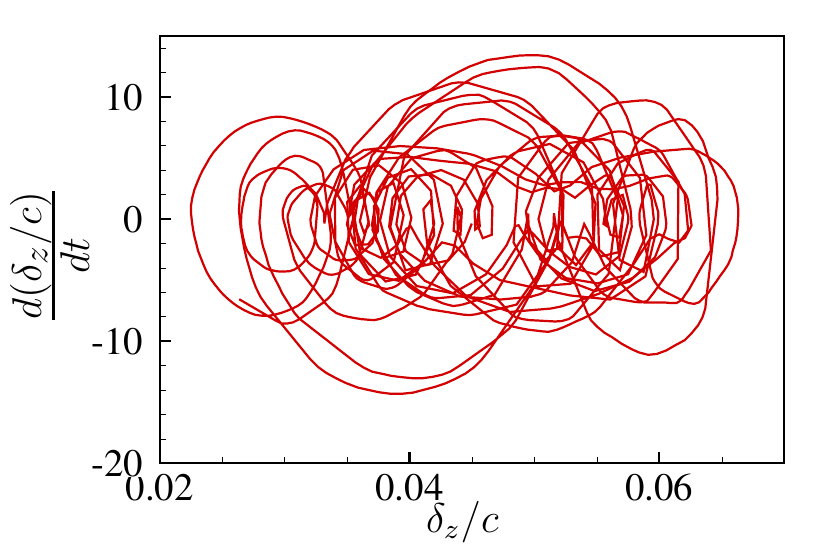}\label{z_vb}}
	\subfloat[][]{\includegraphics[width=0.3\textwidth]{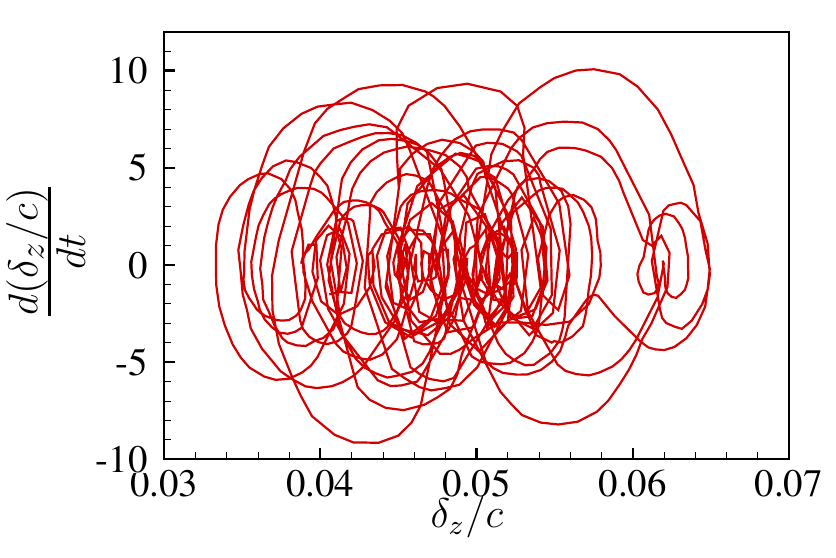}\label{z_vc}}
	\caption{Phase portrait of the membrane center for the 3D flexible membrane at $\alpha$= (a) 15$^\circ$, (b) 20$^\circ$ and (c) 25$^\circ$.}
	\label{z_v}
\end{figure}

Before proceeding with aeroelastic mode decomposition, an overview of the membrane aeroelastic responses is displayed to provide a brief impression of the fluid-membrane interaction problems. In \reffigs{time_history} (a,c,e), we summarize the time histories of the lift coefficient and the normalized membrane displacement normal to the chord at the membrane center over a wide time range for the flexible membrane at three different angles of attack. The gray shaded region in the plots represents the time range selected to collect the snapshots for the mode decomposition, which will be discussed in \refse{mode_decomposition}. It can be seen from \reffig{time_history} that the aeroelastic responses of the flexible membrane at $\alpha=15^\circ$ show almost periodic dynamics. The dominant frequencies of the lift coefficient and the membrane displacement at the center are synchronized. As shown in the phase portrait in \reffig{z_v} \subref{z_va}, the flexible membrane displacement responses at the center exhibit a period 2 state. From the time history responses in \reffig{time_history} (c,e) and the phase portrait in \reffig{z_v} (b,c), the aeroelastic responses tend to be non-periodic at higher angles of attack.

\begin{figure}[H]
	\centering 
	\includegraphics[width=0.6\textwidth]{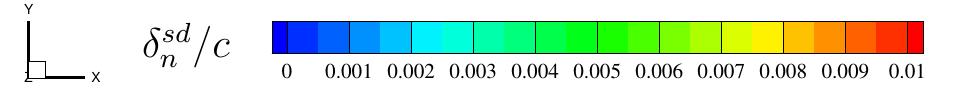}
	\\
	\subfloat[][]{\includegraphics[width=0.22\textwidth]{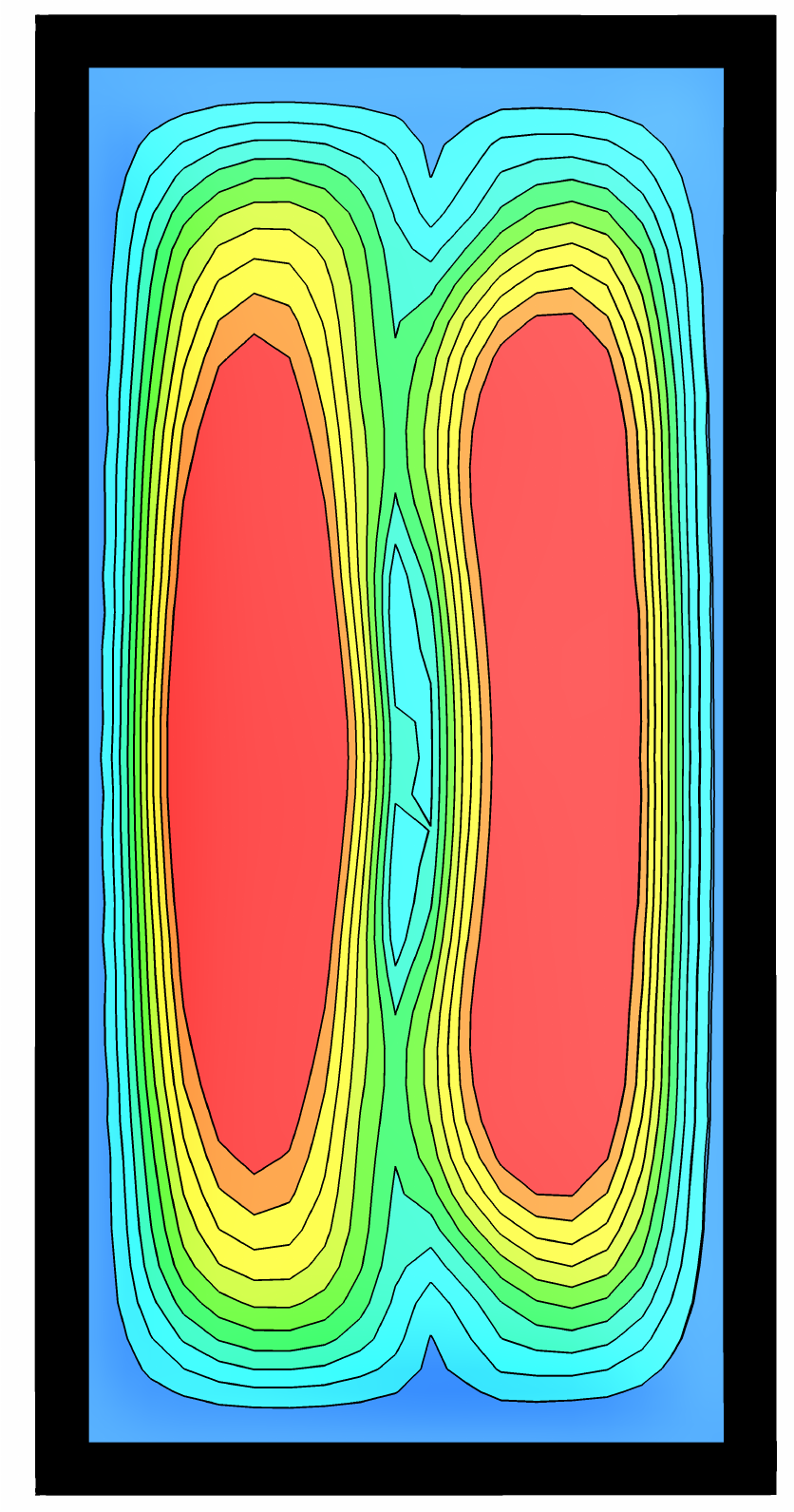}\label{zsd_aoac}}
	\quad
	\subfloat[][]{\includegraphics[width=0.22\textwidth]{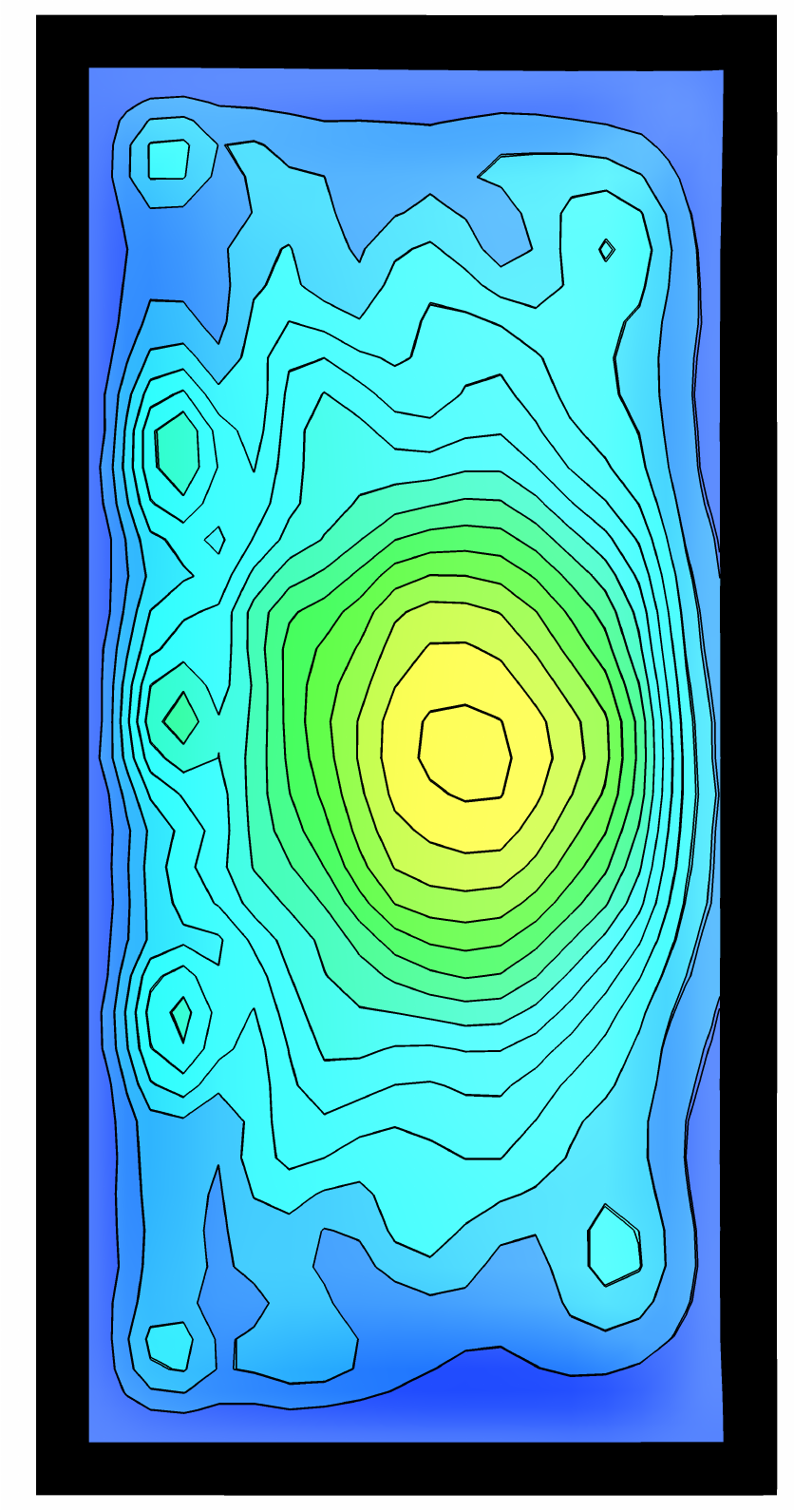}\label{zsd_aoad}}
	\quad
	\subfloat[][]{\includegraphics[width=0.22\textwidth]{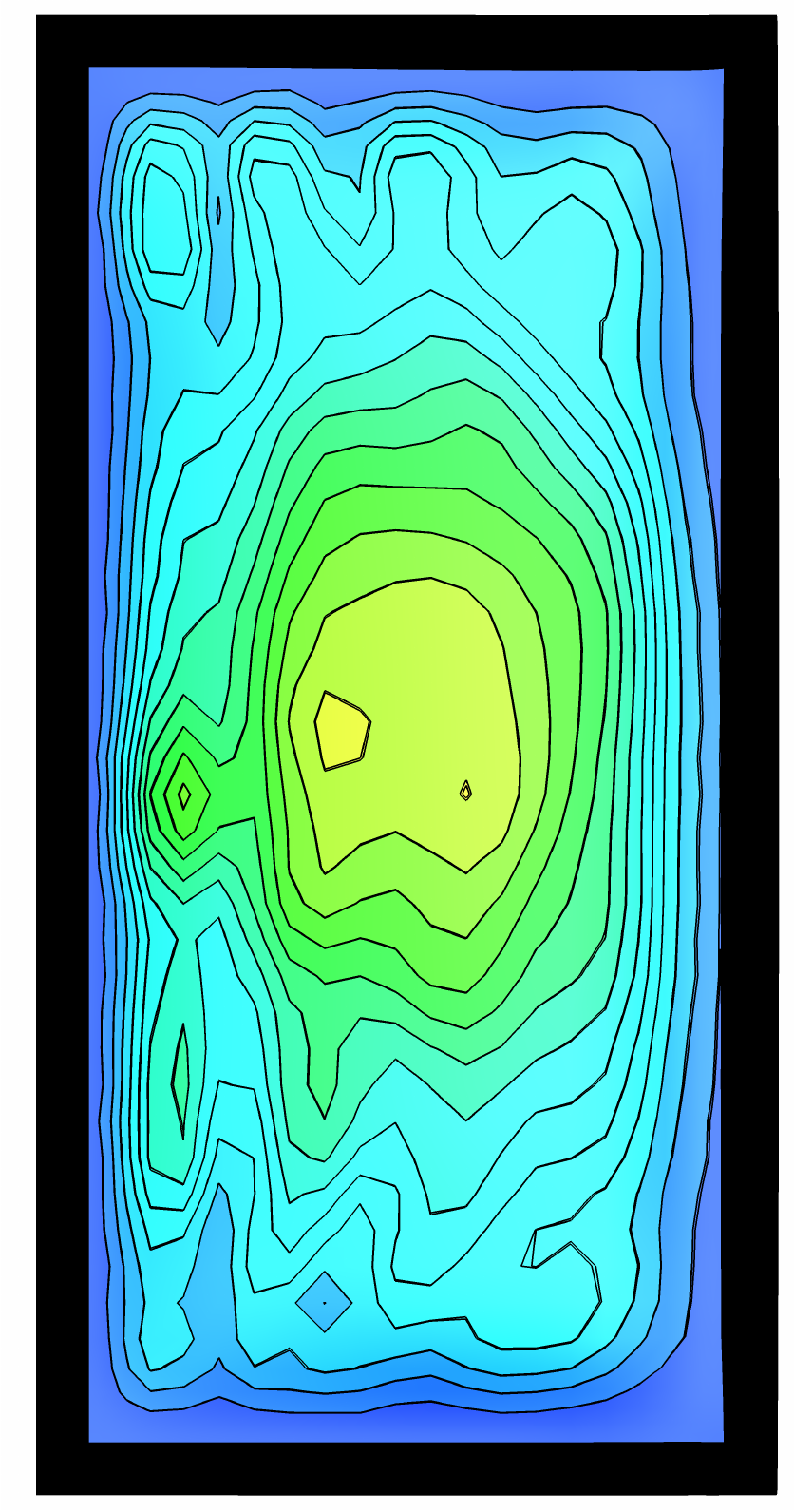}\label{zsd_aoaf}}
	\caption{Standard deviation analysis of normalized membrane displacement normal to the chord $\delta_n^{sd}/c$ for $U_\infty$=5m/s at $\alpha$= (a) 15$^\circ$, (b) 20$^\circ$ and (c) 25$^\circ$.}
	\label{zsd_aoa}
\end{figure}

To gain further insight into the membrane aeroelasticity, we compare the structural vibration characteristics and the flow features of the flexible membrane at different angles of attack. \refFig{zsd_aoa} shows the standard deviation analysis of the normalized membrane displacement $\delta_n^{sd}/c$ over several cycles to reflect the dominant structural vibration modal shapes. A typical chordwise second mode is observed for the elastic membrane at $\alpha=15^\circ$. However, the dominant structural mode of the 3D membrane wing cannot be distinctly identified from the standard deviation analysis at $\alpha=20^\circ$ and $25^\circ$. Some high-order modes can be observed near the leading edge and the wingtip. It is hard to identify and isolate these several influential modes from the time-averaged standard deviation analysis of the multi-modal mixed responses. Similarly, Tregidgo et al. \cite{tregidgo2013unsteady} found that a disturbed membrane in the gusty flow exhibited a chordwise and spanwise first mode based on the standard deviation analysis. However, a chordwise first and spanwise second mode and a chordwise first and spanwise third mode were observed from the instantaneous membrane vibration responses. The standard deviation analysis is not reliable to reflect the structural modes due to the mode overlap in a time-averaged sense.

\begin{figure}[H]
	\centering 
	\subfloat[][]{
	\includegraphics[width=0.4\textwidth]{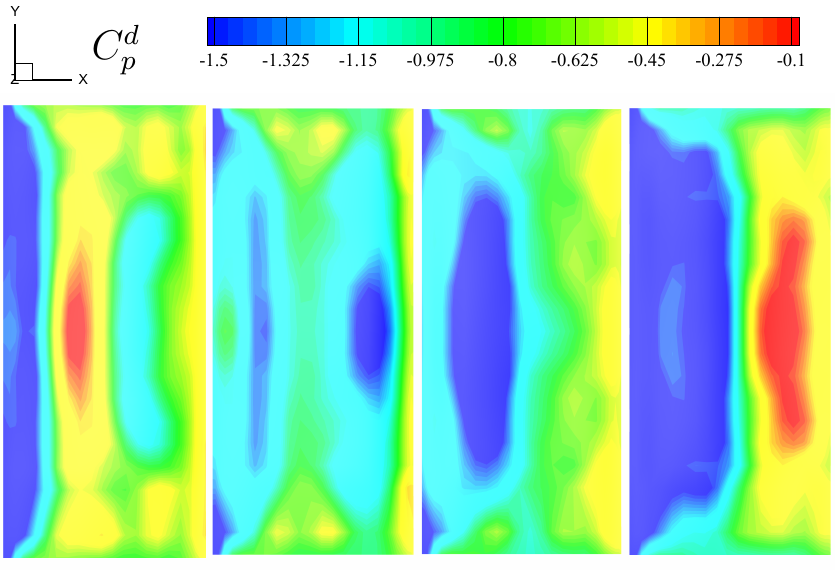}\label{flu_aoaa}}
	\quad
	\subfloat[][]{
	\includegraphics[width=0.4\textwidth]{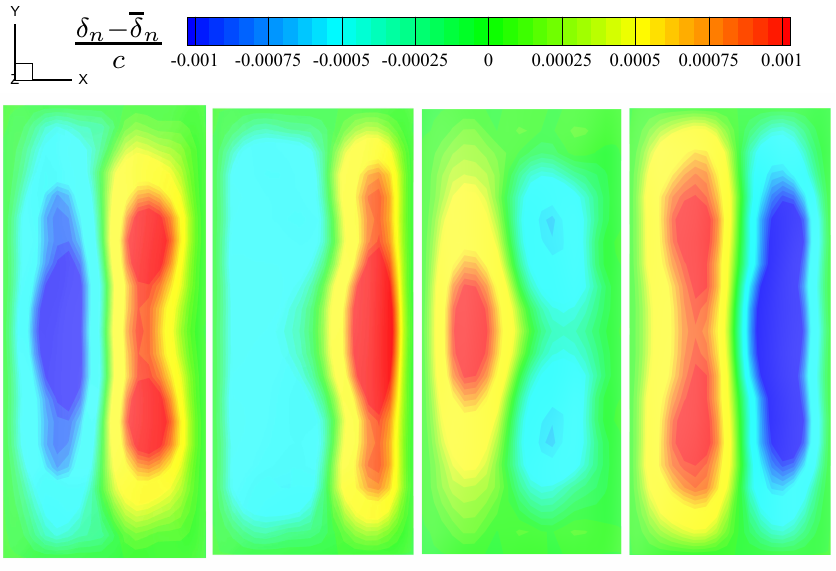}\label{flu_aoab}}
	\\
	\subfloat[][]{
	\includegraphics[width=0.4\textwidth]{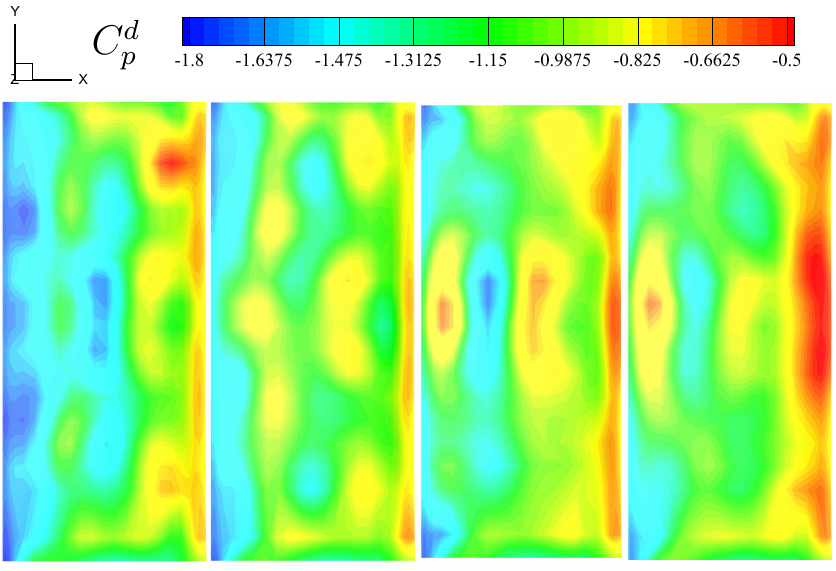}\label{flu_aoac}}
	\quad
	\subfloat[][]{
	\includegraphics[width=0.4\textwidth]{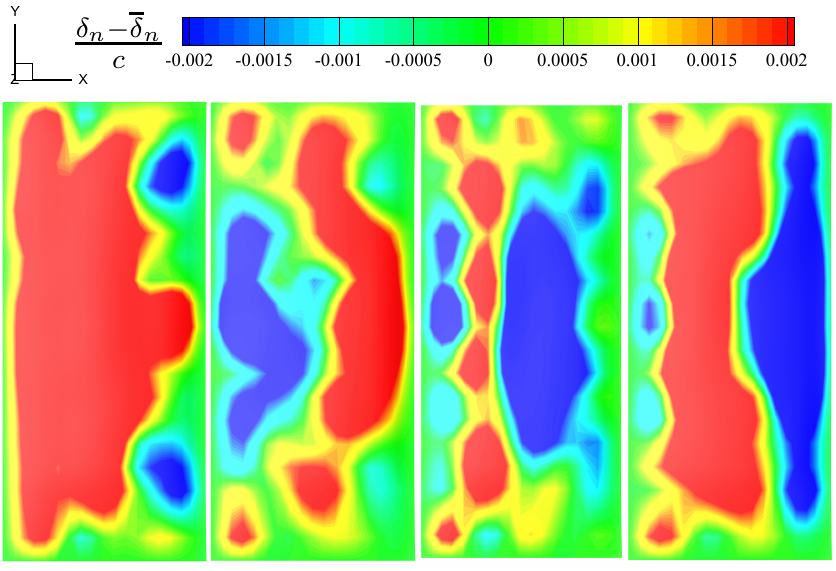}\label{flu_aoad}}
	\\
	\subfloat[][]{
	\includegraphics[width=0.4\textwidth]{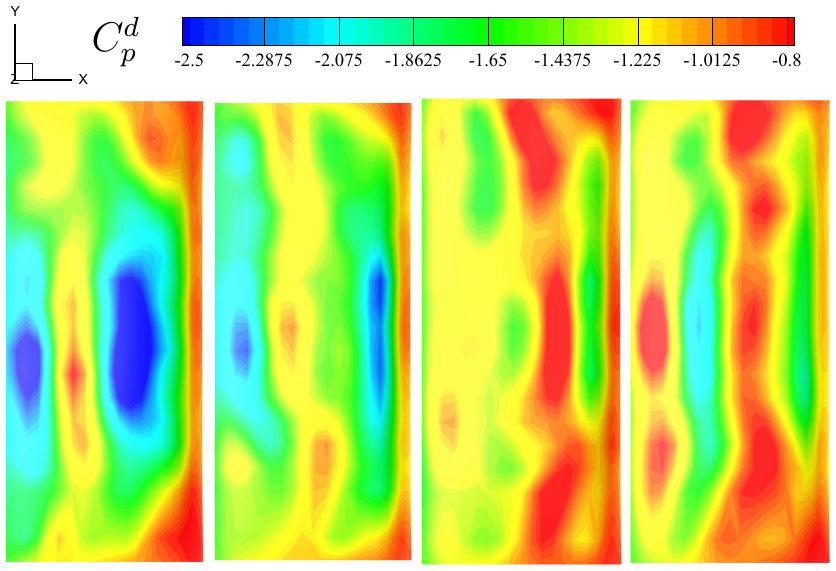}\label{flu_aoae}}
	\quad
	\subfloat[][]{
	\includegraphics[width=0.4\textwidth]{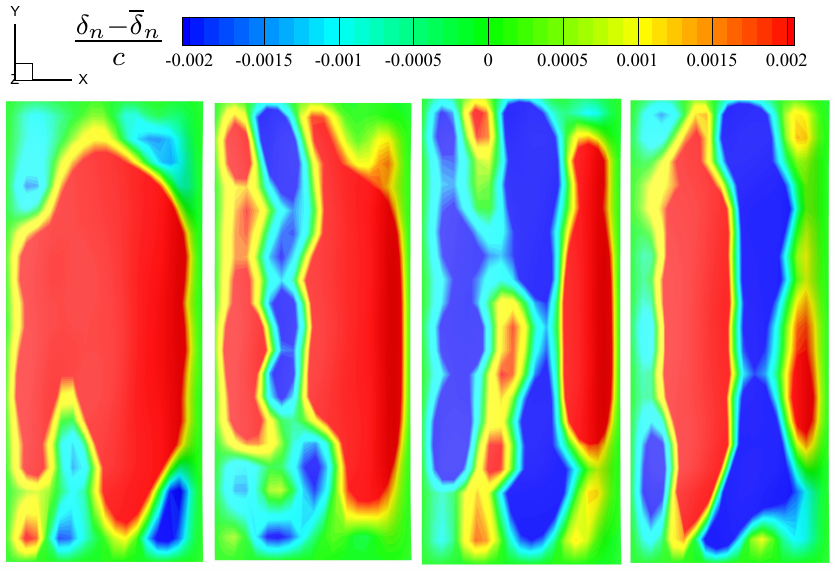}\label{flu_aoaf}}
	\caption{Flow past a 3D rectangular membrane wing: (a,c,e) pressure coefficient difference between the upper and lower surfaces and (b,d,f) fluctuation of membrane displacement at four selected instantaneous time instants plotted in \reffig{time_history} at $\alpha$= (a,b) 15$^\circ$, (c,d) 20$^\circ$ and (e,f) 25$^\circ$.}
	\label{flu_aoa}
\end{figure}

The time-varying instantaneous pressure coefficient difference between the upper and lower surfaces $C_p^{d}$ and the fluctuation contours of the membrane displacement $\frac{\delta_n - \overline{\delta}_n}{c}$ can provide an intuitive understanding of the evolution of the membrane aeroelasticity. For that purpose, we select four equispaced time instants to plot the instantaneous membrane aeroelastic dynamics. These selected time instants are indicated by the black dash dot lines in the time history plots as shown in \reffigs{time_history} (b,d,f). The instantaneous membrane aeroelastic dynamics for flexible membrane at three angles of attack are summarized in \reffig{flu_aoa}. All the pressure difference distributions on the membrane surface show complex evolutions over time and overlapping modal shapes in space. The instantaneous membrane displacement fluctuations at $\alpha=15^\circ$ exhibit an obvious chordwise second mode and varied spanwise modes. We can observe the chordwise second, third and high-order modal shapes that appear occasionally for $\alpha=20^\circ$ and $25^\circ$ as shown in \reffigs{flu_aoa} \subref{flu_aoad} and \subref{flu_aoaf}. Considering the irregular displacement fluctuation under the pressure pulsations, the dominant structural motion will be covered up in the standard deviation contours due to the time-averaged sense of the second and third modes. Thus, the traditional standard deviation analysis is not a suitable indicator to reflect the correct dominant modes of the whole membrane vibrations with temporal-spatial overlapping modal shapes.

\begin{figure}[H]
	\centering 
	\subfloat[][]{
		\includegraphics[width=0.25\textwidth]{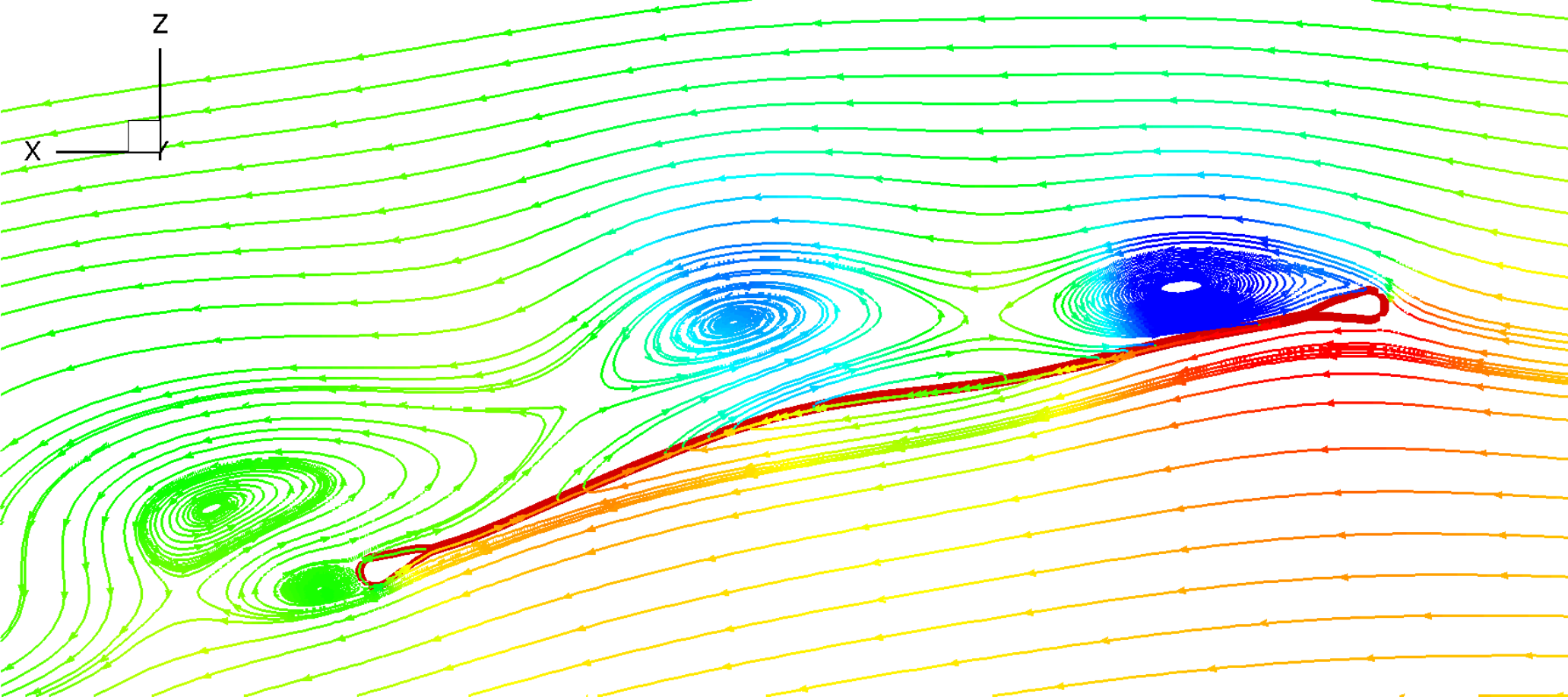}
		\includegraphics[width=0.25\textwidth]{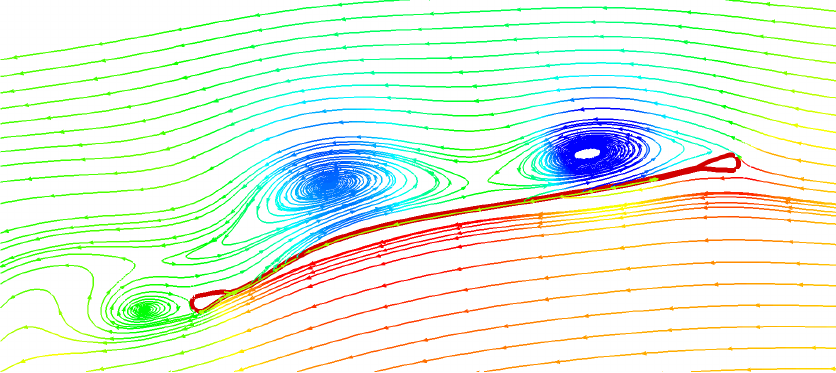}
		\includegraphics[width=0.25\textwidth]{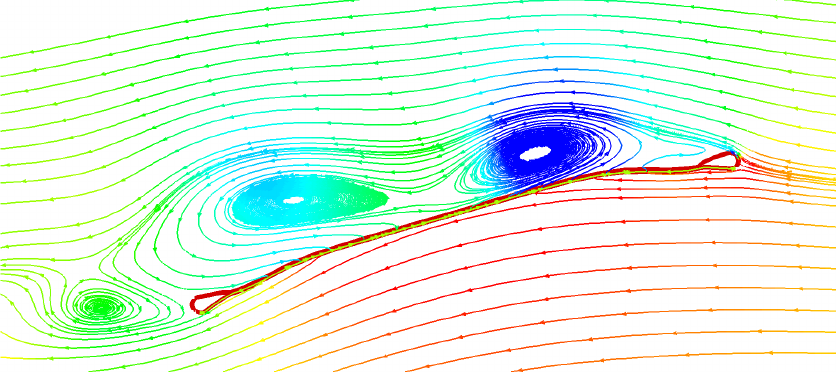}
		\includegraphics[width=0.25\textwidth]{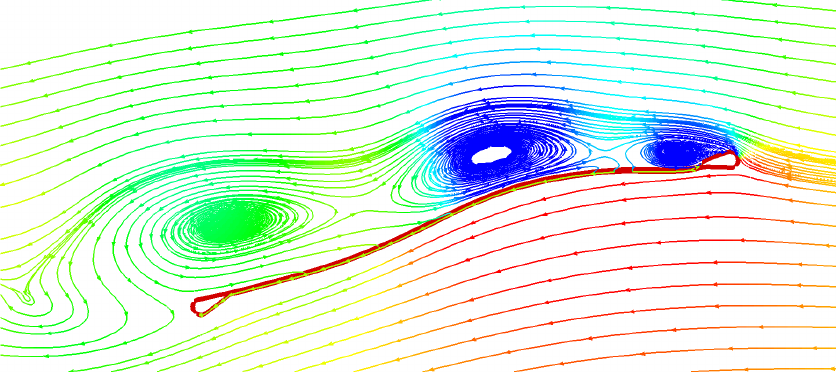}
		\label{streamline_aoac}
	}
	\\
	\subfloat[][]{
		\includegraphics[width=0.25\textwidth]{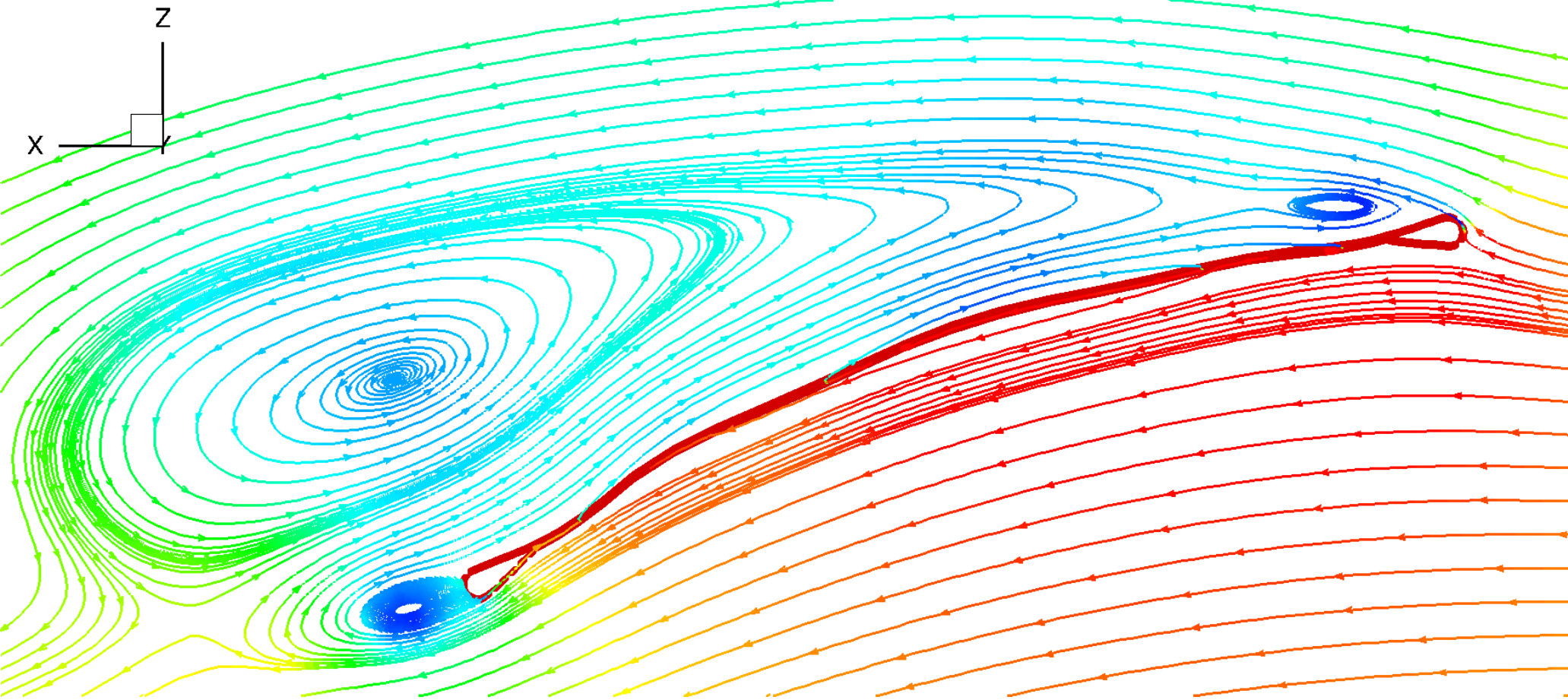}
		\includegraphics[width=0.25\textwidth]{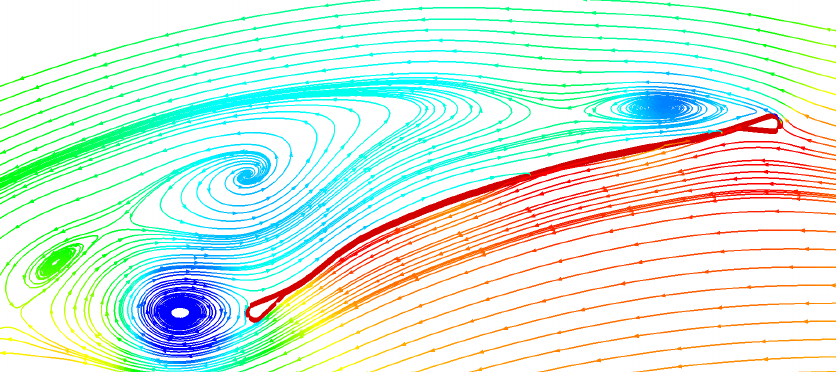}
		\includegraphics[width=0.25\textwidth]{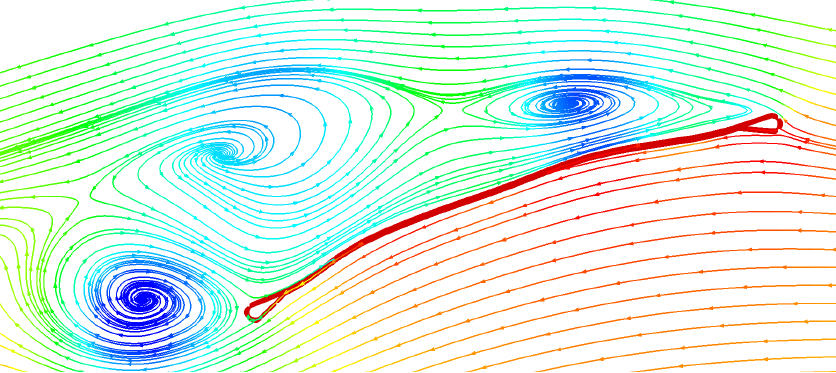}
		\includegraphics[width=0.25\textwidth]{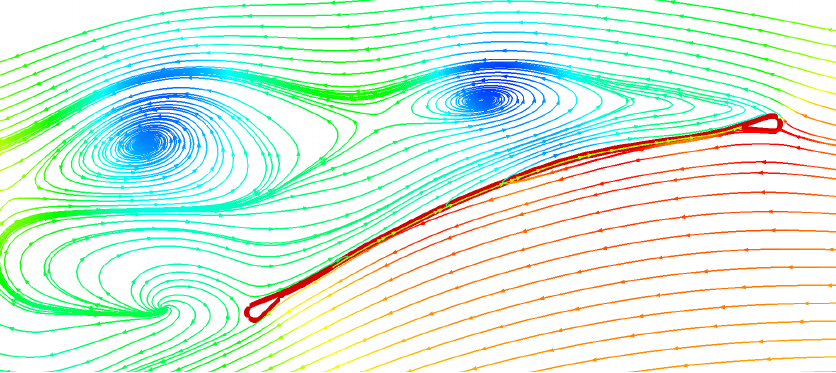}
		\label{streamline_aoad}
	}
	\\
	\subfloat[][]{
		\includegraphics[width=0.25\textwidth]{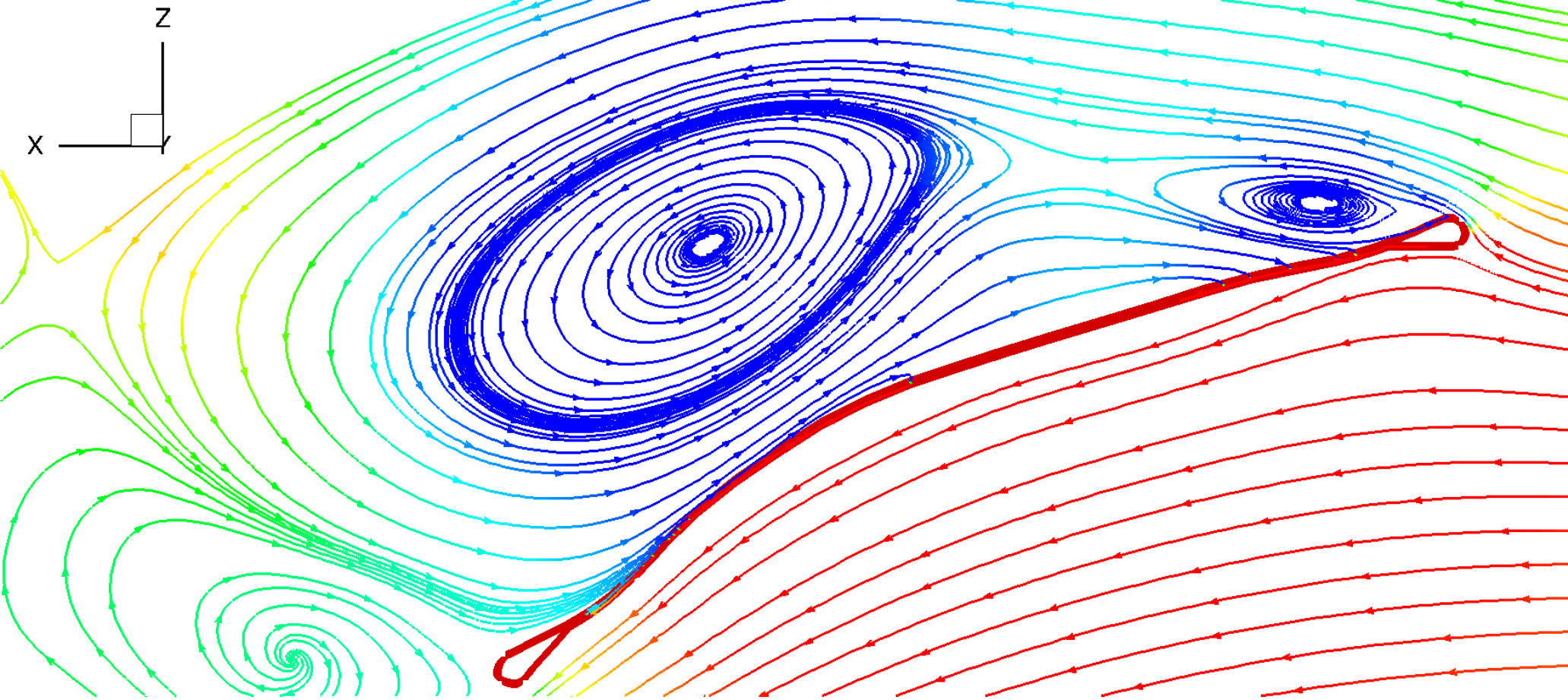}
		\includegraphics[width=0.25\textwidth]{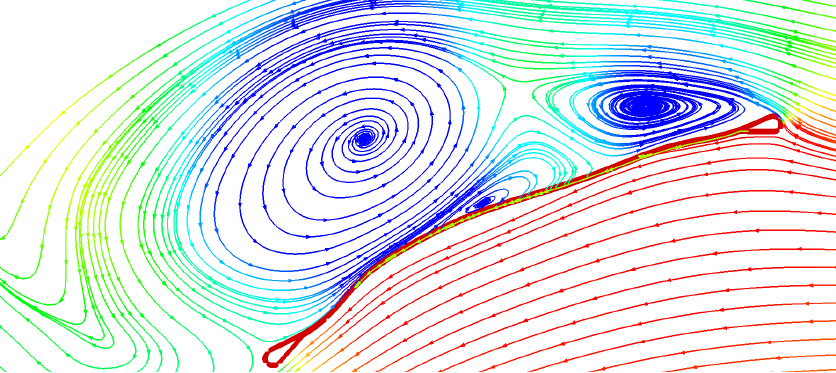}
		\includegraphics[width=0.25\textwidth]{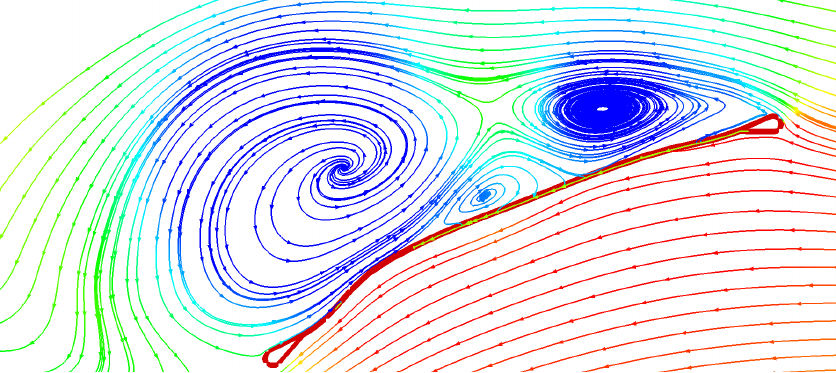}
		\includegraphics[width=0.25\textwidth]{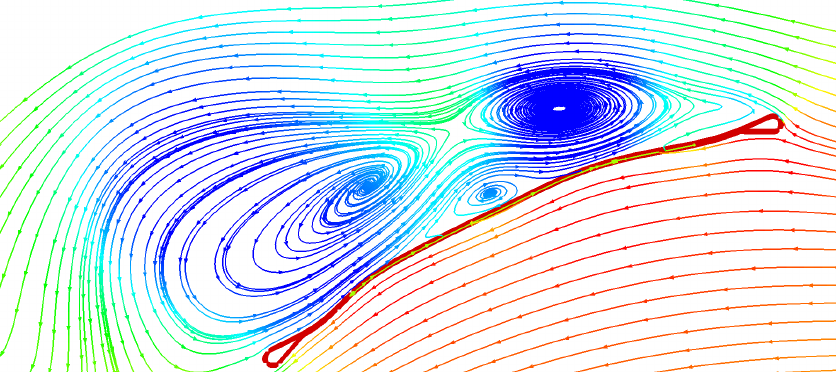}
		\label{streamline_aoaf}
	}
	\\
	\includegraphics[width=0.5\textwidth]{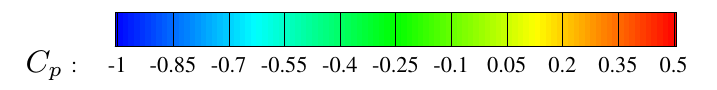}
	\caption{Flow past a 3D rectangular membrane wing: instantaneous streamlines on the mid-span plane colored by pressure coefficient at four selected instantaneous time instants plotted in \reffig{time_history} at $\alpha$= (a) 15$^\circ$, (b) 20$^\circ$ and (c) 25$^\circ$.}
	\label{streamline_aoa}
\end{figure}

We further present the evolutions of the instantaneous streamlines around the membrane on the mid-span plane for different angles of attack at the four selected time instants in \reffig{streamline_aoa}. We observe that large-scale vortices are produced near the leading edge and shed into the wake alternatively at $\alpha=15^\circ$. The scale of the vortices is enlarged as the angle of attack increases to $\alpha=20^\circ$ and $\alpha=25^\circ$. Besides, the vortical structures become more complex at higher angles of attack, compared to $\alpha=15^\circ$. The occasionally occurring chordwise third mode might be related to the small scale vortices containing higher frequency components. From the discussions above, we find that the membrane vibrations and the flow features around the flexible membrane exhibit multi-modal mixed responses both in the temporal and spatial spaces. The aeroelastic responses are overlapped together in the instantaneous plots, which restricts us to isolate the vibration and the flow pattern of interest from the coupled system. To obtain the dominant aeroelastic modes at a specific frequency of interest, the global mode decomposition technique that can separate these time-varying data into frequency-ranked modes is highly desirable.

\begin{figure}[H]
	\centering 
	\includegraphics[width=0.11\textwidth]{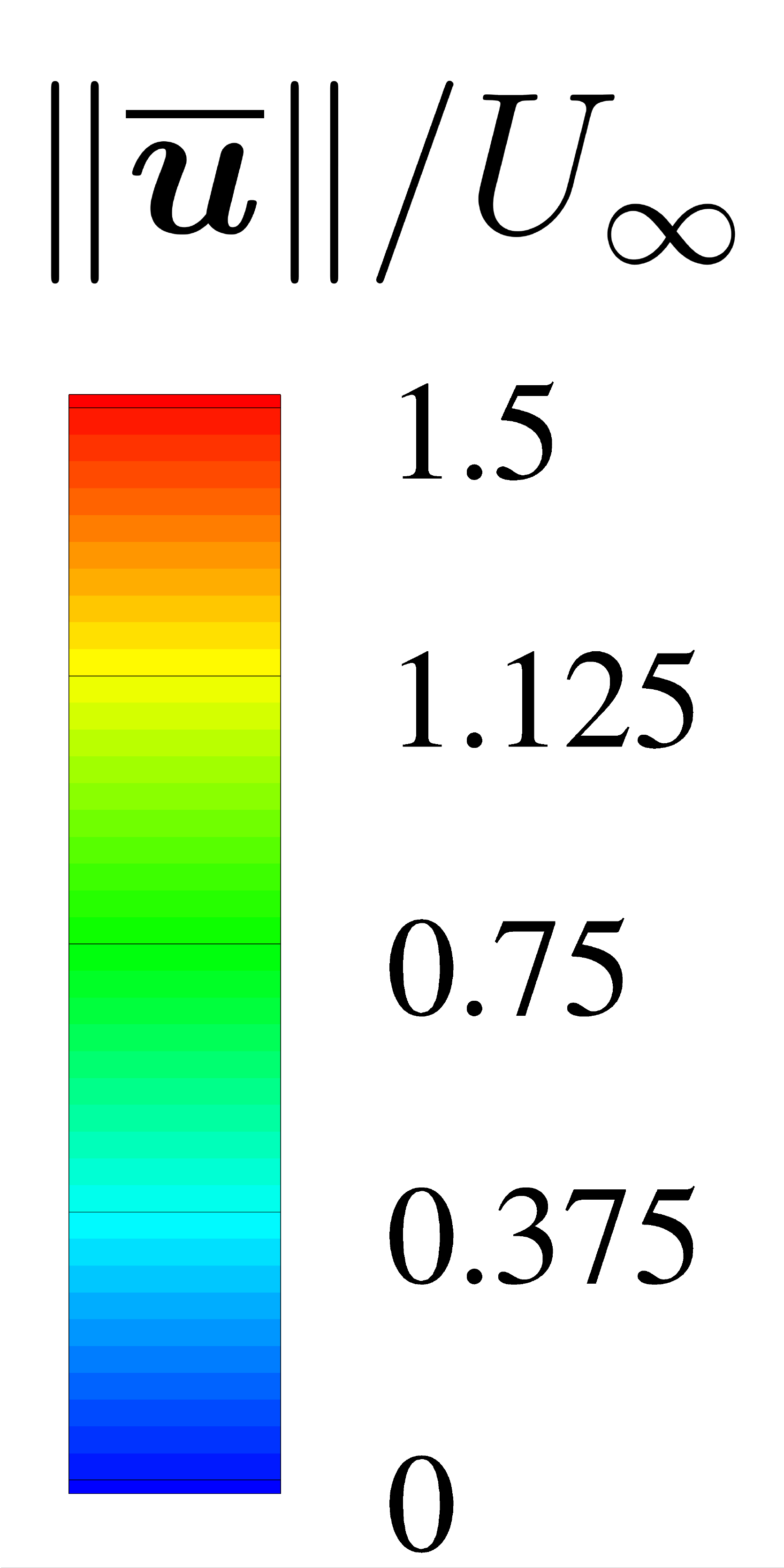}
	\subfloat[][]{
		\includegraphics[width=0.25\textwidth]{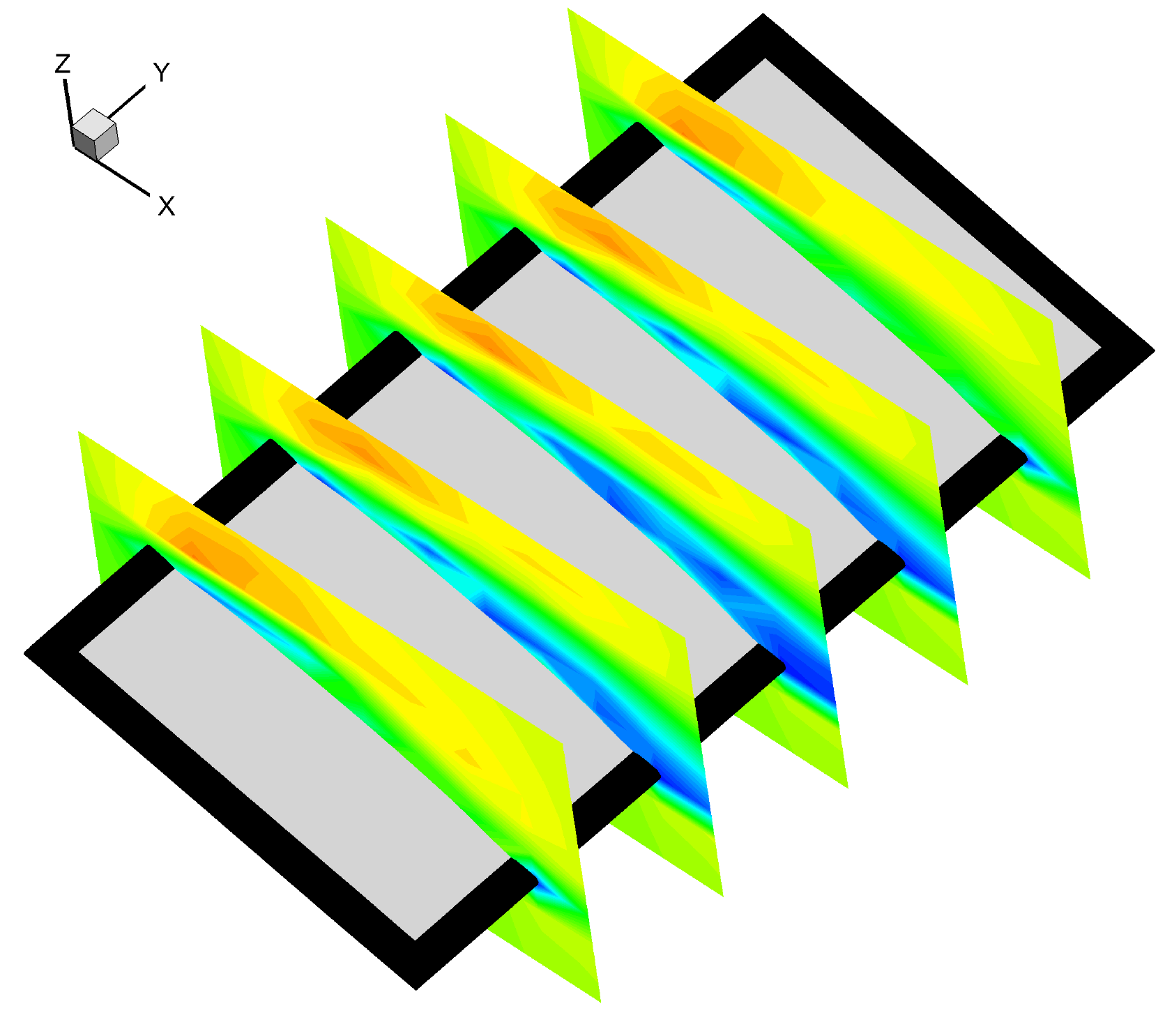}\label{vm_slicea}}
	\subfloat[][]{
		\includegraphics[width=0.25\textwidth]{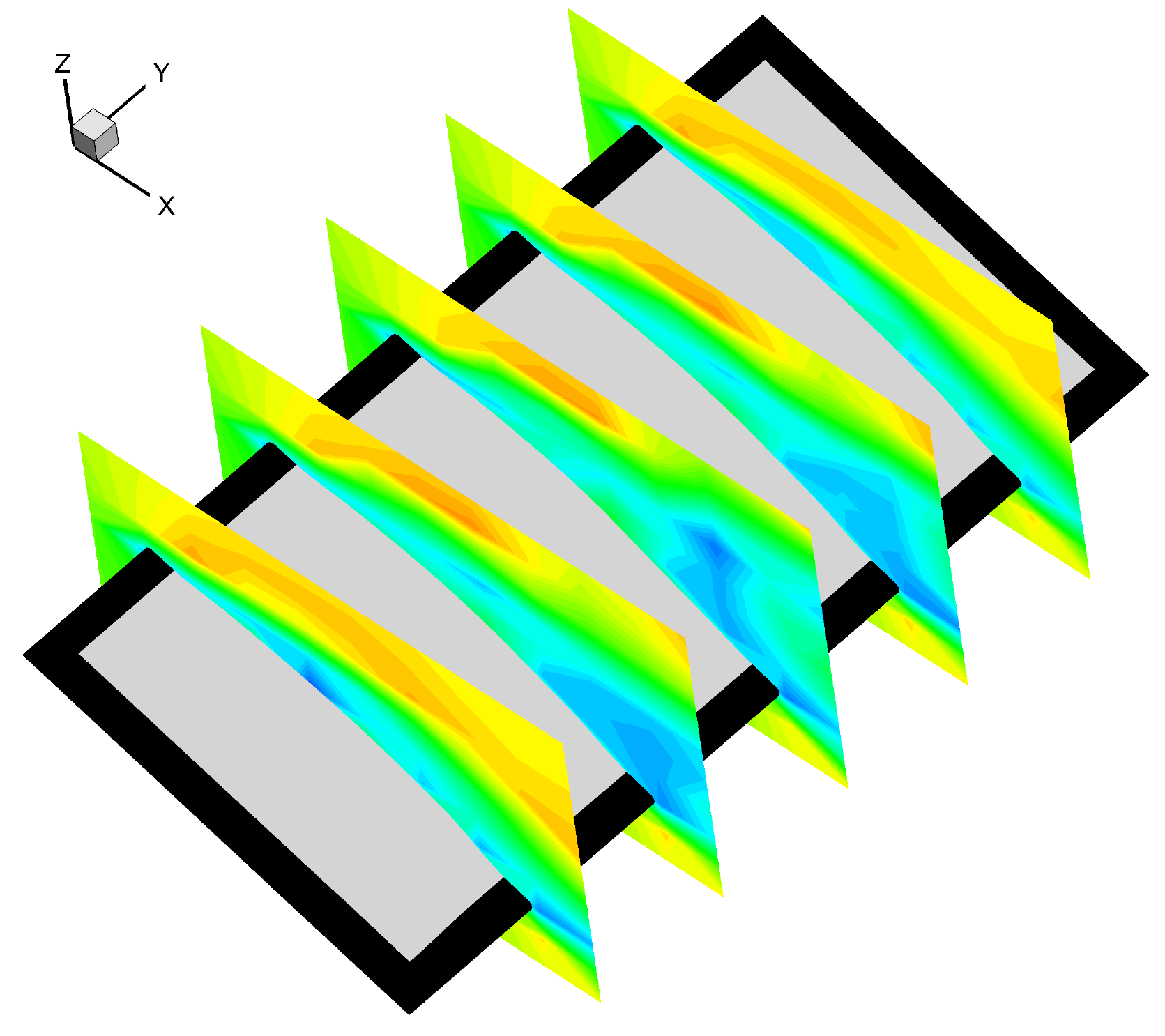}\label{vm_sliceb}}
	\subfloat[][]{
		\includegraphics[width=0.25\textwidth]{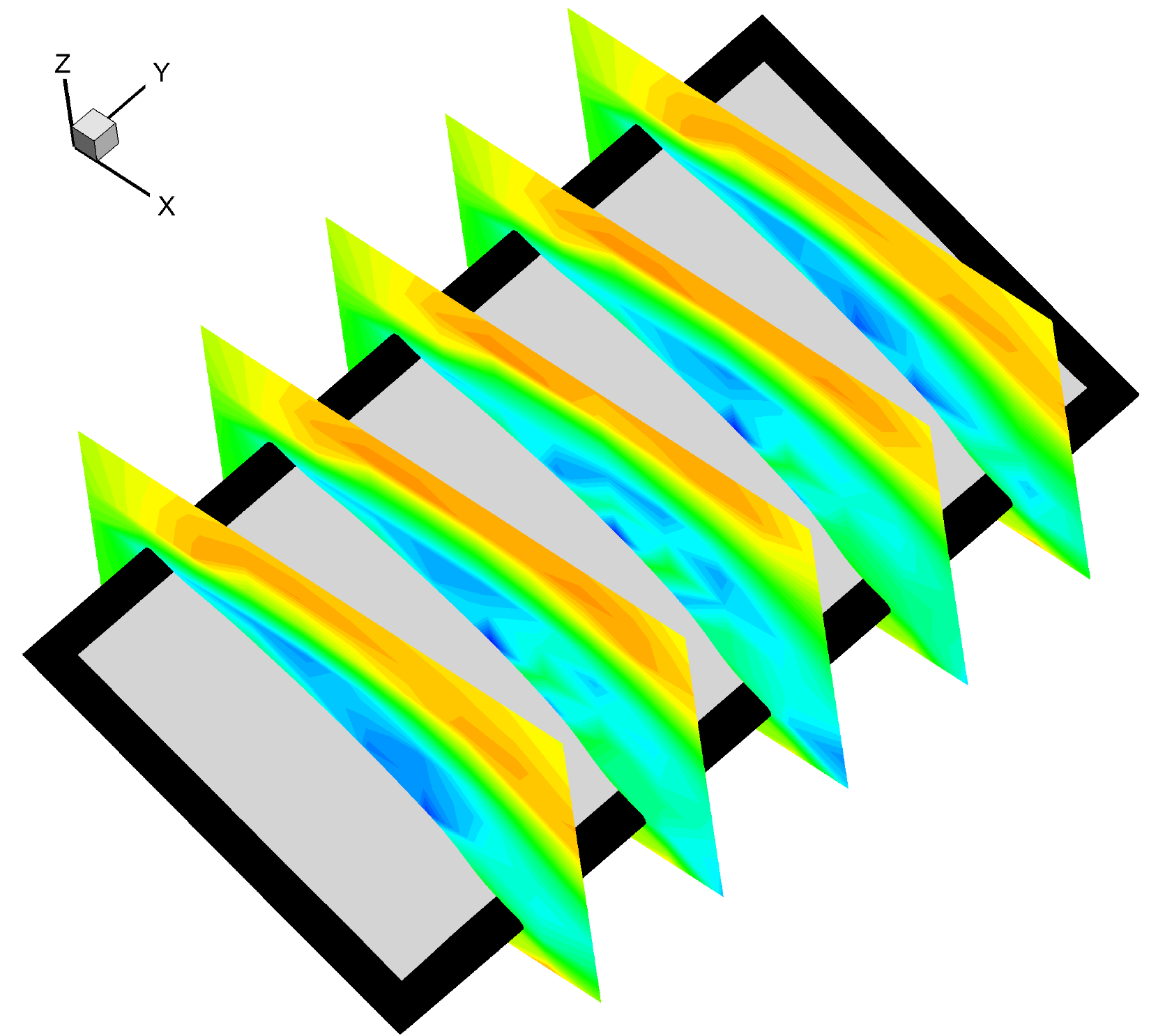}\label{vm_slicec}}
	\caption{Flow past a 3D rectangular membrane wing: time-averaged velocity magnitude on five slices along the spanwise direction at $\alpha$= (a) 15$^\circ$, (b) 20$^\circ$ and (c) 25$^\circ$.}
	\label{vm_slice}
\end{figure}

\begin{figure}[H]
	\centering 
	\includegraphics[width=0.11\textwidth]{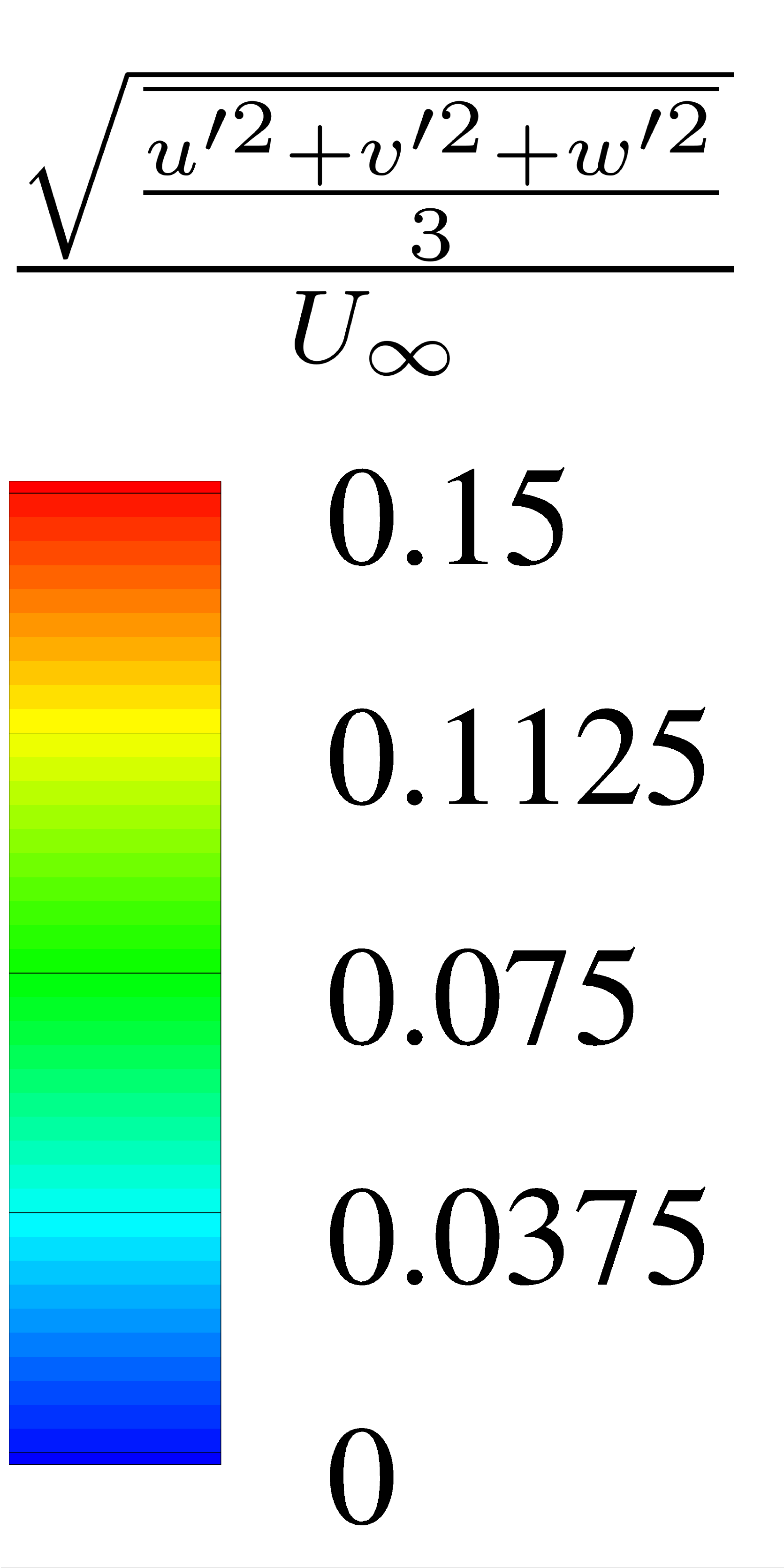}
	\subfloat[][]{
		\includegraphics[width=0.25\textwidth]{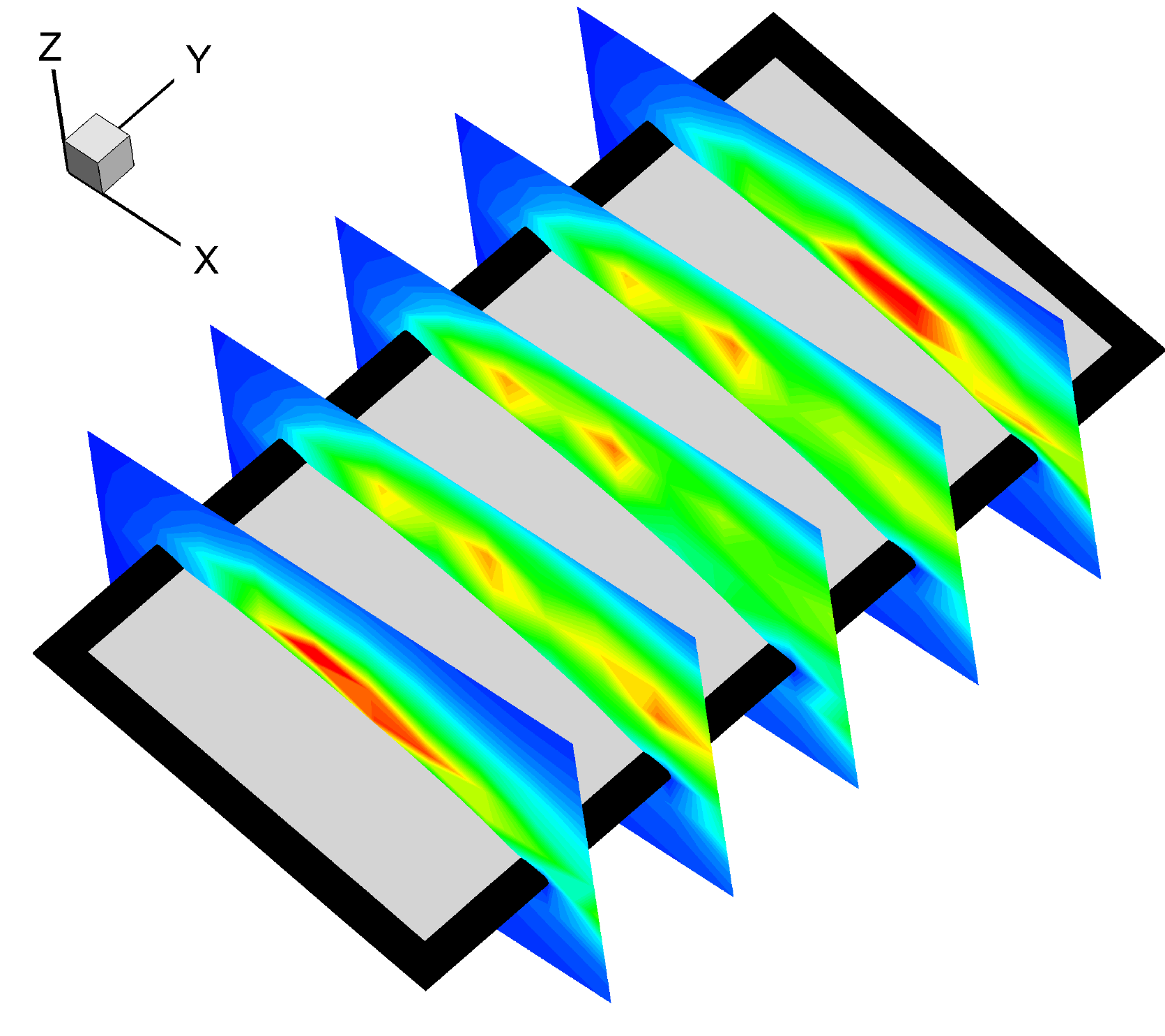}\label{tke_slicea}}
	\subfloat[][]{
		\includegraphics[width=0.25\textwidth]{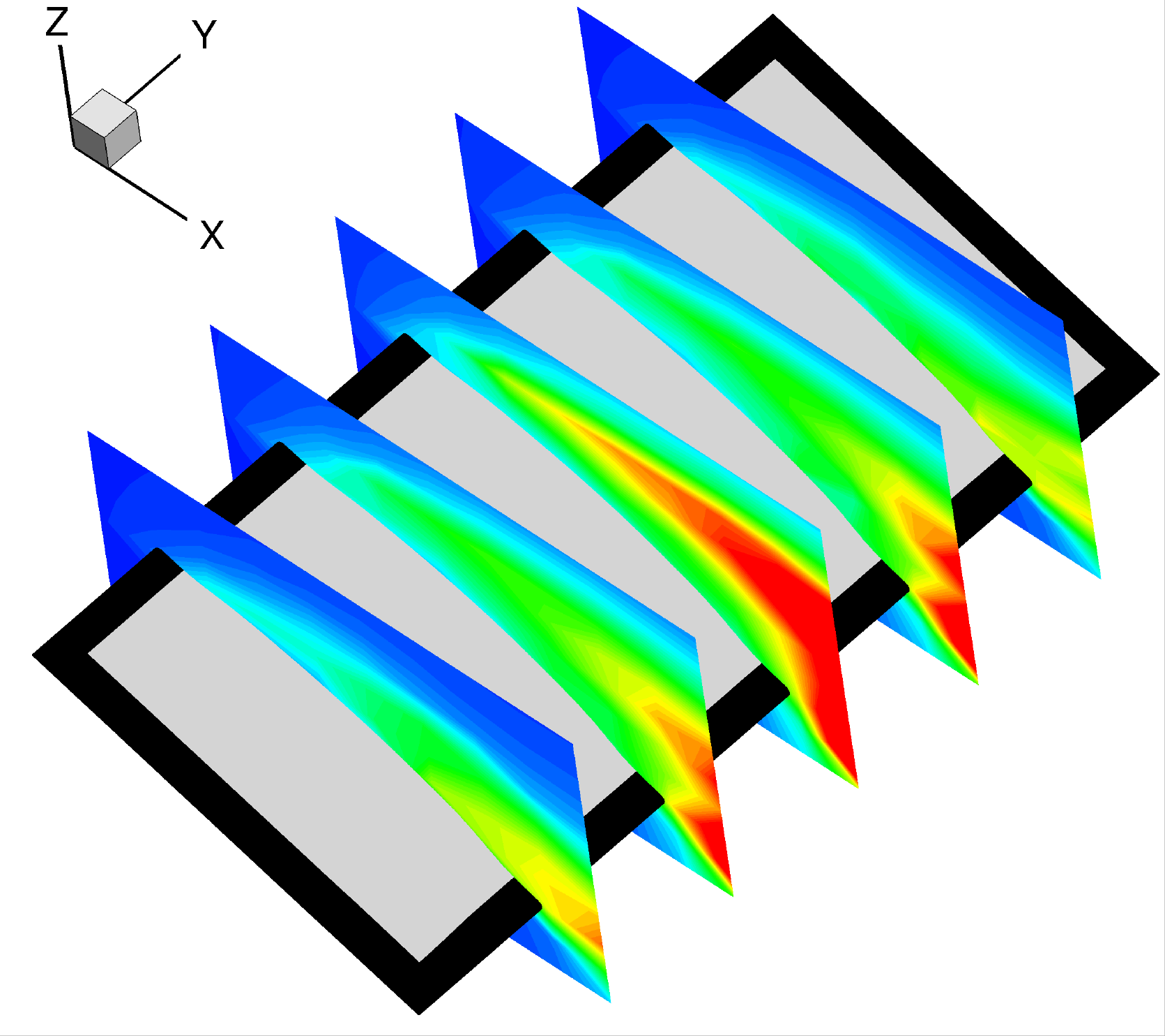}\label{tke_sliceb}}
	\subfloat[][]{
		\includegraphics[width=0.25\textwidth]{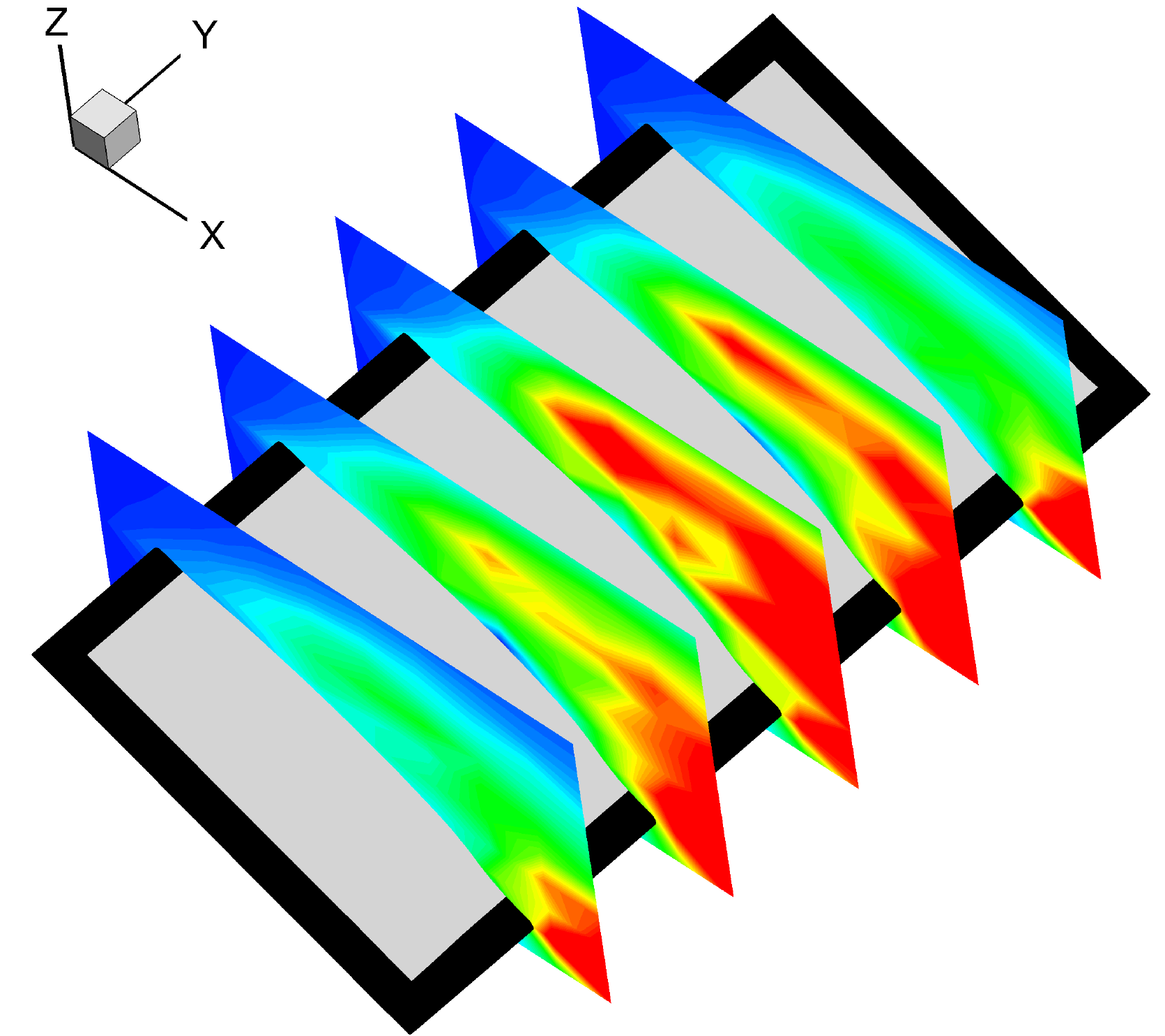}\label{tke_slicec}}
	\caption{Flow past a 3D rectangular membrane wing: turbulent intensity on five slices along the spanwise direction at $\alpha$= (a) 15$^\circ$, (b) 20$^\circ$ and (c) 25$^\circ$.}
	\label{tke_slice}
\end{figure}

\refFigs{vm_slice} and \ref{tke_slice} present the time-averaged velocity magnitude and the turbulent intensity on five equispaced slices along the spanwise direction at three angles of attack, respectively. It can be observed from \reffig{vm_slice} that the low-velocity region is larger on the slice of the mid-span plane than those on the slices near the wingtip. Similarly, the unsteady flow near the mid-span location shows higher turbulent intensity. From \reffigs{zsd_aoa} and \ref{flu_aoa}, we see that the region close to the mid-span location of the membrane has the largest vibration amplitude. Due to the displacement constraints of the membrane at the wingtip, the vibration amplitude near the wingtip becomes smaller. Thus, the flow fluctuations contributed by the membrane vibration is weaker at the wingtip than those near the mid-span location. As the angle of attack increases, both the low-velocity region and the high turbulent intensity region expand further.

\subsection{Aeroelastic mode decomposition and dominant mode identification} \label{mode_decomposition}
In this section, we apply the proposed global FMD method to decompose the coupled system into frequency-ranked aeroelastic modes. The influential modes are identified by detecting the frequency peaks in the mode energy spectrum. For simplicity, we first demonstrate the decomposition process of the RBF-FMD framework for the flexible membrane at the angle of attack of $\alpha=15^\circ$. Detailed explanations of the decomposed aeroelastic modes are then provided. Subsequently, we summarize the influential modes for the flexible membrane at two higher angles of attack.

\subsubsection{Mode decomposition at $\alpha$=15$^\circ$}
\begin{figure}[H]
	\centering 
	\subfloat[][]{\includegraphics[width=0.6\textwidth]{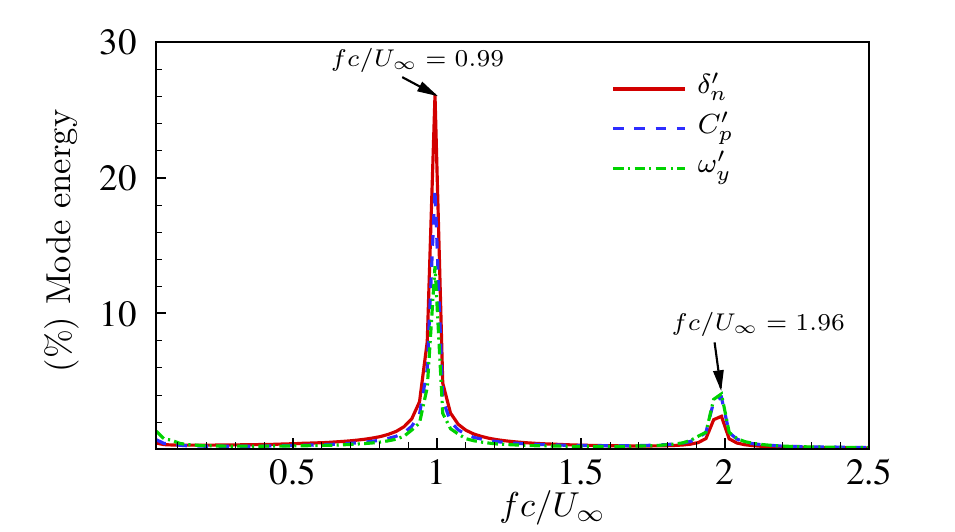}\label{v5a15_surface_mode_fft_slicea}}
	\\
	\subfloat[][]{
		\includegraphics[width=0.35\textwidth]{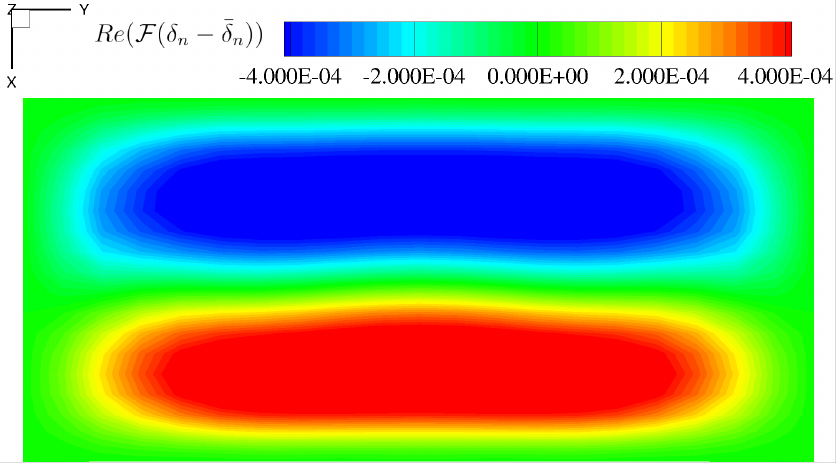}\label{v5a15_surface_mode_fft_sliceb}
	}
	\quad \quad  
	\subfloat[][]{
		\includegraphics[width=0.35\textwidth]{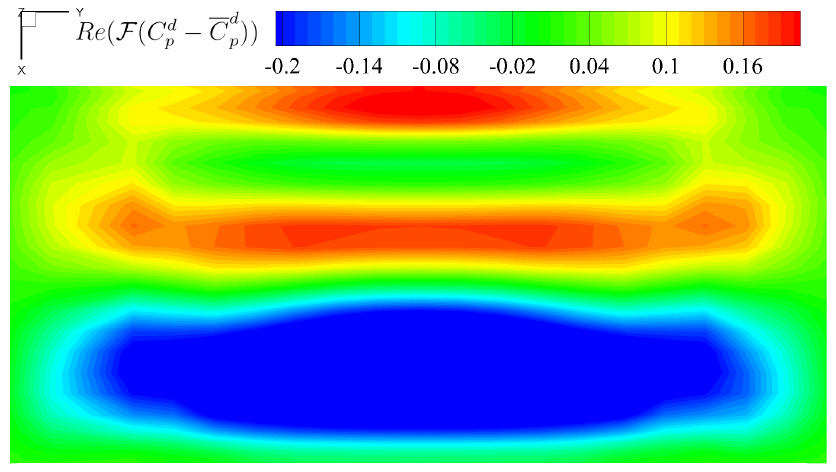}\label{v5a15_surface_mode_fft_slicec}
	}
	\\
	\subfloat[][]{
		\includegraphics[width=0.35\textwidth]{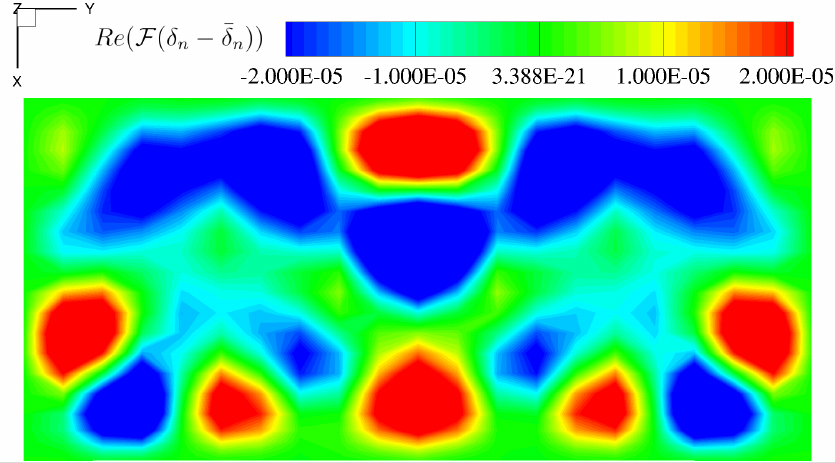}\label{v5a15_surface_mode_fft_sliced}
	}
	\quad \quad
	\subfloat[][]{
		\includegraphics[width=0.35\textwidth]{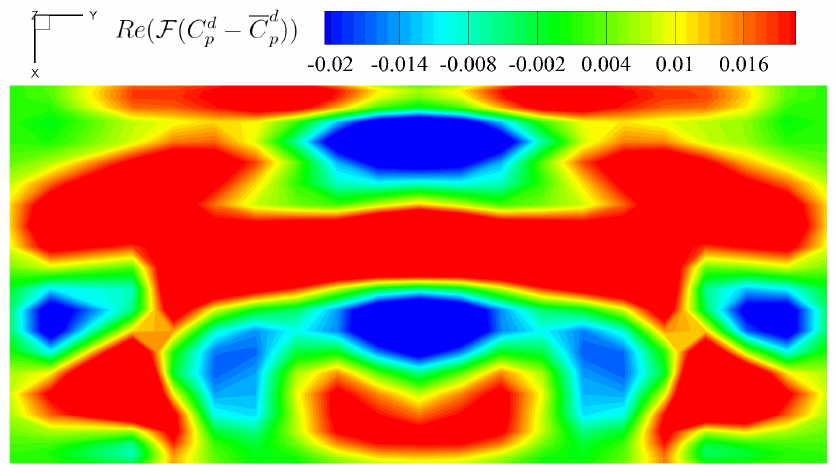}\label{v5a15_surface_mode_fft_slicee}
	}
	\caption{Aeroelastic mode decomposition of 3D flexible membrane at $\alpha$=15$^\circ$: (a) mode energy spectra of the surface displacement fluctuations, the pressure coefficient fluctuations and the $Y$-vorticity fluctuations based on the FMD analysis; the decomposed membrane displacement modes at $f c/ U_{\infty}=$ (b) 0.99 and (d) 1.96 and the surface pressure difference modes at $f c/ U_{\infty}=$ (c) 0.99 and (e) 1.96.}
	\label{v5a15_surface_mode_fft_slice}
\end{figure}

To perform the mode decomposition, we collect 1024 equispaced time-varying samples with a sampling frequency of $f_{sam}$=2000 Hz within the region indicated by the gray color as shown in \reffig{time_history}. The pressure coefficient and the vorticity along the spanwise direction at the body-fitted grids in the fluid domain are projected onto a stationary reference mesh via the RBF method. The membrane displacement and the pressure difference on the membrane surface are collected in the Lagrangian coordinate. All the physical variables are collected simultaneously to ensure the correlation between the modes corresponding to each physical variable at a specific frequency. The global mode energy spectra calculated from \refeq{FMD9} are presented in \reffig{v5a15_surface_mode_fft_slice} \subref{v5a15_surface_mode_fft_slicea}. Two obvious frequency peaks at $f c/ U_{\infty}=$0.99 and 1.96 are observed in the computed mode energy spectra. It is noticed that the energetic frequencies are consistent for the decomposed structural and fluid Fourier modes. This indicates that the membrane vibrations and the flow fluctuations are excited in a frequency synchronized way, resulting in the well-known frequency lock-in phenomenon. The decomposed aeroelastic modes colored by the real part of the Fourier transform coefficients based on the displacements and the pressure difference distributions of the membrane surface at the selected frequency of $f c/ U_{\infty}=$0.99 and 1.96 are plotted in \reffig{v5a15_surface_mode_fft_slice} (b-e), respectively. We notice that a typical chordwise second mode is excited at $f c/ U_{\infty}=$0.99 and a high-order mode both in the chordwise and spanwise directions is observed at a higher frequency of $f c/ U_{\infty}=$1.96. Except for the decomposed surface pressure modal shapes near the leading edge, the overall modes present similar modal shapes as the decomposed surface displacement modes at both energetic frequencies.

\begin{figure}[H]
	\centering 
	\includegraphics[width=0.4\textwidth]{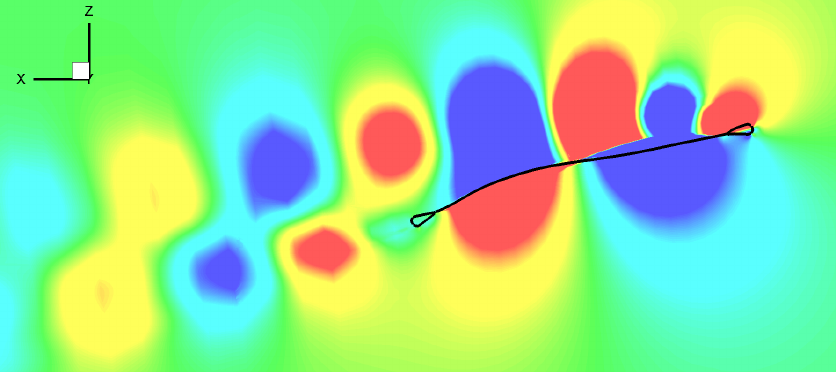}
	\quad
	\includegraphics[width=0.4\textwidth]{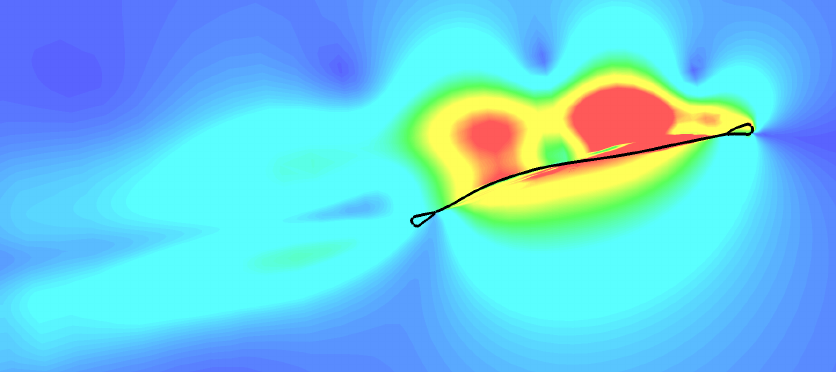}
	\\
	\subfloat[][]{\includegraphics[width=0.4\textwidth]{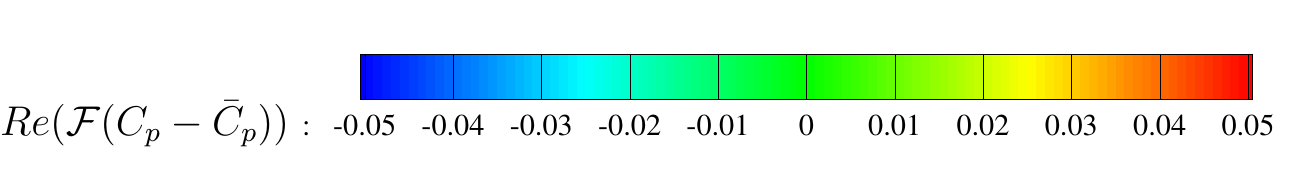}\label{v5a15_pressure_fft_slicea}}
	\quad
	\subfloat[][]{\includegraphics[width=0.4\textwidth]{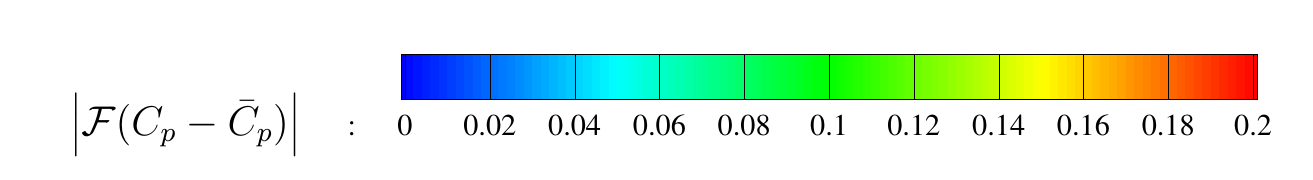}\label{v5a15_pressure_fft_sliceb}}
	\\
	\includegraphics[width=0.4\textwidth]{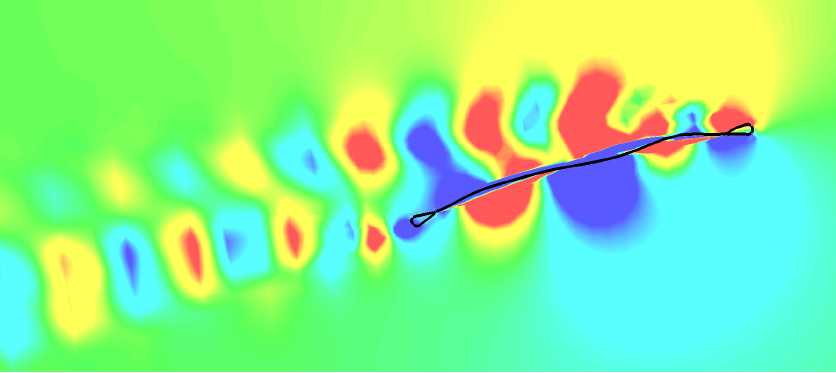}
	\quad
	\includegraphics[width=0.4\textwidth]{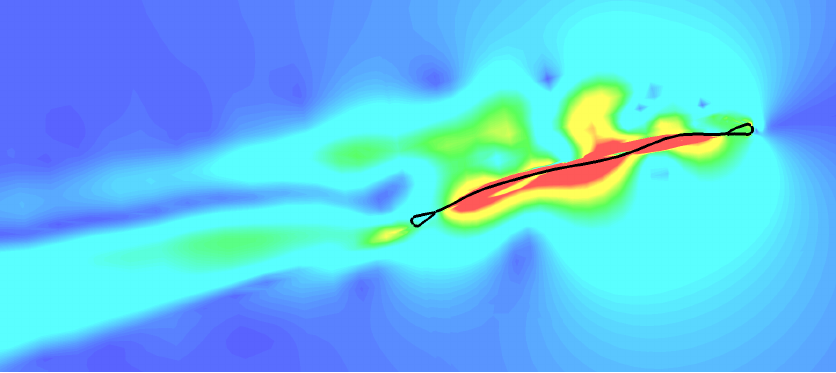}
	\\
	\subfloat[][]{\includegraphics[width=0.4\textwidth]{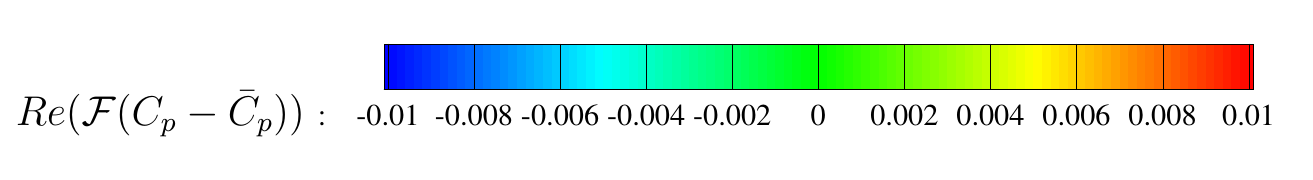}\label{v5a15_pressure_fft_slicec}}
	\quad
	\subfloat[][]{\includegraphics[width=0.4\textwidth]{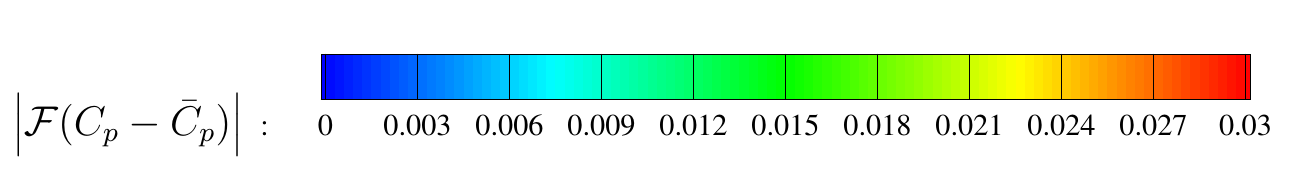}\label{v5a15_pressure_fft_sliced}}
	\caption{Aeroelastic mode decomposition of 3D flexible membrane at $\alpha$=15$^\circ$: contours of the real part (a,c) and the amplitude (b,d) of the Fourier transform coefficients of the pressure coefficient fluctuation field corresponding to the non-dimensional frequency of $f c/ U_{\infty}=$ (a,b) $0.99$ and (c,d) $1.96$.}
	\label{v5a15_pressure_fft_slice}
\end{figure}

\begin{figure}[H]
	\centering 
	\includegraphics[width=0.4\textwidth]{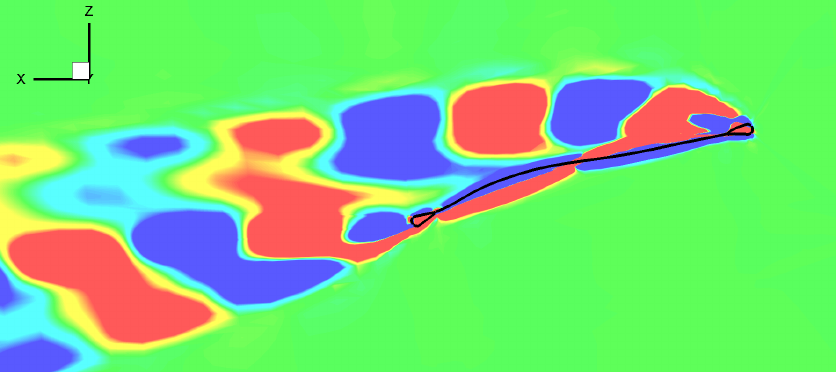}
	\quad
	\includegraphics[width=0.4\textwidth]{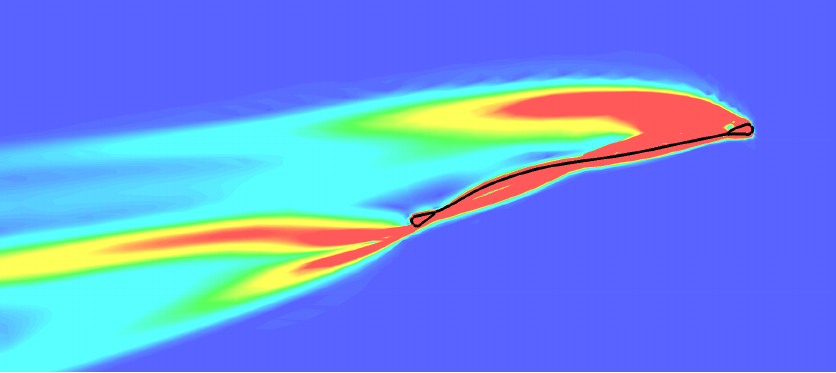}
	\\
	\subfloat[][]{\includegraphics[width=0.4\textwidth]{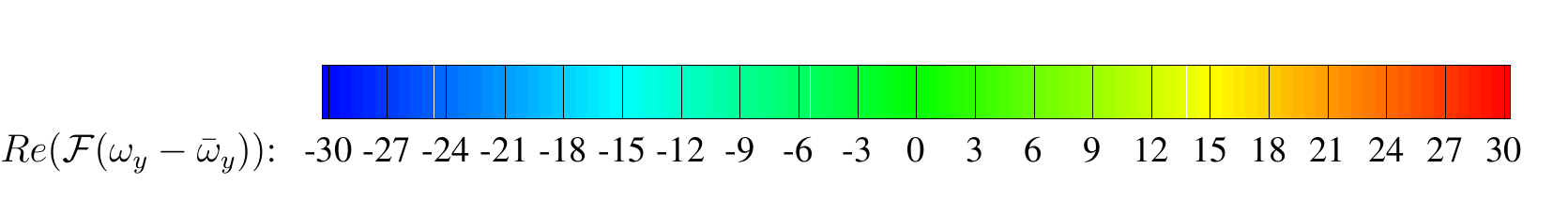}\label{v5a15_yv_fft_slicea}}
	\quad
	\subfloat[][]{\includegraphics[width=0.4\textwidth]{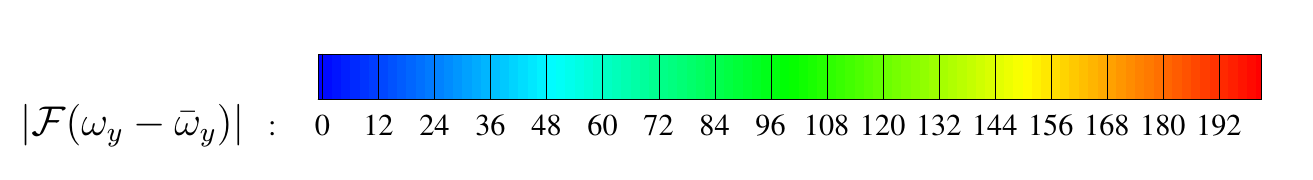}\label{v5a15_yv_fft_sliceb}}
	\\
	\includegraphics[width=0.4\textwidth]{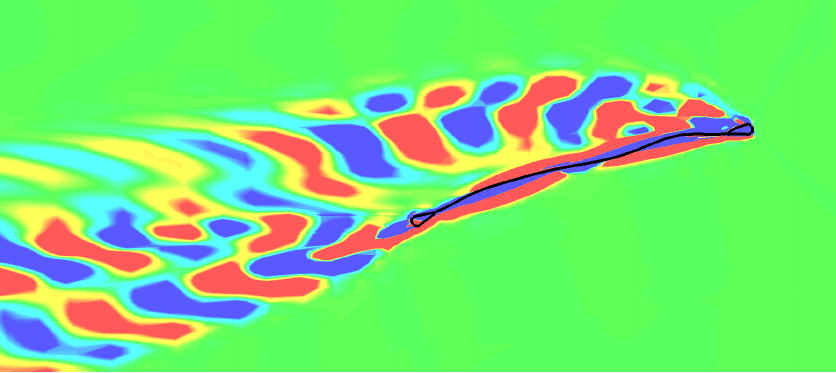}
	\quad
	\includegraphics[width=0.4\textwidth]{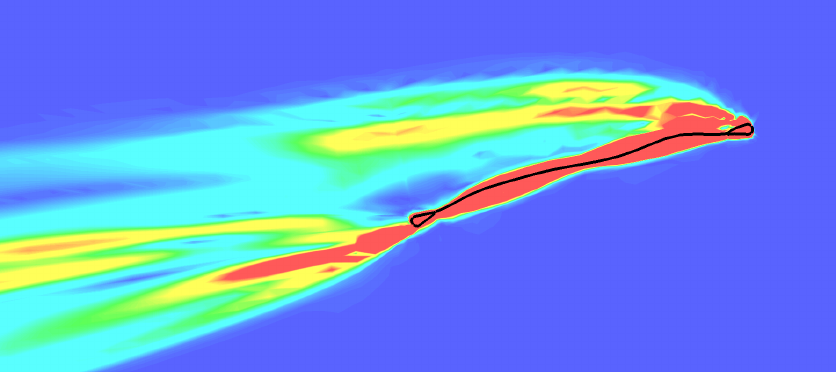}
	\\
	\subfloat[][]{\includegraphics[width=0.4\textwidth]{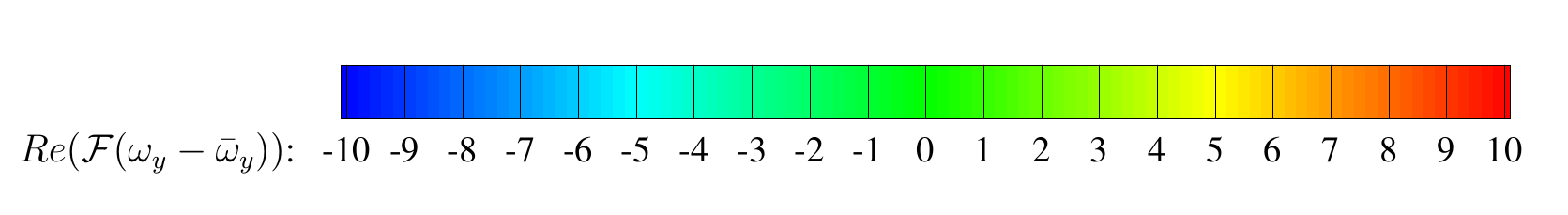}\label{v5a15_yv_fft_slicec}}
	\quad
	\subfloat[][]{\includegraphics[width=0.4\textwidth]{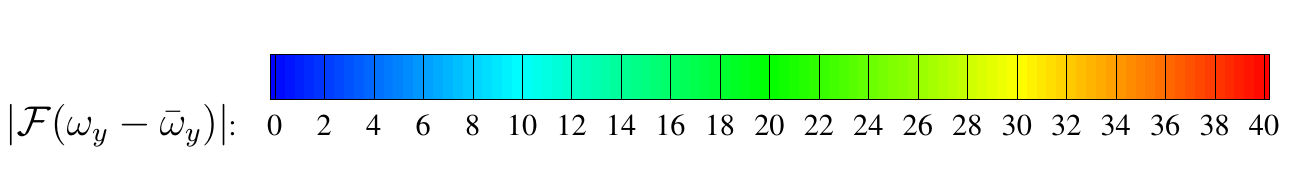}\label{v5a15_yv_fft_sliced}}
	\caption{Aeroelastic mode decomposition of 3D flexible membrane at $\alpha$=15$^\circ$: contours of the real part (a,c) and the amplitude (b,d) of the Fourier transform coefficients of the $Y-$vorticity fluctuation field corresponding to the non-dimensional frequency of $f c/ U_{\infty}=$ (a,b) $0.99$ and (c,d) $1.96$.}
	\label{v5a15_yv_fft_slice}
\end{figure}

To study the spatial flow structures correlated with the membrane vibration, we extract the dynamic Fourier modes in the spatial pressure and $Y$-vorticity fields on the mid-span plane. In the plots of the decomposed fluid Fourier modes, the structural modal shape corresponding to the same frequency is added to help understand the correlation between the Fourier modes in the fluid and structural domains. Due to the small values of the structural Fourier modes, these structural modal shapes indicated by the black line are constructed by amplifying the corresponding structural Fourier modes based on the time-averaged membrane shape for visualization purposes. The real part $Re(\mathcal{F}(C_p - \bar{C}_p))$ and the amplitude $\left|\mathcal{F}(C_p - \bar{C}_p) \right|$ of the decomposed pressure fluctuation fields at the non-dimensional frequency of $f c/ U_{\infty}=$0.99 corresponding to the chordwise second mode are shown in \reffig{v5a15_pressure_fft_slice} (a,b), respectively. The real part of the transformed coefficient reflects the spatial structure of the mode. The amplitude represents the intensity distributions of the decomposed physical variables. Two small-scale pressure fluctuation regions are observed near the leading edge on the upper membrane surface. These pressure fluctuations are mainly caused by the rolled-up vortices at the leading edge in \reffig{v5a15_yv_fft_slice} \subref{v5a15_yv_fft_slicea}. Two larger pressure fluctuation regions on the upper surface are generated during the periodic leading edge vortex shedding process. From the amplitude contour of the decomposed pressure field in \reffig{v5a15_pressure_fft_slice} \subref{v5a15_pressure_fft_sliceb}, the large-scale pressure pulsations with high values are noticed on the upper surface. The severe vorticity fluctuations are mainly formed at the periodic vortex shedding regions near the leading and trailing edges in \reffig{v5a15_yv_fft_slice} \subref{v5a15_yv_fft_sliceb}.

As the increase of the non-dimensional frequency to $f c/ U_{\infty}=$1.96 with the high-order mode, the pressure wavelength and the flow scales become smaller. The high-intensity pressure pulsations still keep close to the membrane surface as shown in \reffig{v5a15_pressure_fft_slice} \subref{v5a15_pressure_fft_sliced}. However, the amplitude values in this region are far less than those at $f c/ U_{\infty}=$0.99 due to the weaker mode energy of the high-order mode. Meanwhile, the small-scale vortices originating from the leading edge move backwards to merge with the trailing edge vortices behind the membrane in \reffig{v5a15_yv_fft_slice} \subref{v5a15_yv_fft_slicec}. It can be observed from \reffig{v5a15_yv_fft_slice} \subref{v5a15_yv_fft_sliced} that the high-intensity vorticity fluctuation region shrinks and the amplitude value in this region is reduced, compared to the decomposed vorticity field at the dominant frequency of $f c/ U_{\infty}=$0.99.

\subsubsection{Mode decomposition at $\alpha$=20$^\circ$ and 25$^\circ$}
The aeroelastic responses at $\alpha=20^\circ$ and $25^\circ$ are also decomposed via the proposed FMD technique. It can be seen from \reffigs{time_history} and \ref{z_v} that the aeroelastic responses are non-periodic and contain some noise components at $\alpha=20^\circ$ and $25^\circ$. To improve the detection performance, Welch's method is employed in the mode energy spectrum estimation for these two cases. As shown in \reffigs{time_history} \subref{time_historyc} and \subref{time_historye}, we extract 1024 equispaced snapshots within the selected time range indicated by the gray color from the coupled fluid-membrane system. The membrane displacements, the surface pressure coefficient, the spatial pressure coefficient and the spanwise $Y$-vorticity are collected in the same sampling frequency of $f_{sam}$=2000 Hz. The data sequence is split into two overlapping data segments and each segment is windowed with a Hamming window. 

\begin{figure}[H]
	\centering 
	\subfloat[][]{\includegraphics[width=0.6\textwidth]{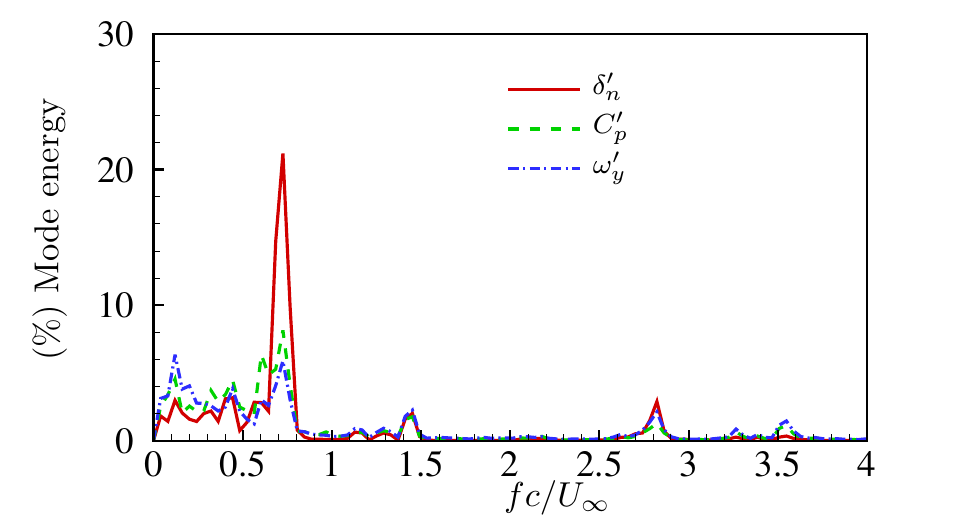}\label{v5a20_surface_mode_fft_slicea}}
	\\
	\subfloat[][]{
		\includegraphics[width=0.35\textwidth]{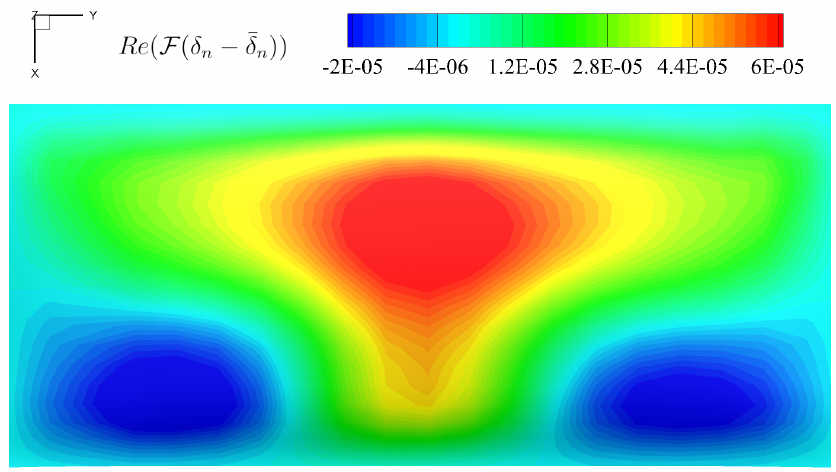}\label{v5a20_surface_mode_fft_sliceb}
	}
	\quad \quad  
	\subfloat[][]{
		\includegraphics[width=0.35\textwidth]{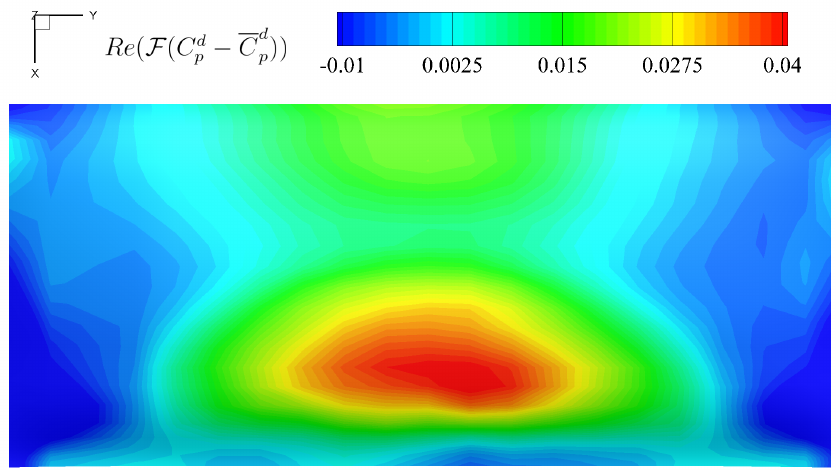}\label{v5a20_surface_mode_fft_slicec}
	}
   \\
	\subfloat[][]{
		\includegraphics[width=0.35\textwidth]{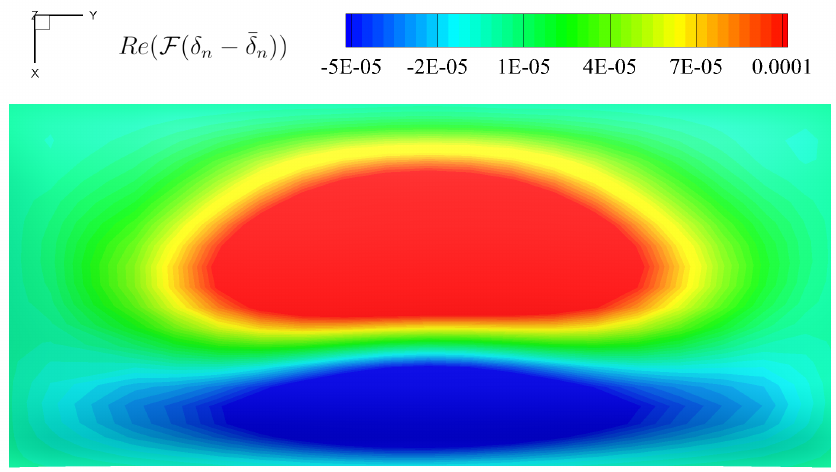}\label{v5a20_surface_mode_fft_sliced}
	}
	\quad \quad  
	\subfloat[][]{
		\includegraphics[width=0.35\textwidth]{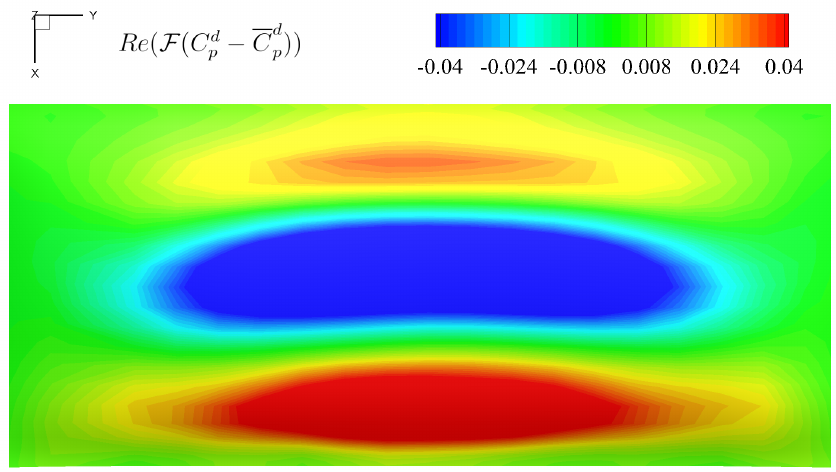}\label{v5a20_surface_mode_fft_slicee}
	}
	\\
	\subfloat[][]{
		\includegraphics[width=0.35\textwidth]{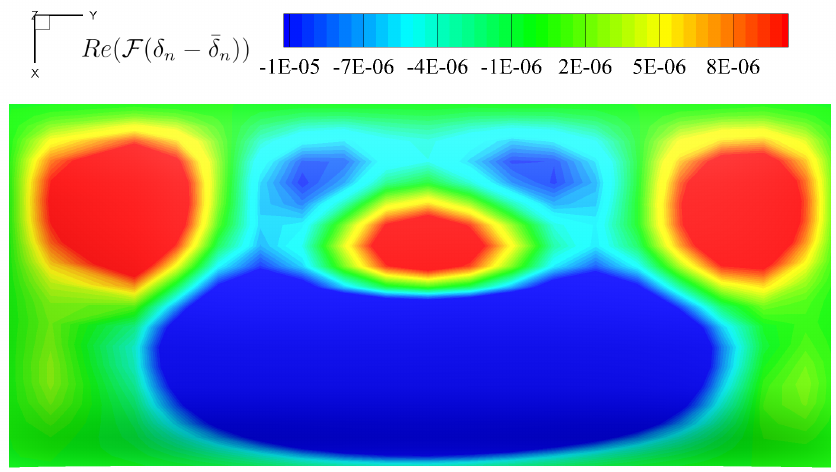}\label{v5a20_surface_mode_fft_slicef}
	}
	\quad \quad
	\subfloat[][]{
		\includegraphics[width=0.35\textwidth]{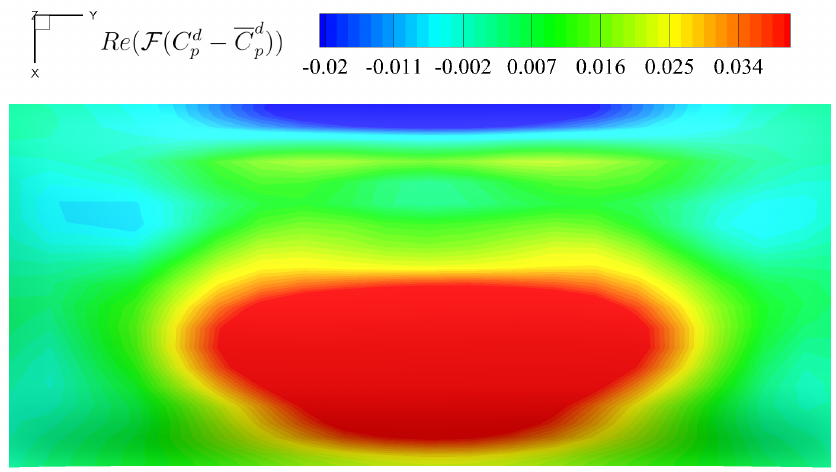}\label{v5a20_surface_mode_fft_sliceg}
	}
    \\
    \subfloat[][]{
    	\includegraphics[width=0.35\textwidth]{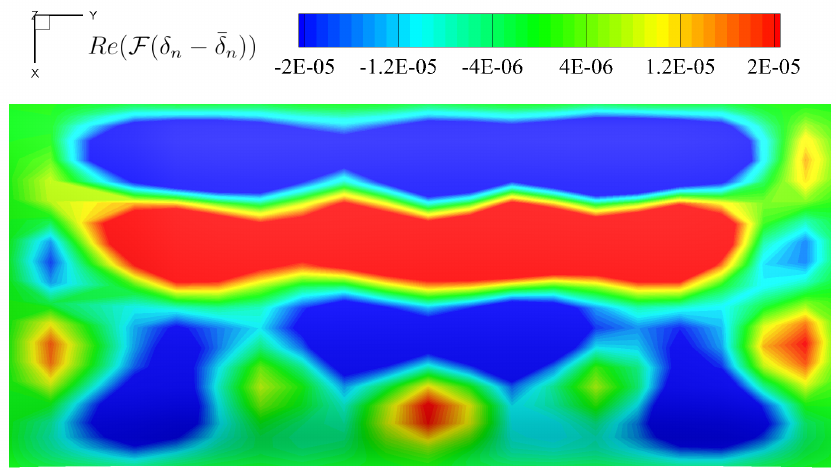}\label{v5a20_surface_mode_fft_sliceh}
    }
    \quad \quad
    \subfloat[][]{
    	\includegraphics[width=0.35\textwidth]{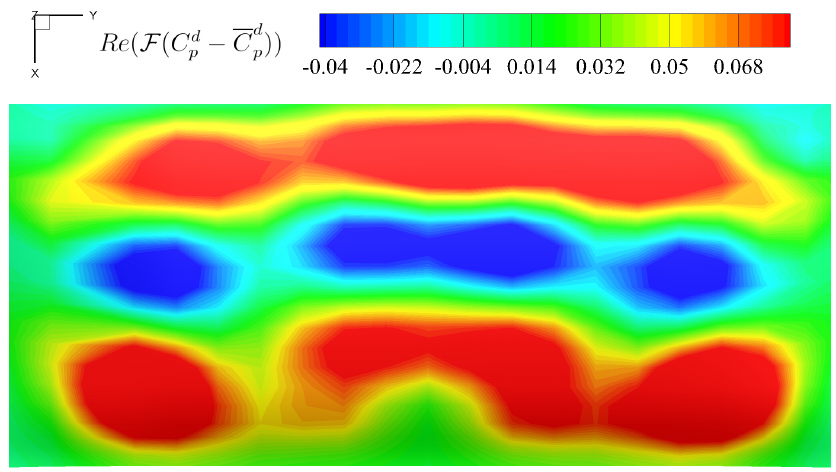}\label{v5a20_surface_mode_fft_slicei}
    }
	\caption{Aeroelastic mode decomposition of 3D flexible membrane at $\alpha$=20$^\circ$: (a) mode energy spectra of the surface displacement fluctuations, the pressure coefficient fluctuations and the $Y$-vorticity fluctuations based on the FMD analysis; (b,d,f,h) the decomposed membrane displacement modes and (c,e,g,i) the surface pressure difference modes at $f c/ U_{\infty}=$ (b,c) 0.122, (d,e) 0.727, (f,h) 1.45 and (h,i) 2.82.}
	\label{v5a20_surface_mode_fft_slice}
\end{figure}

\refFig{v5a20_surface_mode_fft_slice} presents the aeroelastic mode spectra and energetic structural modes at four selected frequencies. It can be seen from \reffig{v5a20_surface_mode_fft_slice} \subref{v5a20_surface_mode_fft_slicea} that the flexible membrane is highly coupled with the unsteady flow and responds in a frequency-synchronized manner, resulting a frequency-lock phenomenon. Different from the mode energy spectra at $\alpha=15^\circ$, some low frequency components within the range of $f c / U_{\infty} \in [0, 0.6]$ are observed from the mode energy spectra at $\alpha=20^\circ$. The aeroelastic modes at $f c / U_{\infty}$=0.122 are chosen to examine the aeroelastic characteristics caused by the low frequency components. The aeroelastic modes at $f c / U_{\infty}$=0.727 are the dominant modes in the coupled system, which exhibit the largest mode energies. A second harmonic frequency of the dominant frequency is observed at $f c / U_{\infty}$=1.45. We also investigate the aeroelastic modes at a higher frequency of 2.82. As shown in \reffig{v5a20_surface_mode_fft_slice} \subref{v5a20_surface_mode_fft_sliced}, the dominant structural mode is a chordwise second and spanwise first mode. In \reffigs{v5a20_surface_mode_fft_slice} \subref{v5a20_surface_mode_fft_slicef} and \subref{v5a20_surface_mode_fft_sliceh}, higher order modes both in the chordwise and spanwise directions are noticed. We observe some occasionally occurring high-order modes from the instantaneous structural displacement fluctuations shown in \reffig{flu_aoa} \subref{flu_aoad}. However, these high-order modes are covered by the dominant modes. With the aid of the mode decomposition techniques, these overlapping modes with lower mode energies can be separated from the coupled system. The correlated modal shapes based on the surface pressure difference are also extracted together with the structural modes at the selected frequencies. The surface pressure difference presents overall similar modal shapes as the structural displacement fluctuation except for some differences near the leading edge.

\begin{figure}[H]
	\centering 
	\includegraphics[width=0.4\textwidth]{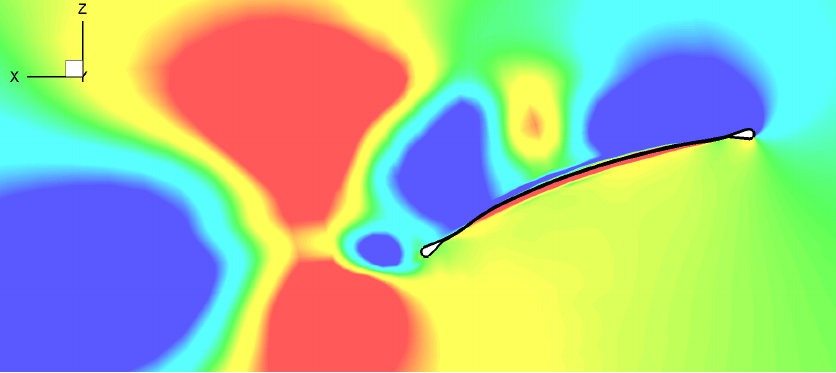}
	\quad
	\includegraphics[width=0.4\textwidth]{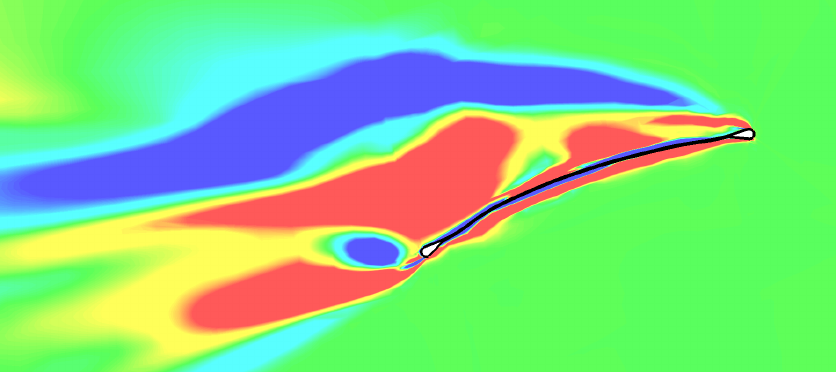}
	\\
	\subfloat[][]{\includegraphics[width=0.4\textwidth]{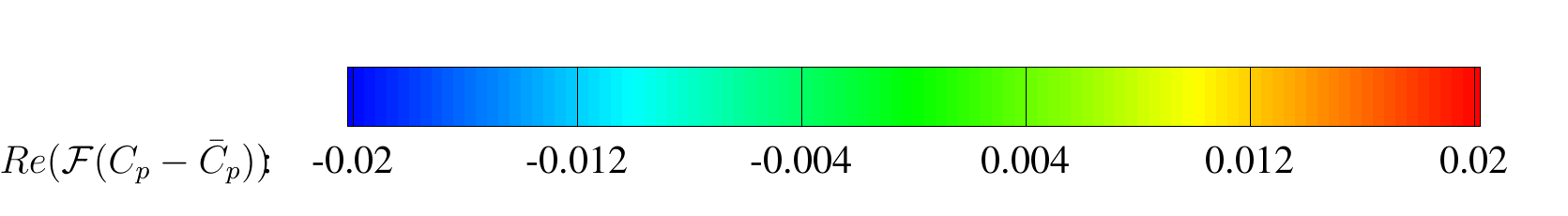}\label{v5a20_yv_fft_slicea}}
	\quad
	\subfloat[][]{\includegraphics[width=0.4\textwidth]{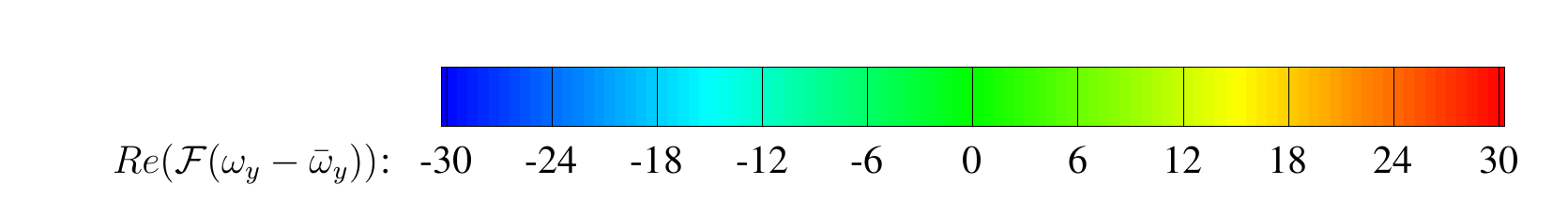}\label{v5a20_yv_fft_sliceb}}
	\\
	\includegraphics[width=0.4\textwidth]{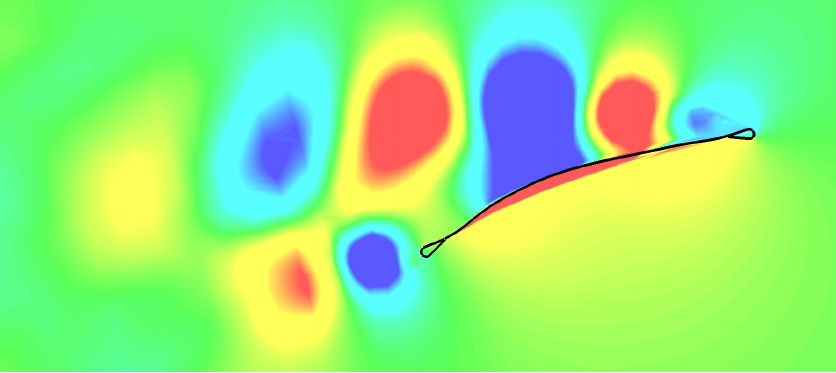}
	\quad
	\includegraphics[width=0.4\textwidth]{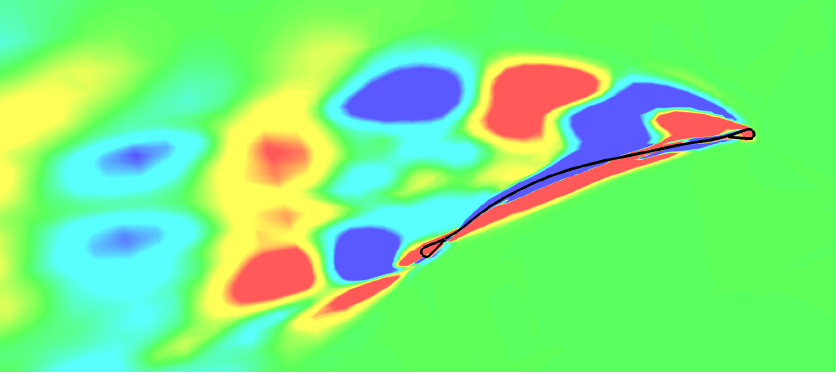}
	\\
	\subfloat[][]{\includegraphics[width=0.4\textwidth]{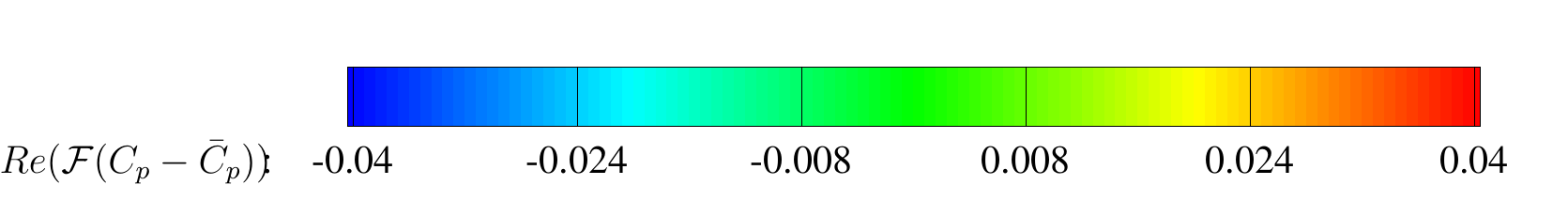}\label{v5a20_yv_fft_slicec}}
	\quad
	\subfloat[][]{\includegraphics[width=0.4\textwidth]{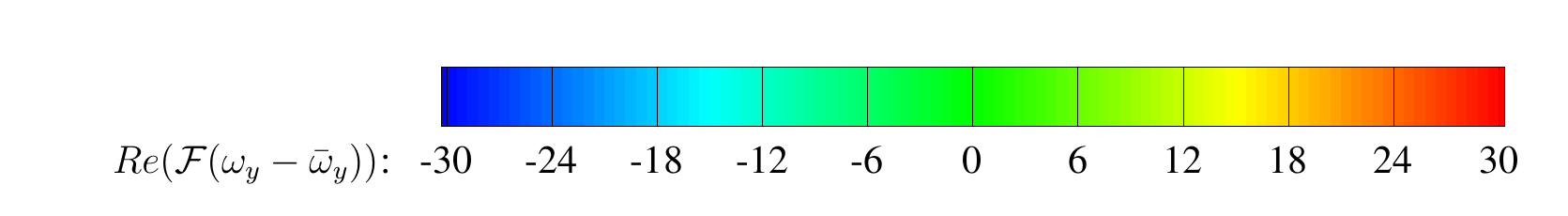}\label{v5a20_yv_fft_sliced}}
	\\
	\includegraphics[width=0.4\textwidth]{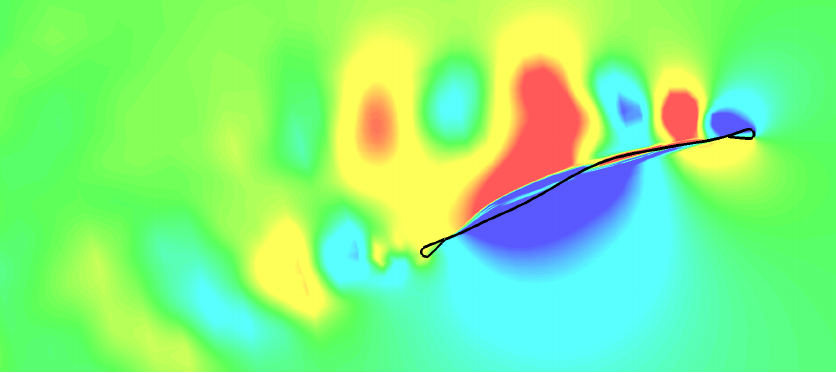}
	\quad
	\includegraphics[width=0.4\textwidth]{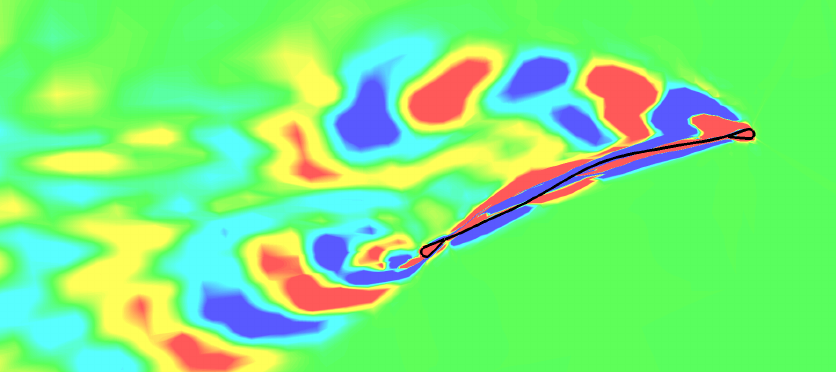}
	\\
	\subfloat[][]{\includegraphics[width=0.4\textwidth]{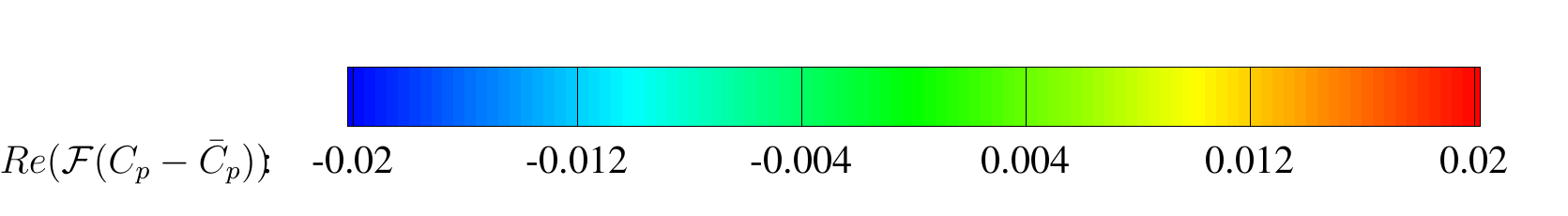}\label{v5a20_yv_fft_slicee}}
	\quad
	\subfloat[][]{\includegraphics[width=0.4\textwidth]{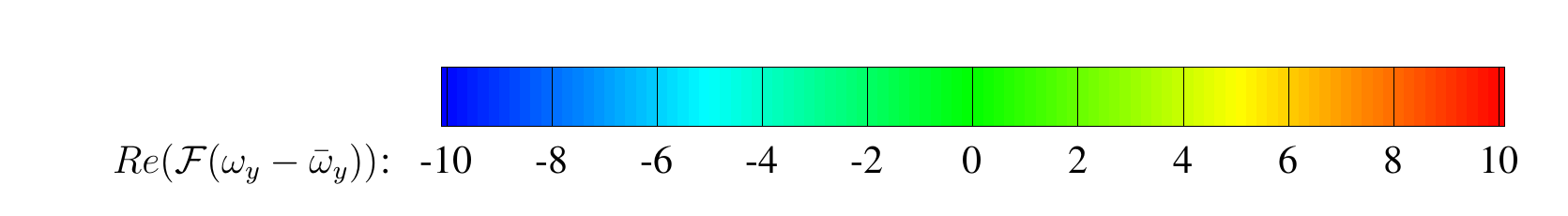}\label{v5a20_yv_fft_slicef}}
	\\
	\includegraphics[width=0.4\textwidth]{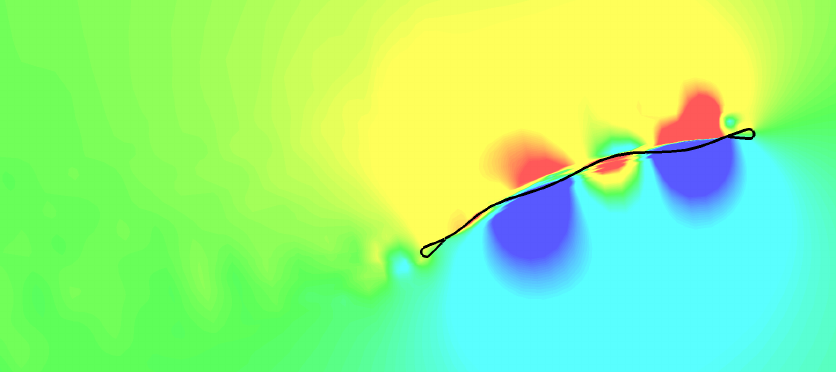}
	\quad
	\includegraphics[width=0.4\textwidth]{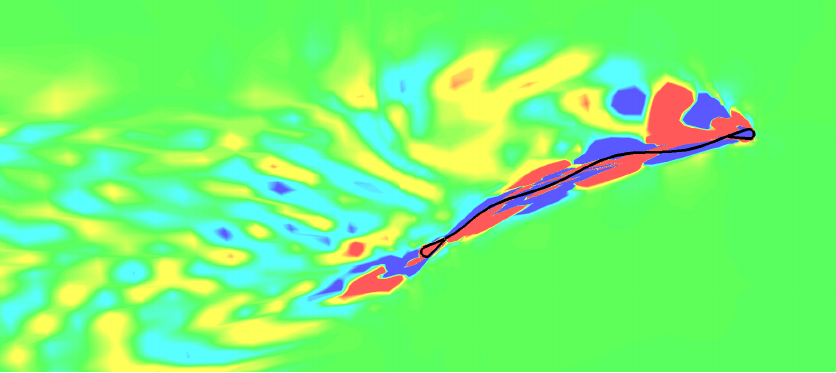}
	\\
	\subfloat[][]{\includegraphics[width=0.4\textwidth]{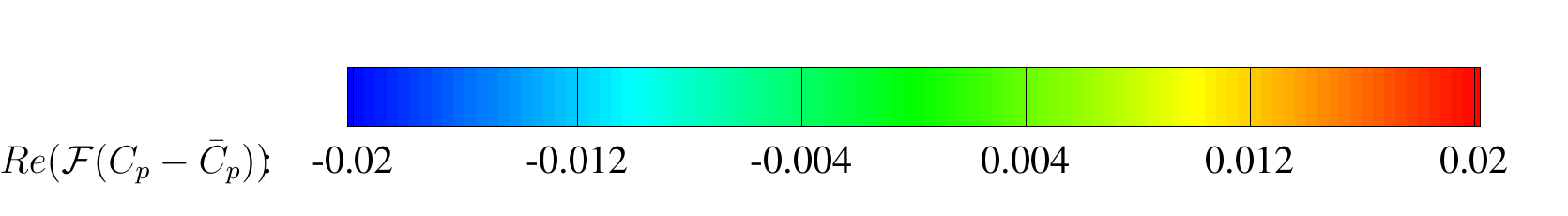}\label{v5a20_yv_fft_sliceg}}
	\quad
	\subfloat[][]{\includegraphics[width=0.4\textwidth]{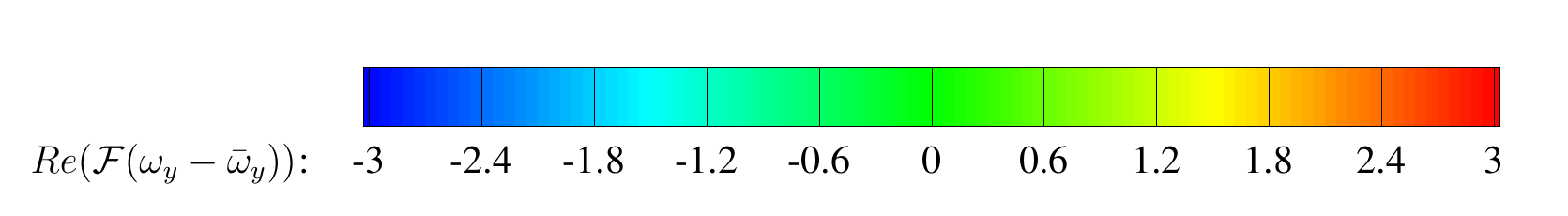}\label{v5a20_yv_fft_sliceh}}
	\caption{Aeroelastic mode decomposition of 3D flexible membrane at $\alpha$=20$^\circ$: contours of the real part of the Fourier transform coefficients of (a,c,e,g) the pressure coefficient fluctuation field and (b,d,f,h) the $Y-$vorticity fluctuation field corresponding to the non-dimensional frequency of $f c/ U_{\infty}=$ (a,b) 0.122, (c,d) 0.727, (e,f) 1.45 and (g,h) 2.82.}
	\label{v5a20_fft_slice}
\end{figure}

The fluid modes based on the real parts of the Fourier transform coefficients of the pressure coefficient fluctuation $Re(\mathcal{F}(C_p - \bar{C}_p))$ and the $Y$-vorticity fluctuation $Re(\mathcal{F}(\omega_y - \bar{\omega}_y))$ at the four selected frequencies are presented in \reffig{v5a20_fft_slice}. Large size vortices are noticed on the whole membrane surface at $f c/ U_{\infty}=$0.122 in \reffig{v5a20_fft_slice} \subref{v5a20_yv_fft_sliceb}. Periodic vortex shedding can be observed near the leading and trailing edges at the dominant frequencies of $f c/ U_{\infty}=$0.727 from \reffig{v5a20_fft_slice} \subref{v5a20_yv_fft_sliced}. The vortices at the second harmonics are shed in a similar way to those at the dominant frequency but with smaller sizes. In the decomposed vorticity modes shown in \reffig{v5a20_fft_slice} \subref{v5a20_yv_fft_sliceh}, some complex vortex structures are noticed, which are different from the vortex structures at the dominant frequency and its second harmonics. When the vortices flow past through the flexible membrane, pressure pulsations are induced. The spatial pressure modes frequency-synchronized with the vortex shedding modes are plotted in \reffig{v5a20_fft_slice} (a,c,e,g). These pressure pulsations are coupled with the membrane vibrations with similar modal shapes.

\begin{figure}[H]
	\centering 
	\subfloat[][]{\includegraphics[width=0.6\textwidth]{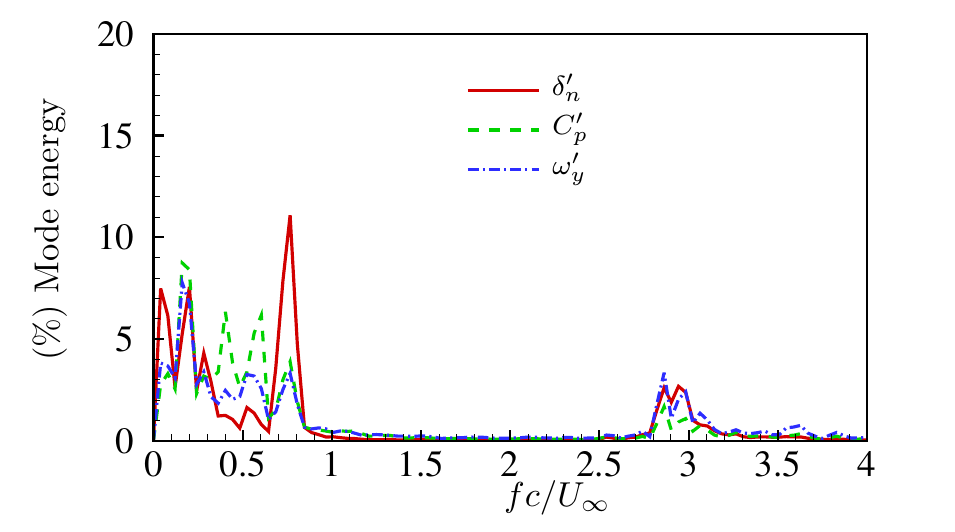}\label{v5a25_surface_mode_fft_slicea}}
	\\
	\subfloat[][]{
		\includegraphics[width=0.35\textwidth]{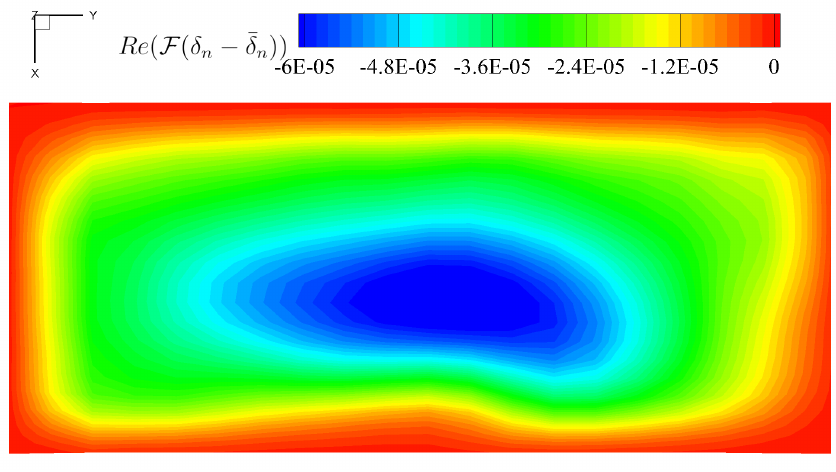}\label{v5a25_surface_mode_fft_sliceb}
	}
	\quad \quad  
	\subfloat[][]{
		\includegraphics[width=0.35\textwidth]{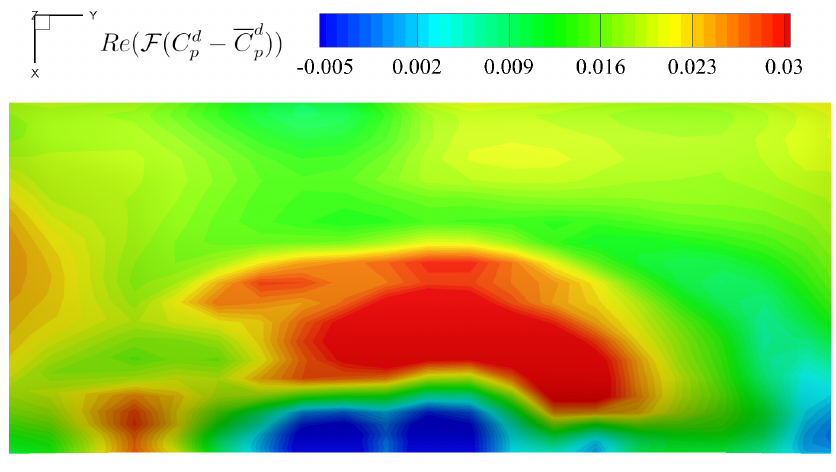}\label{v5a25_surface_mode_fft_slicec}
	}
	\\
	\subfloat[][]{
		\includegraphics[width=0.35\textwidth]{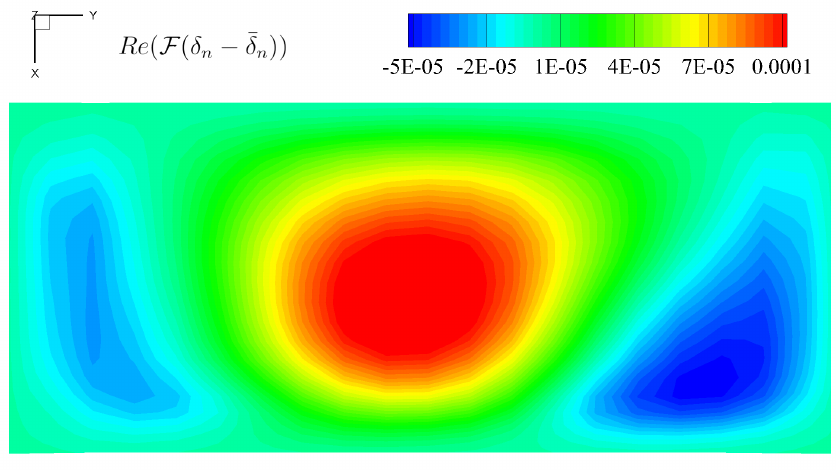}\label{v5a25_surface_mode_fft_sliced}
	}
	\quad \quad  
	\subfloat[][]{
		\includegraphics[width=0.35\textwidth]{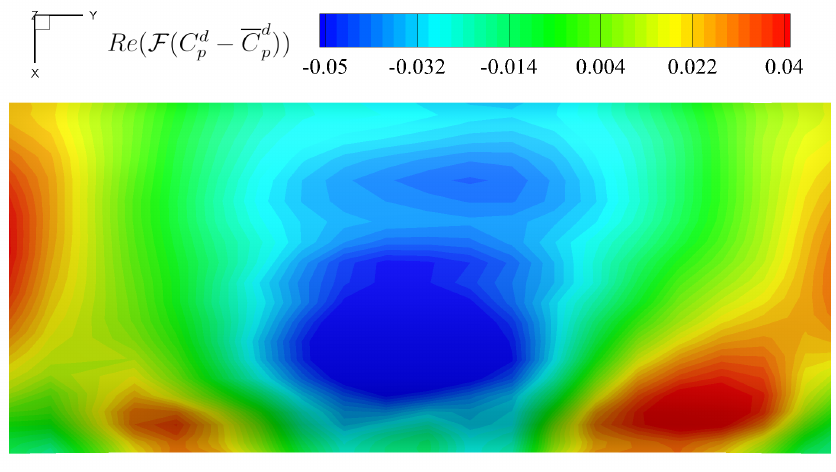}\label{v5a25_surface_mode_fft_slicee}
	}
	\\
	\subfloat[][]{
		\includegraphics[width=0.35\textwidth]{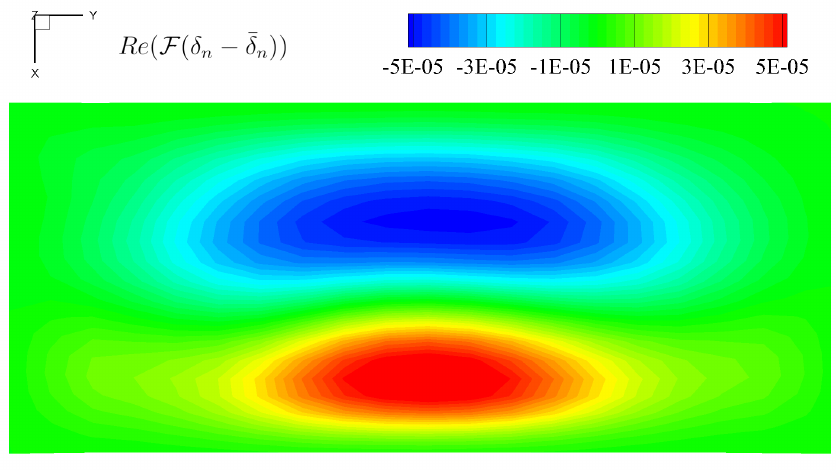}\label{v5a25_surface_mode_fft_slicef}
	}
	\quad \quad
	\subfloat[][]{
		\includegraphics[width=0.35\textwidth]{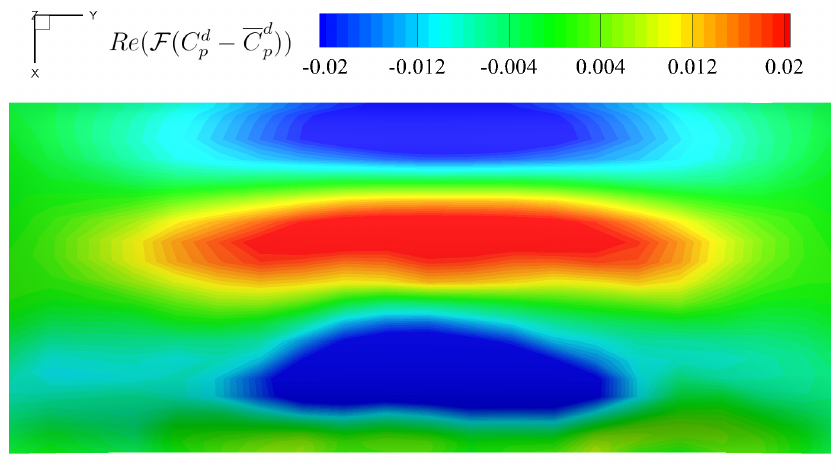}\label{v5a25_surface_mode_fft_sliceg}
	}
	\\
	\subfloat[][]{
		\includegraphics[width=0.35\textwidth]{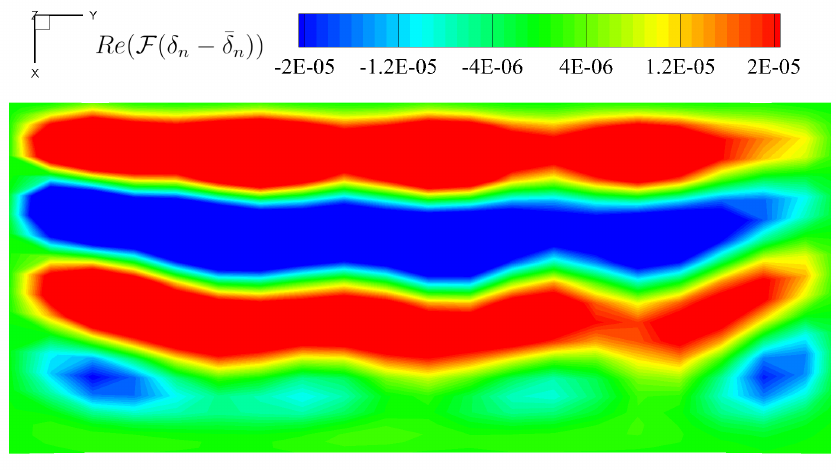}\label{v5a25_surface_mode_fft_sliceh}
	}
	\quad \quad
	\subfloat[][]{
		\includegraphics[width=0.35\textwidth]{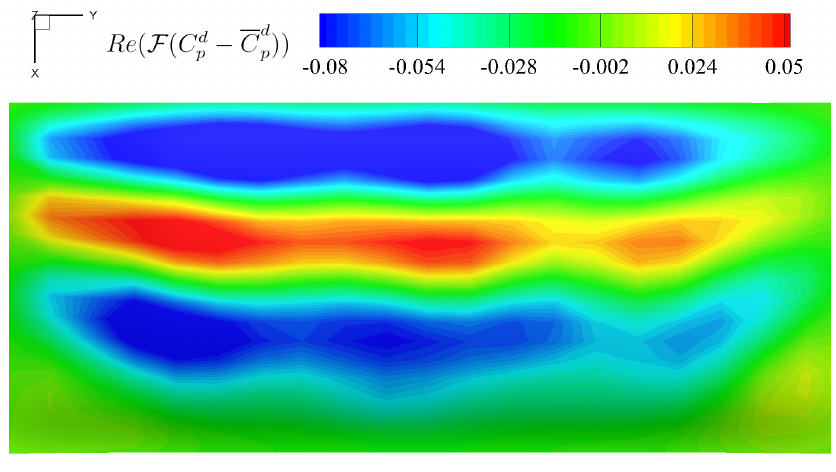}\label{v5a25_surface_mode_fft_slicei}
	}
	\caption{Aeroelastic mode decomposition of 3D flexible membrane at $\alpha$=25$^\circ$: (a) mode energy spectra of the surface displacement fluctuations, the pressure coefficient fluctuations and the $Y$-vorticity fluctuations based on the FMD analysis; (b,d,f,h) the decomposed membrane displacement modes and (c,e,g,i) the surface pressure difference modes at $f c/ U_{\infty}=$ (b,c) 0.0403, (d,e) 0.2, (f,h) 0.764 and (h,i) 2.866.}
	\label{v5a25_surface_mode_fft_slice}
\end{figure}

The mode energy spectra and the decomposed modes in the structural domain for the flexible membrane at $\alpha$=25$^\circ$ are summarized in \reffig{v5a25_surface_mode_fft_slice}. The coupled system responds over a broadband frequency range. Similar to the coupled system at $\alpha$=20$^\circ$, some non-harmonics of the dominant structural frequency are also observed in the mode energy spectra as shown in \reffig{v5a25_surface_mode_fft_slice} \subref{v5a25_surface_mode_fft_slicea}. To understand the non-periodic aeroelastic responses, we select two energetic modes at $f c/ U_{\infty}=$0.0403 and 0.2. The aeroelastic modes at the dominant frequency of 0.764 and a higher frequency of 2.866 are also extracted from the coupled system. The modal shapes of the structural displacement fluctuations and the surface pressure difference fluctuations are presented in \reffig{v5a25_surface_mode_fft_slice} (b-i), respectively. The dominant structural mode exhibits a chordwise second and spanwise first modal shape. The structural vibration modes and the correlated surface pressure difference modes at the same frequency show overall similar modal shapes.

\begin{figure}[H]
	\centering 
	\includegraphics[width=0.4\textwidth]{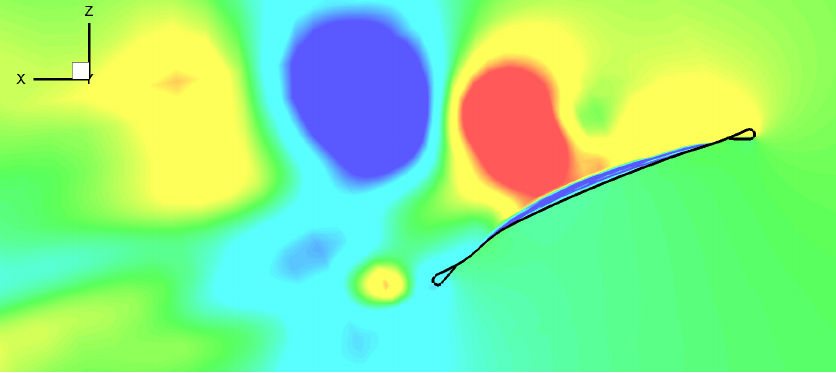}
	\quad
	\includegraphics[width=0.4\textwidth]{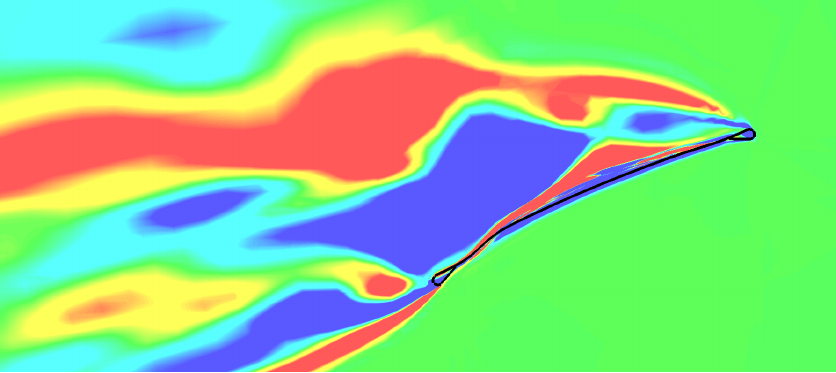}
	\\
	\subfloat[][]{\includegraphics[width=0.4\textwidth]{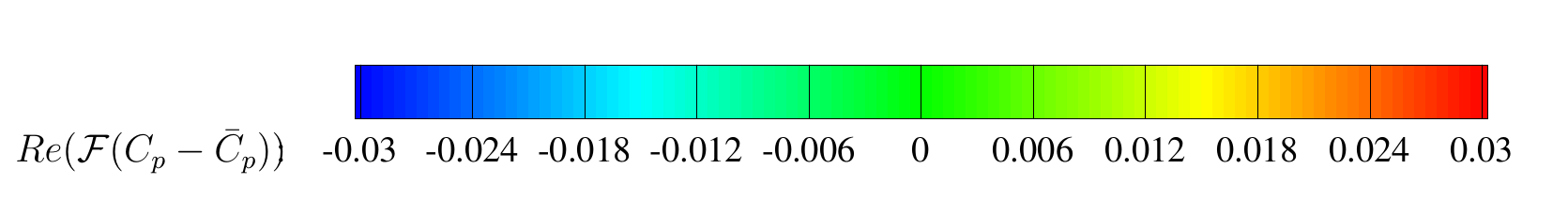}\label{v5a25_yv_fft_slicea}}
	\quad
	\subfloat[][]{\includegraphics[width=0.4\textwidth]{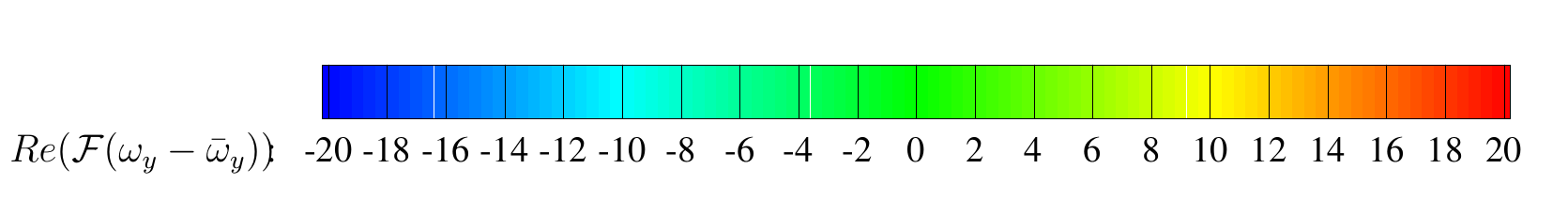}\label{v5a25_yv_fft_sliceb}}
	\\
	\includegraphics[width=0.4\textwidth]{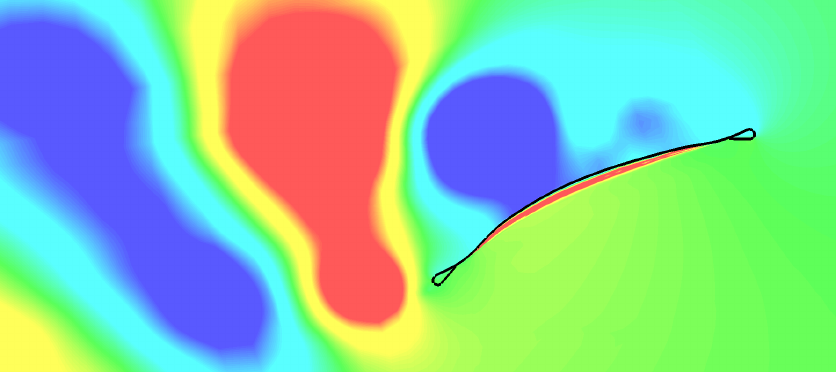}
	\quad
	\includegraphics[width=0.4\textwidth]{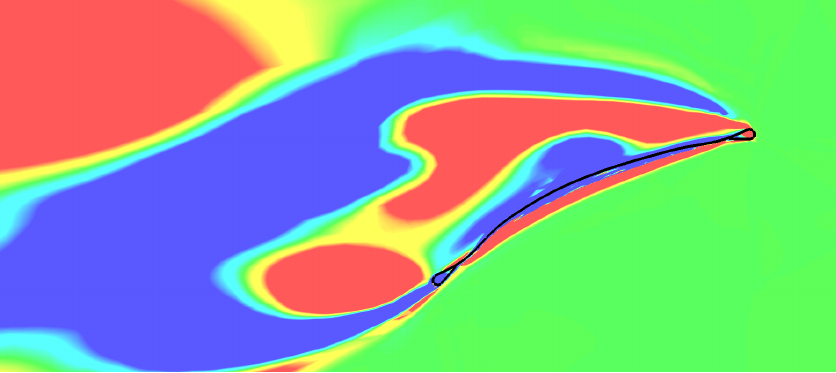}
	\\
	\subfloat[][]{\includegraphics[width=0.4\textwidth]{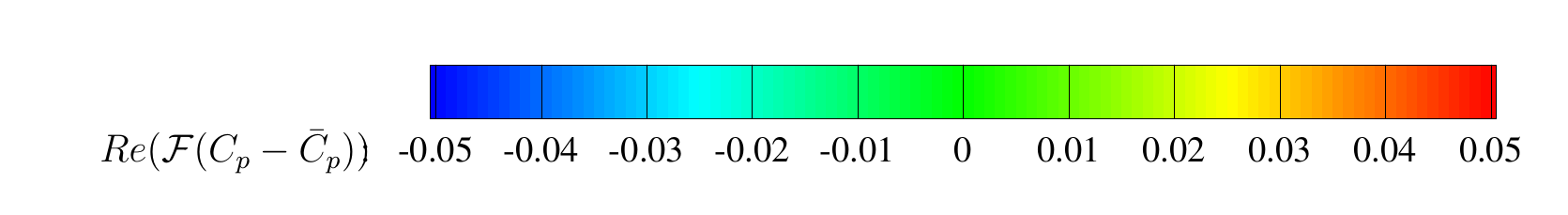}\label{v5a25_yv_fft_slicec}}
	\quad
	\subfloat[][]{\includegraphics[width=0.4\textwidth]{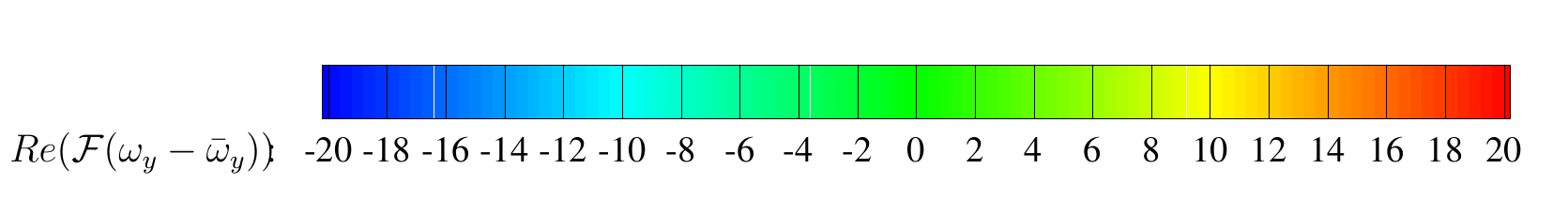}\label{v5a25_yv_fft_sliced}}
	\\
	\includegraphics[width=0.4\textwidth]{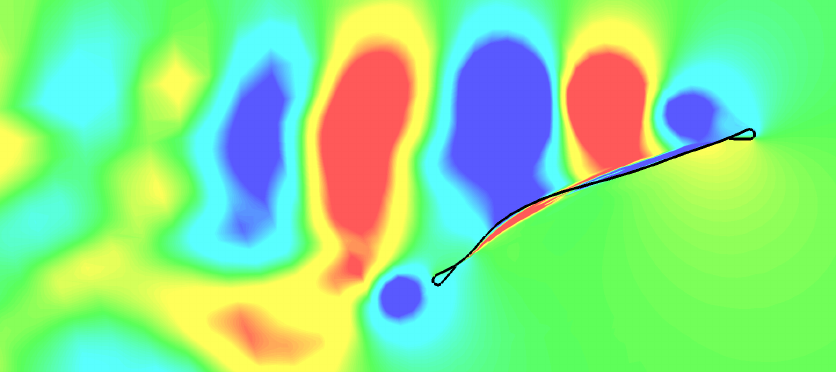}
	\quad
	\includegraphics[width=0.4\textwidth]{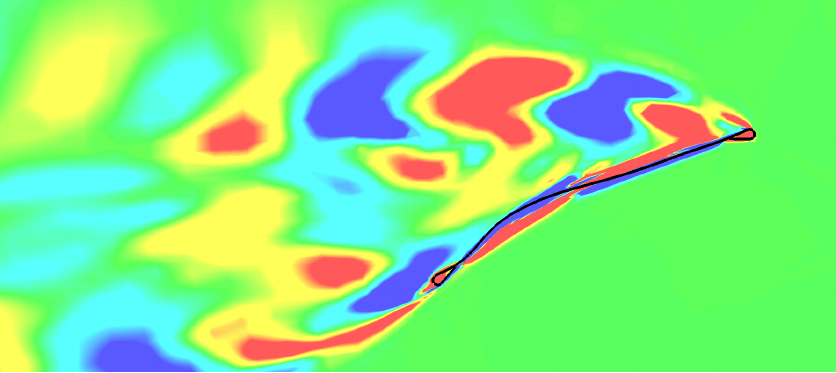}
	\\
	\subfloat[][]{\includegraphics[width=0.4\textwidth]{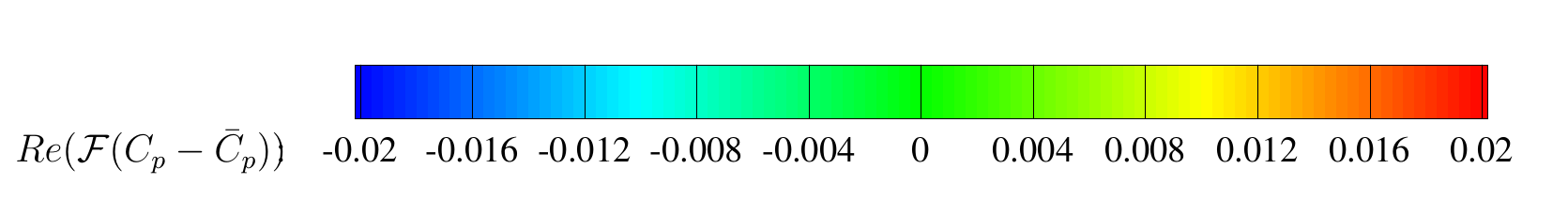}\label{v5a25_yv_fft_slicee}}
	\quad
	\subfloat[][]{\includegraphics[width=0.4\textwidth]{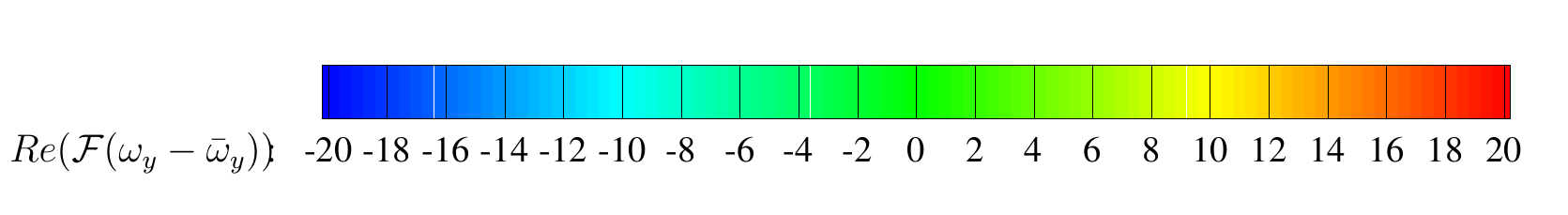}\label{v5a25_yv_fft_slicef}}
	\\
	\includegraphics[width=0.4\textwidth]{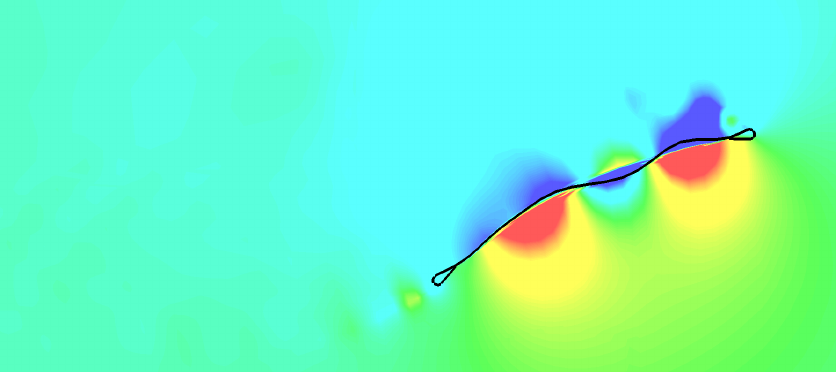}
	\quad
	\includegraphics[width=0.4\textwidth]{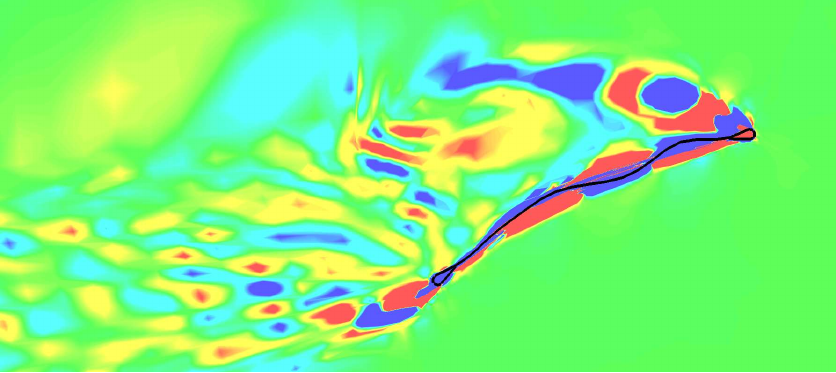}
	\\
	\subfloat[][]{\includegraphics[width=0.4\textwidth]{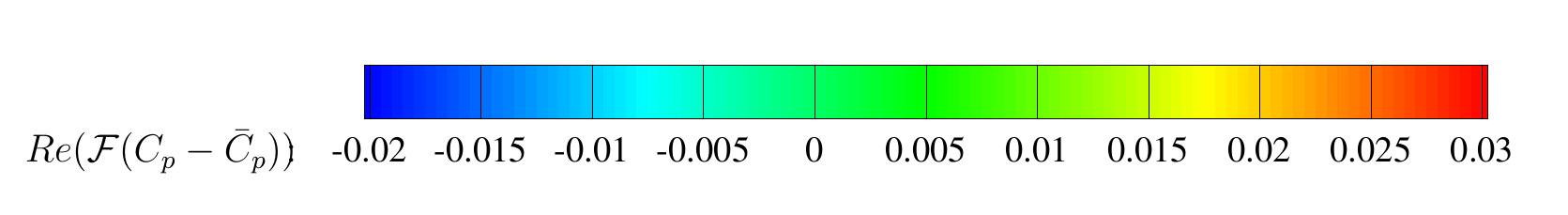}\label{v5a25_yv_fft_sliceg}}
	\quad
	\subfloat[][]{\includegraphics[width=0.4\textwidth]{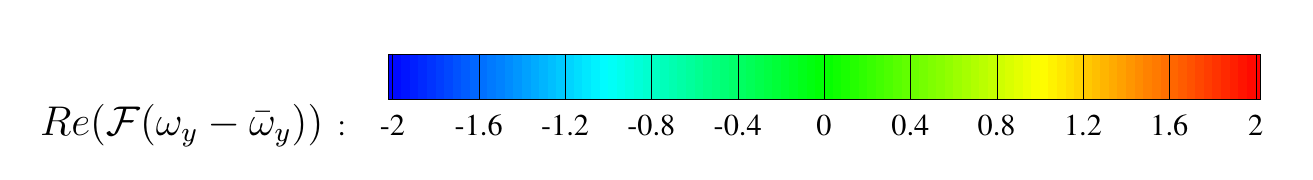}\label{v5a25_yv_fft_sliceh}}
	\caption{Aeroelastic mode decomposition of 3D flexible membrane at $\alpha$=25$^\circ$: contours of the real part of the Fourier transform coefficients of (a,c,e,g) the pressure coefficient fluctuation field and (b,d,f,h) the $Y-$vorticity fluctuation field corresponding to the non-dimensional frequency of $f c/ U_{\infty}=$ (a,b) 0.0403, (c,d) 0.2, (e,f) 0.764 and (g,h) 2.866.}
	\label{v5a25_fft_slice}
\end{figure}

The Fourier modes of the spatial pressure and the $Y$-vorticity in the fluid domain at four selected frequencies are presented in \reffig{v5a25_fft_slice}. The vorticity modes at $f c/ U_{\infty}=$0.0403 and 0.2 consist of large size vortices shed from the whole surface, which can also be seen in the instantaneous streamlines in \reffig{streamline_aoa} \subref{streamline_aoaf}. These vortex shedding patterns are coupled with the chordwise first structural mode. The vortices are shed from the leading edge at the dominant frequency of 0.764 and a chordwise second structural mode is observed. The vortex structures become complex and the vortex sizes are reduced significantly at $f c/ U_{\infty}=$2.866, which are associated with a chordwise third structural mode.

\subsection{Effect of flexibility and aeroelastic mode selection}
From the observation of the aeroelastic response and the spatial flow structure of the flexible membrane, several intertwined modes are excited through fluid-membrane interaction at higher angles of attack. The vortex shedding frequency is synchronized with the membrane vibration frequency, leading to the frequency lock-in phenomenon. As the angle of attack increases, the aeroelastic response tends to be non-periodic. The results indicate that membrane flexibility plays an important role in selecting particular aeroelastic modes via an underlying mechanism. To explore the role of flexibility in the aeroelastic mode selection process, we further simulate rigid flat wings and rigid cambered wings at three angles of attack. 

\begin{figure}[H]
	\centering 
	\subfloat[][]{\includegraphics[width=0.3\textwidth,height=0.15\textwidth]{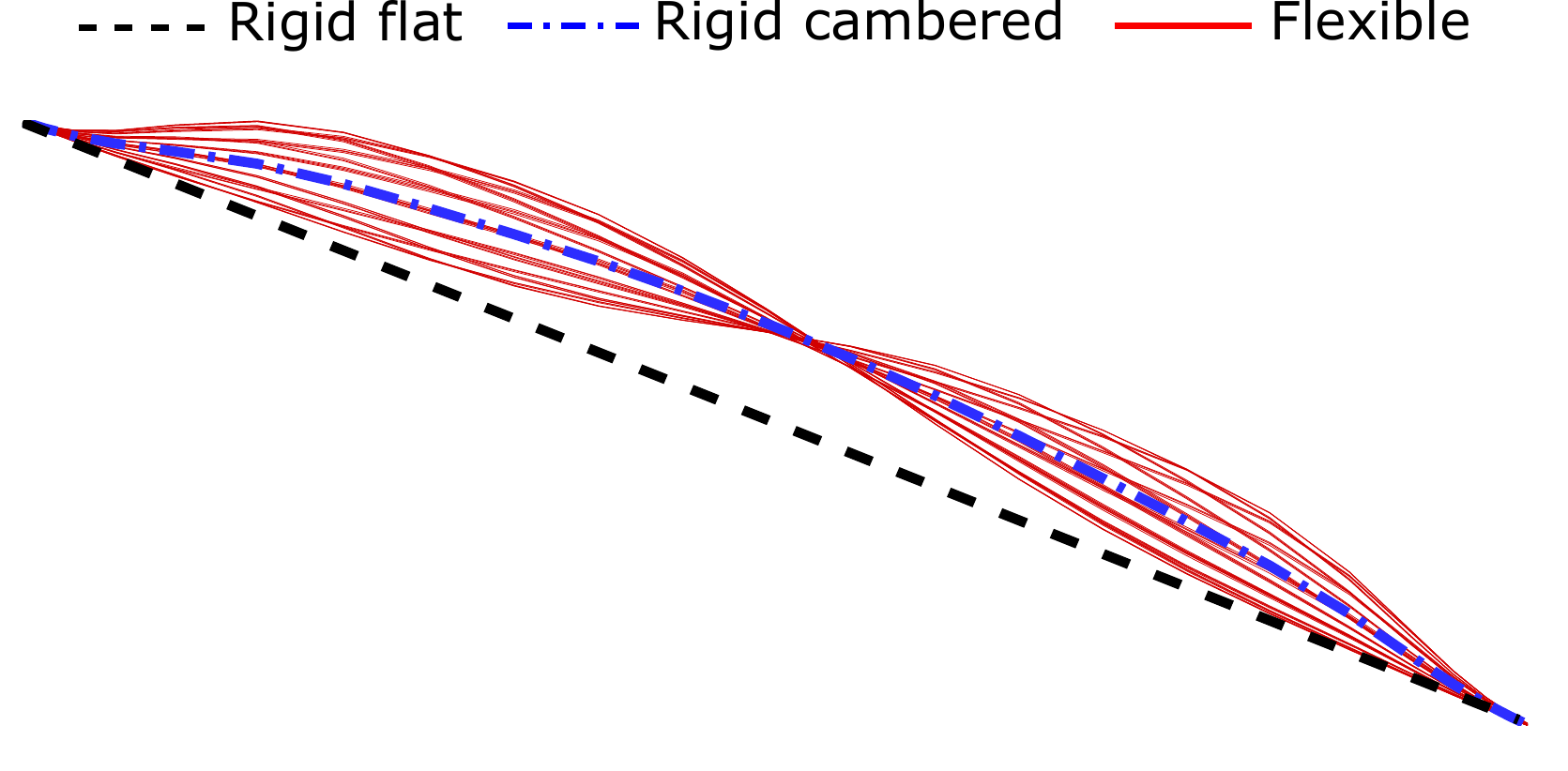}\label{rigid_flexiblea}}
	\
	\subfloat[][]{\includegraphics[width=0.3\textwidth,height=0.15\textwidth]{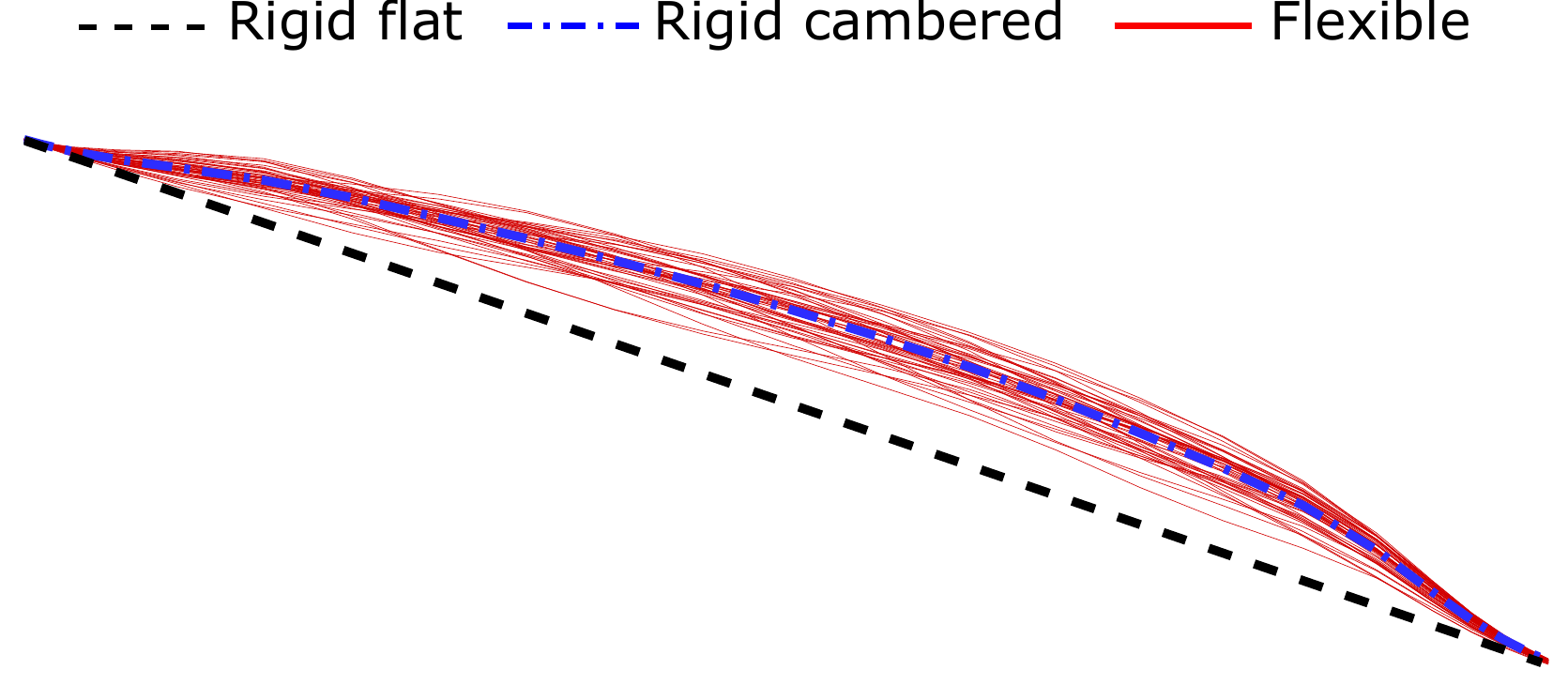}\label{rigid_flexibleb}}
	\
	\subfloat[][]{\includegraphics[width=0.3\textwidth,height=0.15\textwidth]{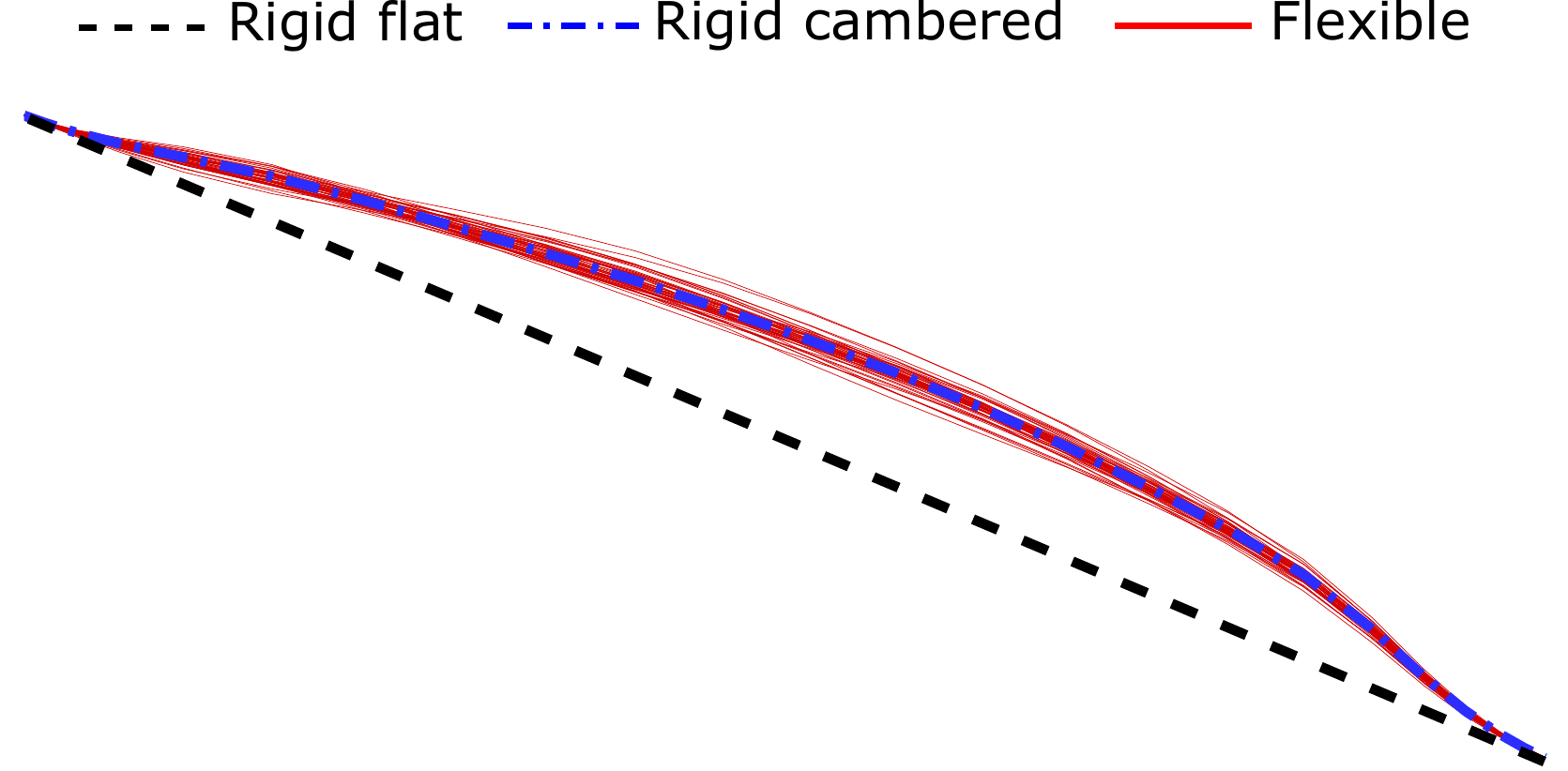}\label{rigid_flexiblec}}
	\\
	\subfloat[][]{\includegraphics[width=0.3\textwidth]{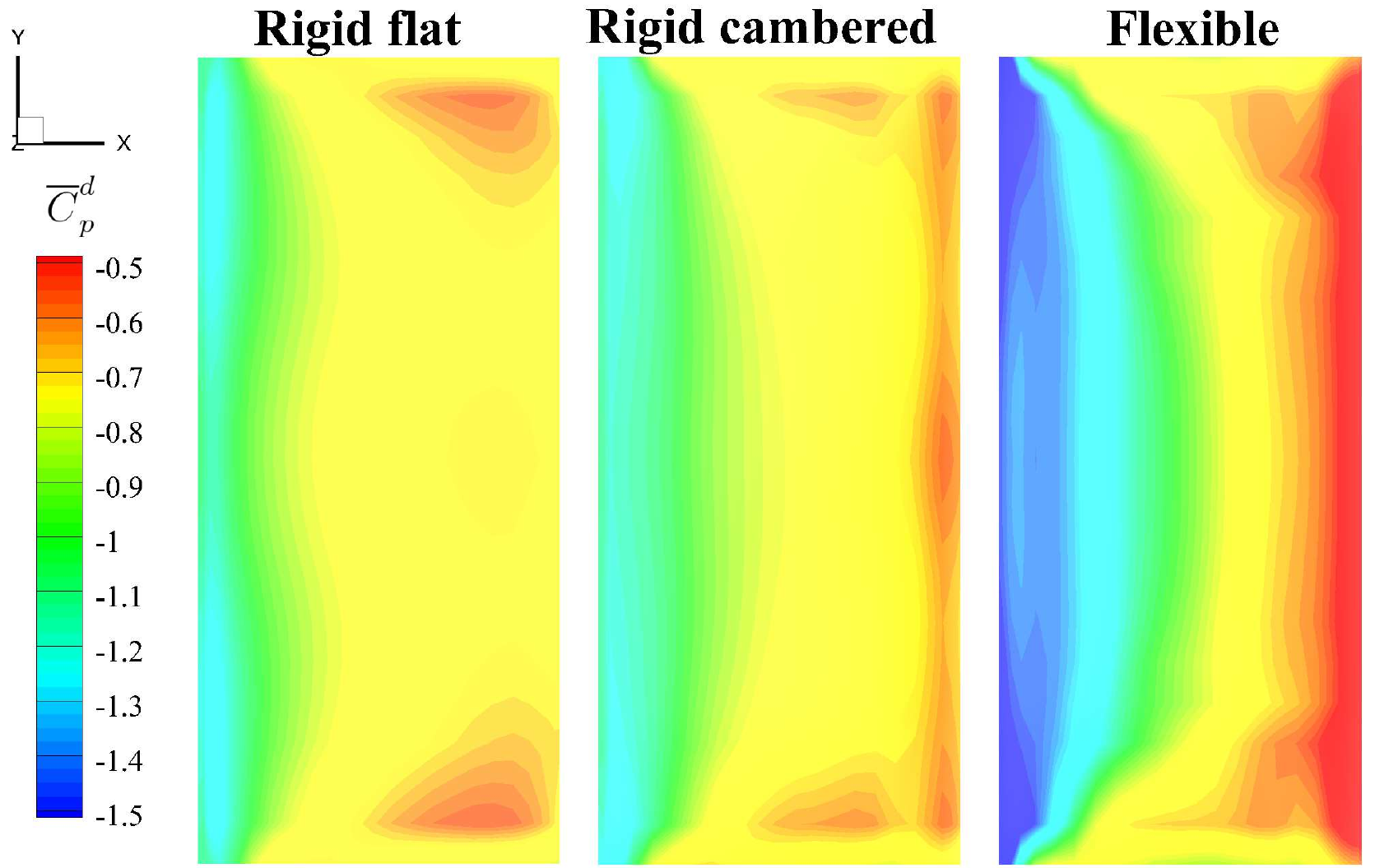}\label{rigid_flexibled}}
	\
	\subfloat[][]{\includegraphics[width=0.3\textwidth]{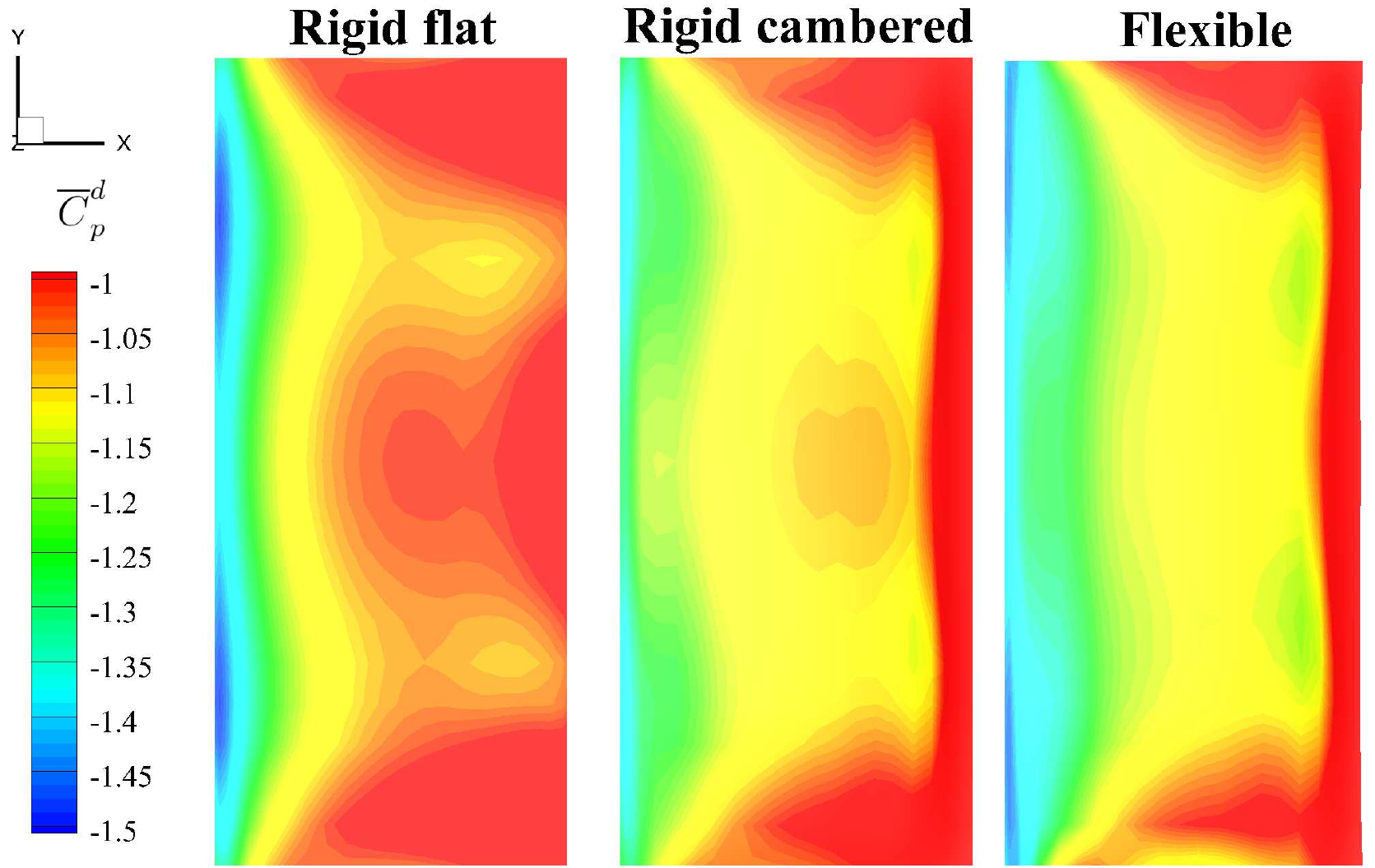}\label{rigid_flexiblee}}
	\
	\subfloat[][]{\includegraphics[width=0.3\textwidth]{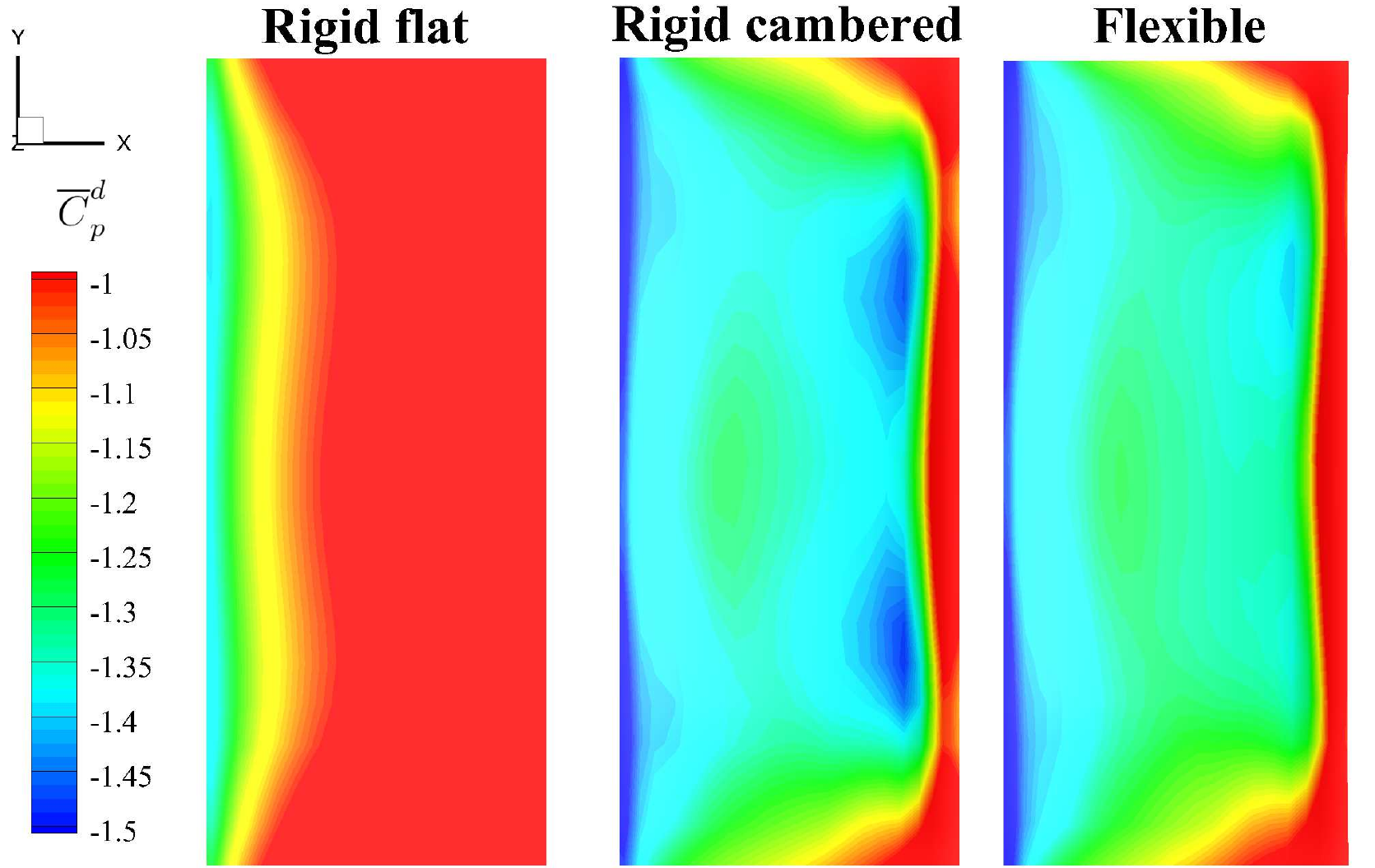}\label{rigid_flexiblef}}
	\\
	\subfloat[][]{\includegraphics[width=0.3\textwidth]{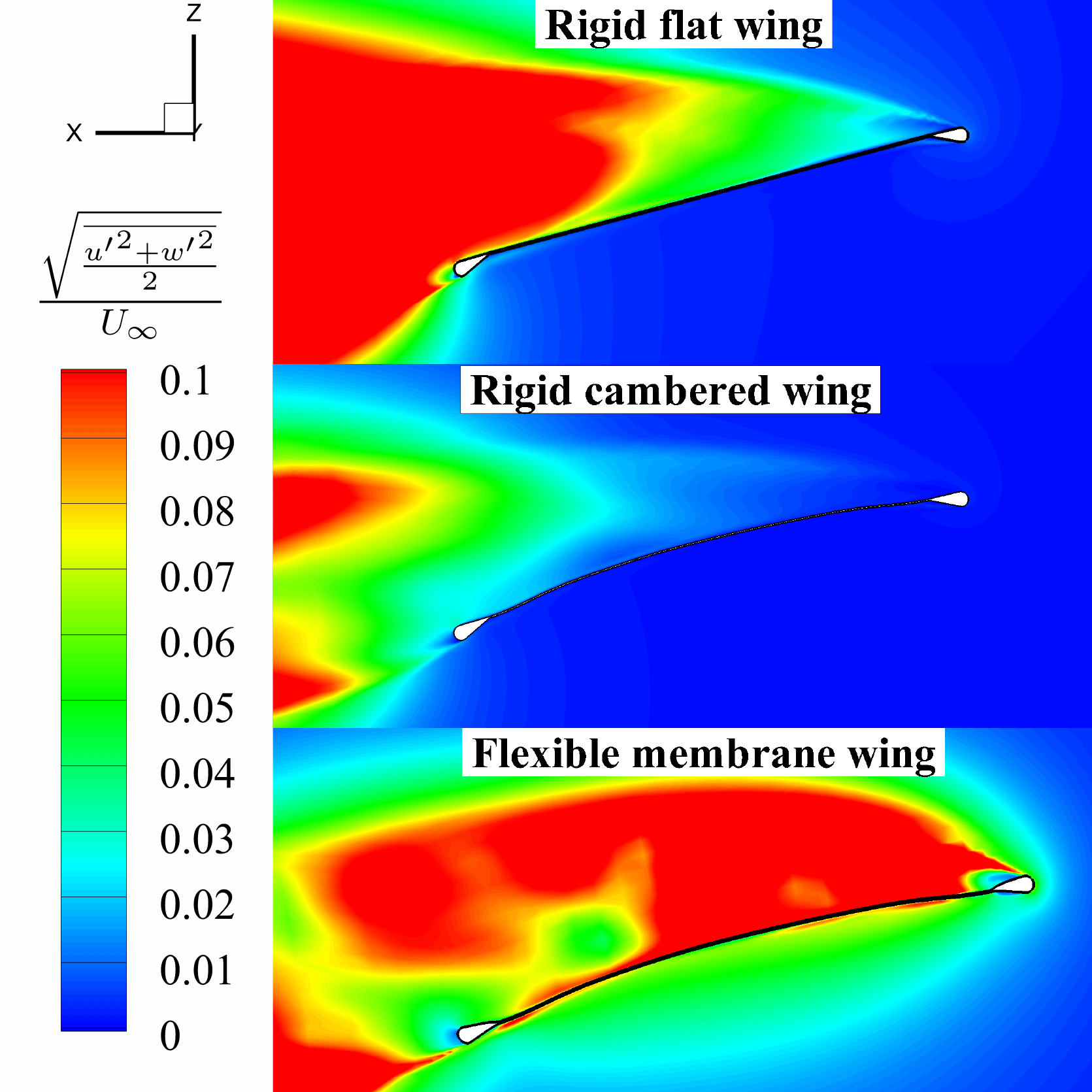}\label{rigid_flexibleg}}
	\
	\subfloat[][]{\includegraphics[width=0.3\textwidth]{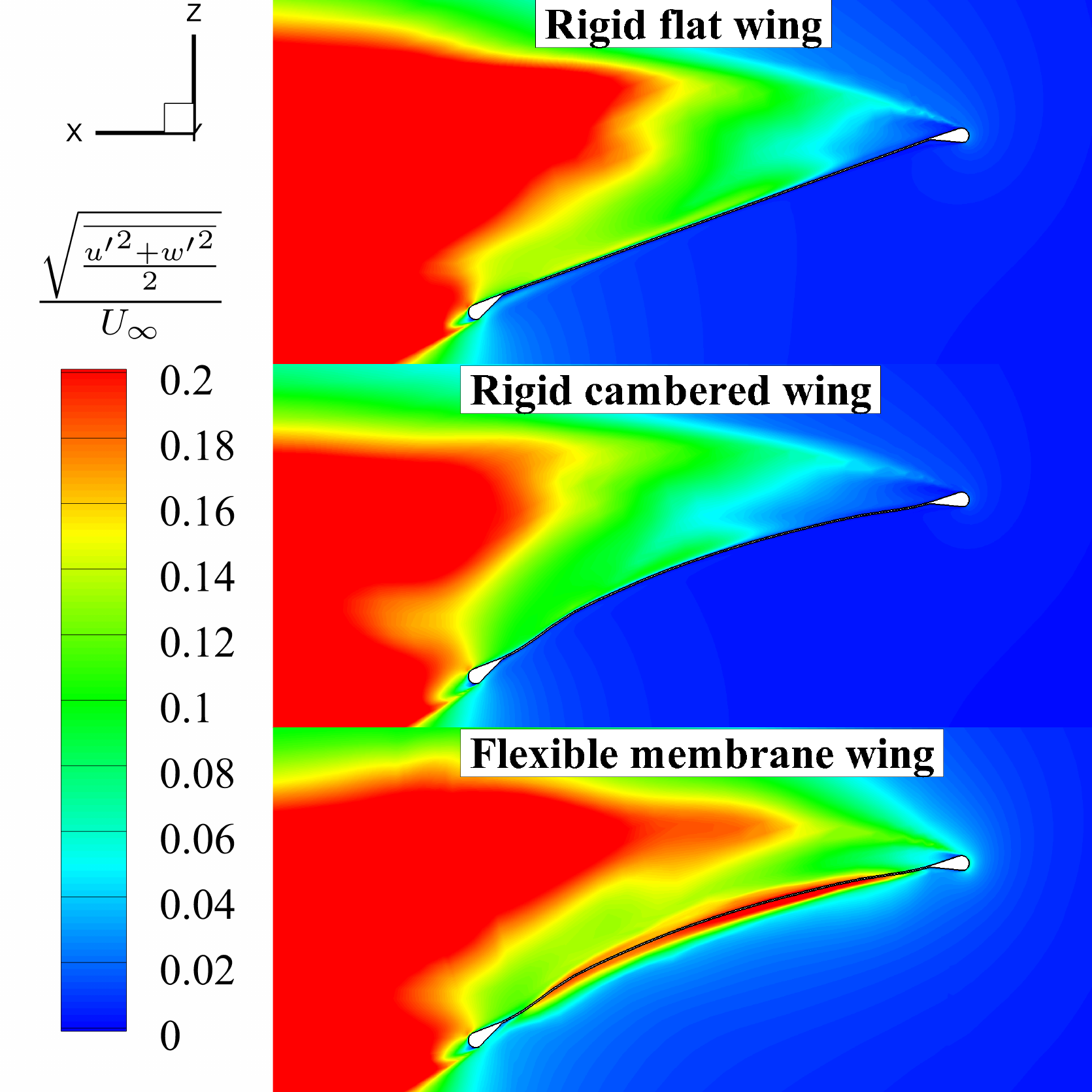}\label{rigid_flexibleh}}
	\
	\subfloat[][]{\includegraphics[width=0.3\textwidth]{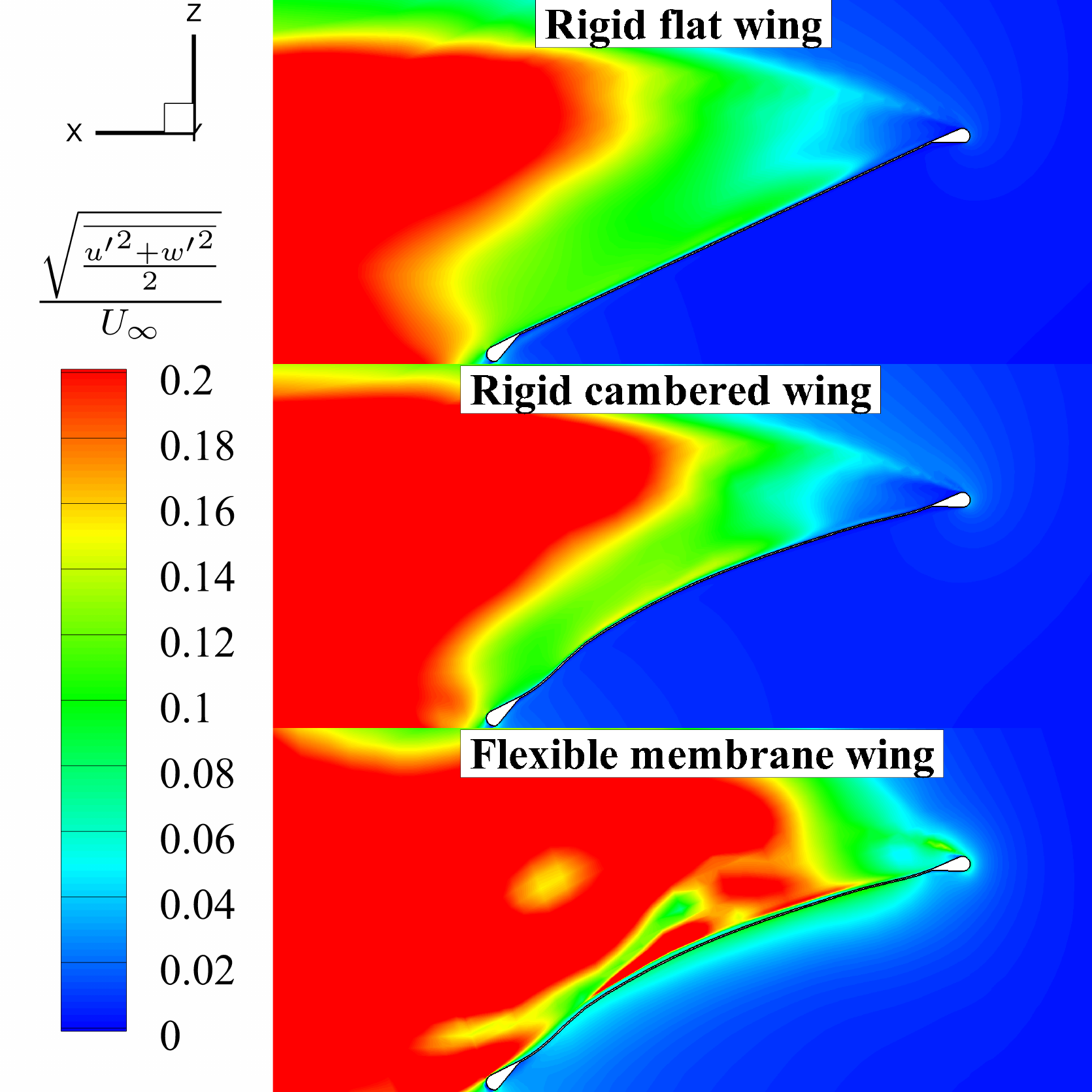}\label{rigid_flexiblei}}
	\caption{Comparison of flow features between rigid flat wings, rigid cambered wings and flexible membrane wings at $\alpha$= (a,d,g) 15$^\circ$, (b,e,h) 20$^\circ$ and (c,f,i) 25$^\circ$: (a,b,c) membrane profiles along the chord at the mid-span location, (d,e,f) time-averaged pressure coefficient difference between the upper surface and the lower surface and (g,h,i) turbulent intensity on the mid-span plane.}
	\label{rigid_flexible}
\end{figure}

\refFigs{rigid_flexible} (a-c) present the comparison of the membrane profiles along the chord at the mid-span location between these three types of wings at different angles of attack, respectively. The rigid flat wing has the same wing geometry as the undeformed geometry of the flexible membrane. The rigid cambered wing shares the same wing shape as the mean wing shape of the flexible membrane under aerodynamic loads. The flexibility affects the membrane dynamics from two aspects, namely (i) camber effect and (ii) flow-excited vibration. The camber effect can be investigated by comparing the flow features between the rigid flat wing and the rigid cambered wing. When the flexibility is introduced into the flexible membrane and coupled with the unsteady flow, the effect of the flow-excited vibration can be examined. This investigation provides a perspective to explore how the flexible structure is coupled with the unsteady flow to alter the flow features and excite particular structural modes. 

\begin{figure}[H]
	\centering 
	\subfloat[][Rigid flat wing at $\alpha=15^\circ$]{
		\includegraphics[width=0.24\textwidth]{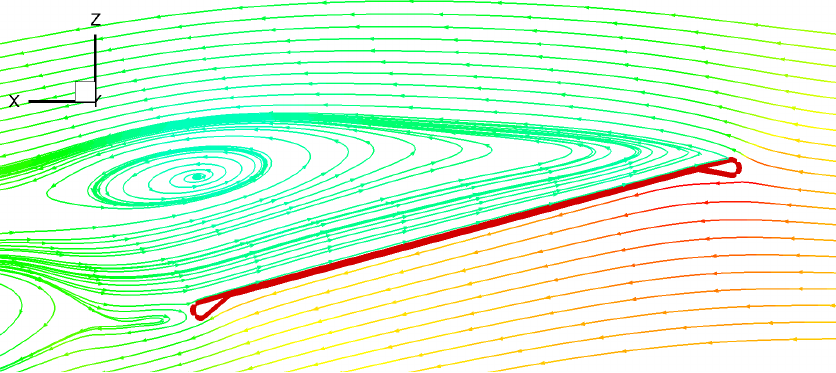}
		\includegraphics[width=0.24\textwidth]{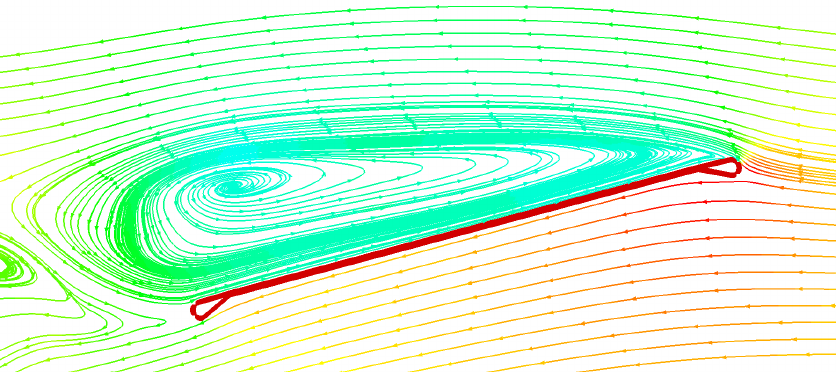}
		\includegraphics[width=0.24\textwidth]{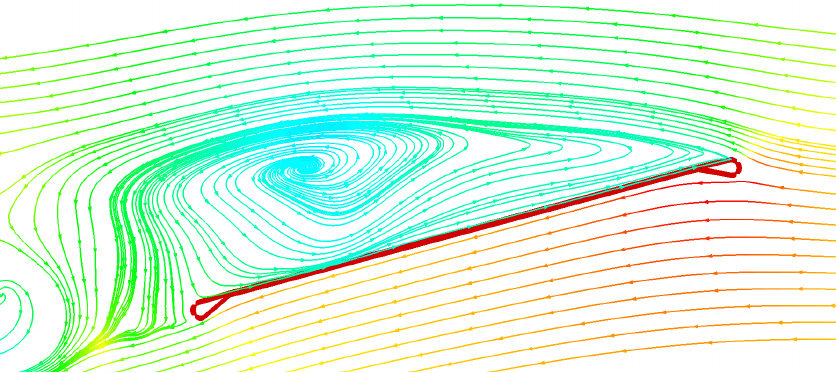}
		\includegraphics[width=0.24\textwidth]{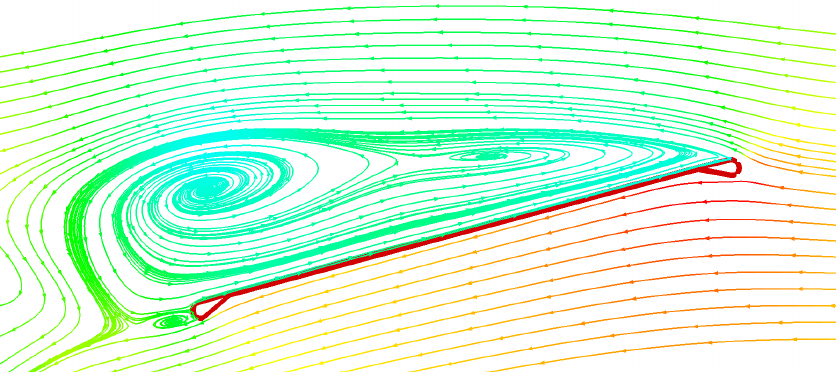}
		\label{streamline_rigid_a15}
	}
	\\
	\subfloat[][Rigid flat wing at $\alpha=20^\circ$]{
		\includegraphics[width=0.24\textwidth]{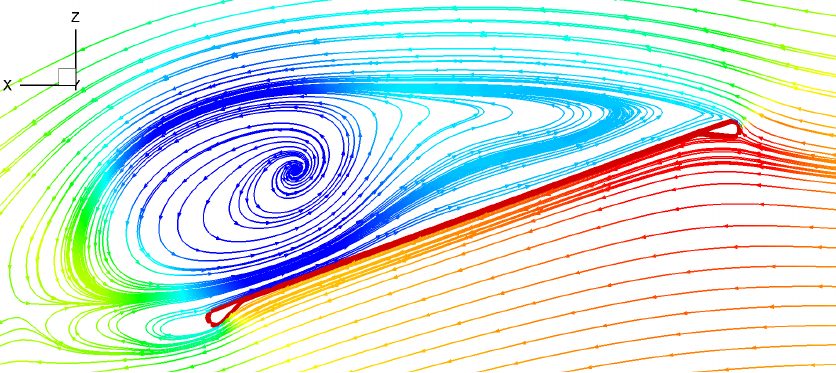}
		\includegraphics[width=0.24\textwidth]{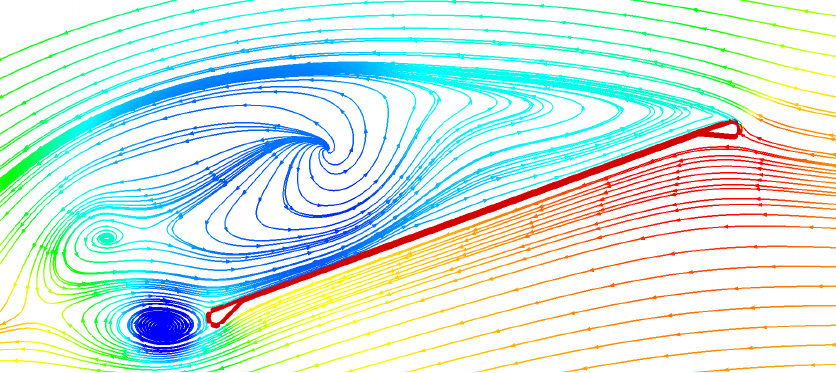}
		\includegraphics[width=0.24\textwidth]{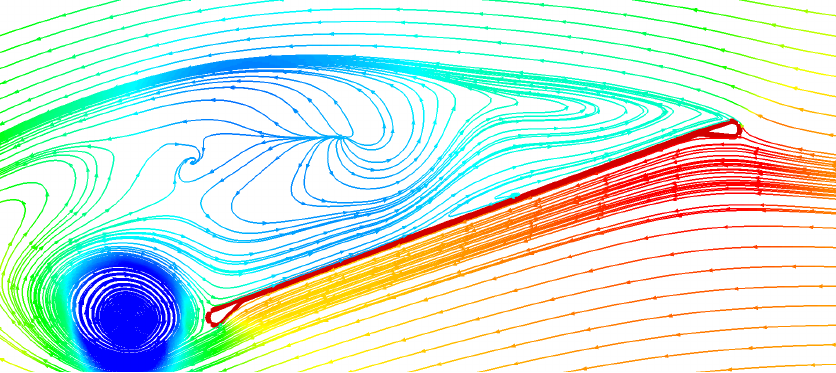}
		\includegraphics[width=0.24\textwidth]{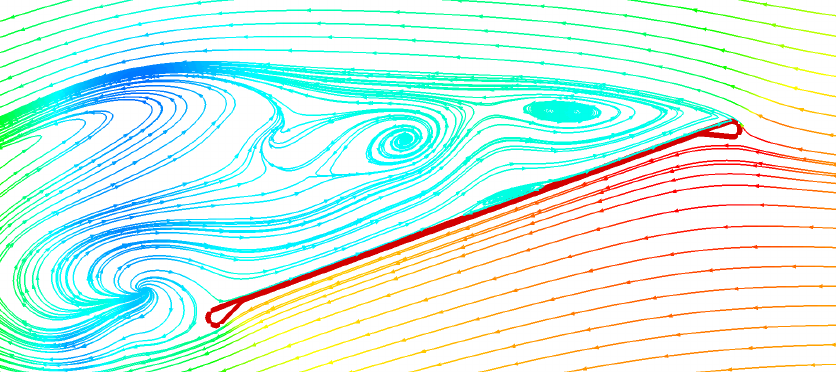}
		\label{streamline_rigid_a20}
	}
    \\
   \subfloat[][Rigid flat wing at $\alpha=25^\circ$]{
   	\includegraphics[width=0.24\textwidth]{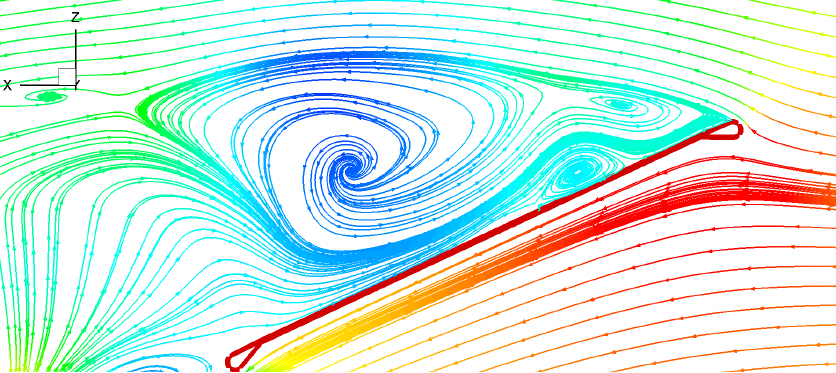}
   	\includegraphics[width=0.24\textwidth]{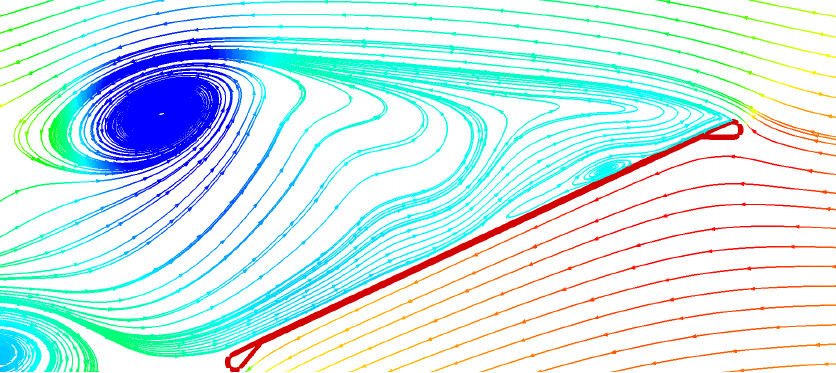}
   	\includegraphics[width=0.24\textwidth]{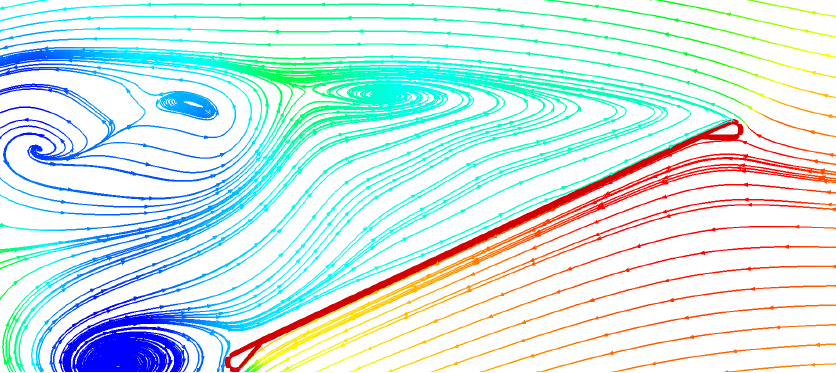}
   	\includegraphics[width=0.24\textwidth]{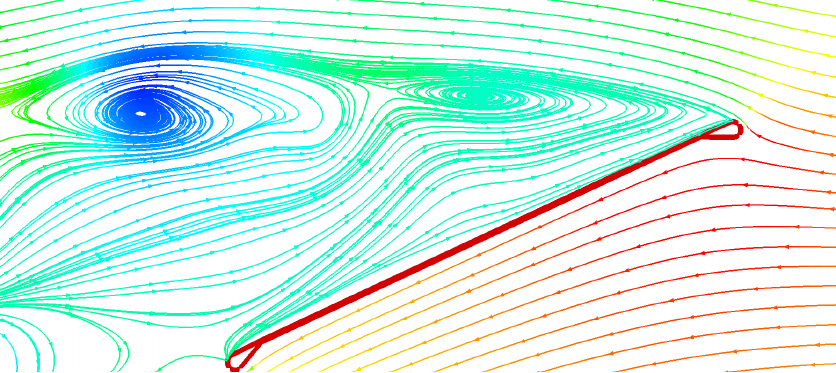}
   	\label{streamline_rigid_a25}
   }
   \\
	\subfloat[][Rigid cambered wing at $\alpha=15^\circ$]{
		\includegraphics[width=0.24\textwidth]{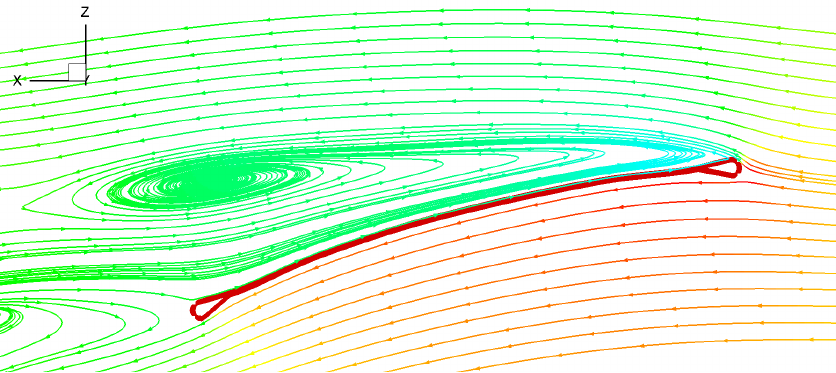}
		\includegraphics[width=0.24\textwidth]{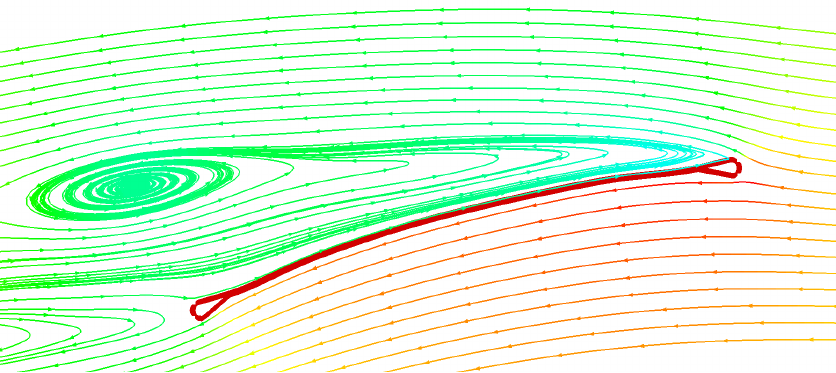}
		\includegraphics[width=0.24\textwidth]{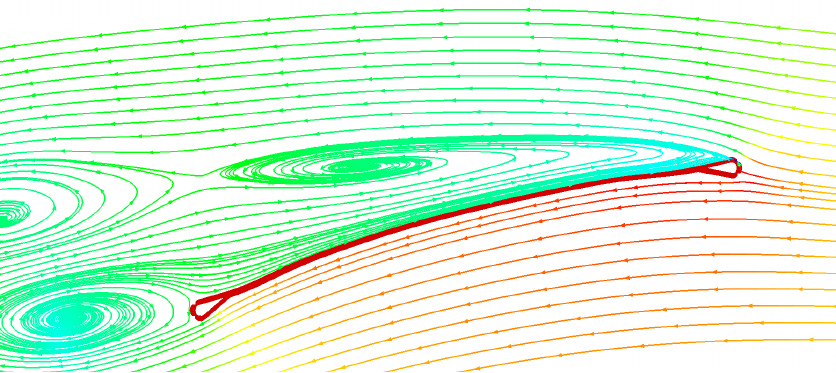}
		\includegraphics[width=0.24\textwidth]{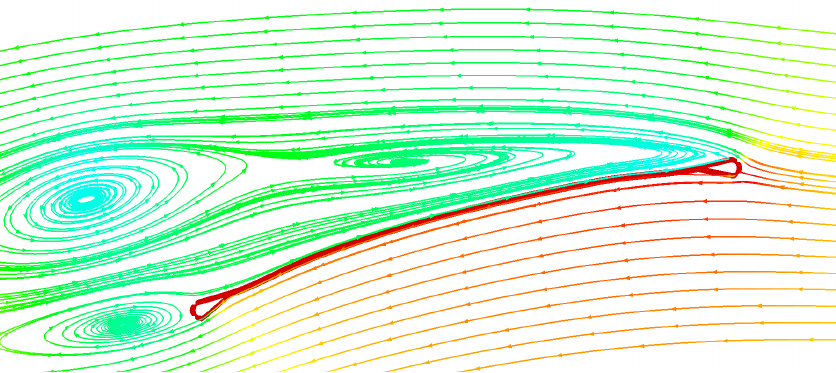}
		\label{streamline_camber_a15}
	}
	\\
	\subfloat[][Rigid cambered wing at $\alpha=20^\circ$]{
		\includegraphics[width=0.24\textwidth]{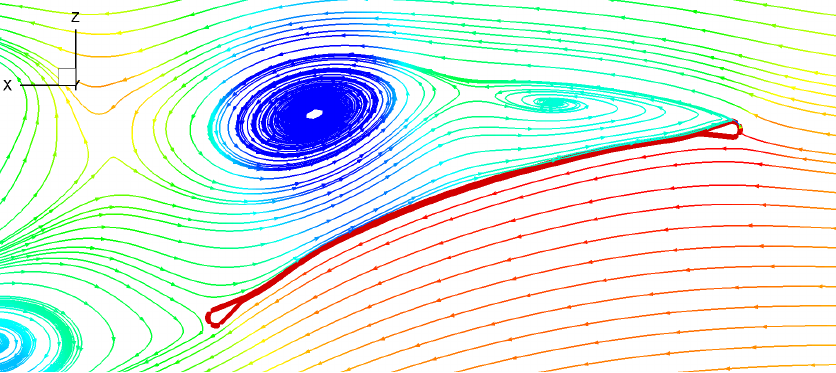}
		\includegraphics[width=0.24\textwidth]{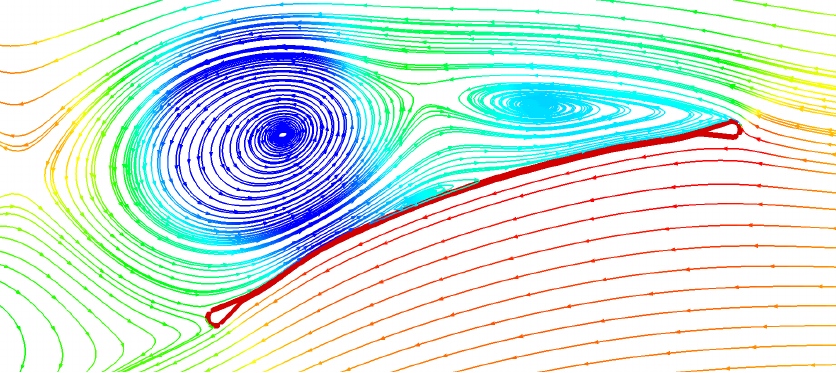}
		\includegraphics[width=0.24\textwidth]{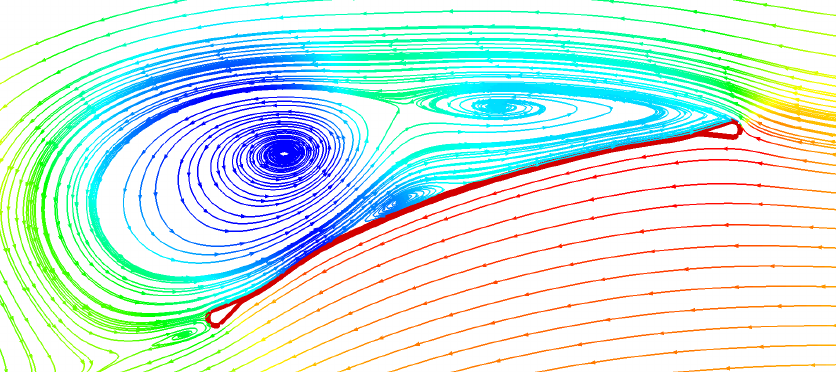}
		\includegraphics[width=0.24\textwidth]{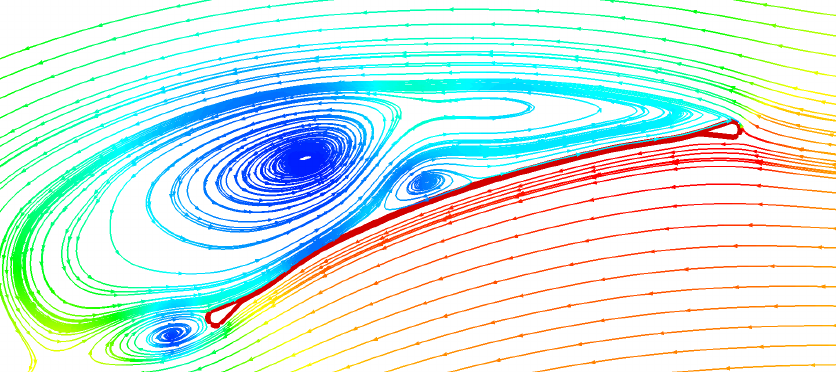}
		\label{streamline_camber_a20}
	}
   \\
   \subfloat[][Rigid cambered wing at $\alpha=25^\circ$]{
   	\includegraphics[width=0.24\textwidth]{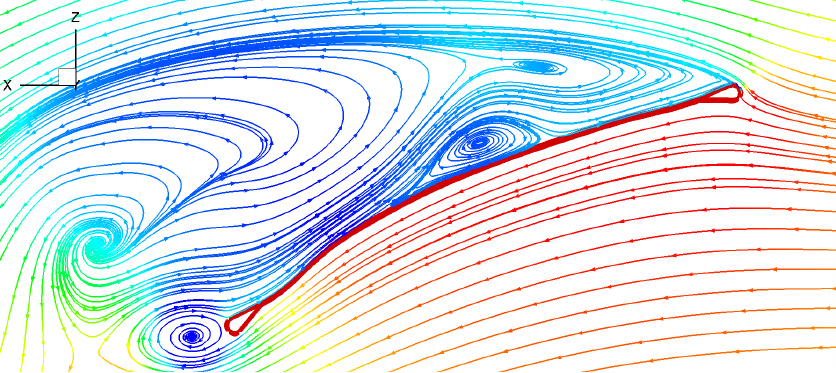}
   	\includegraphics[width=0.24\textwidth]{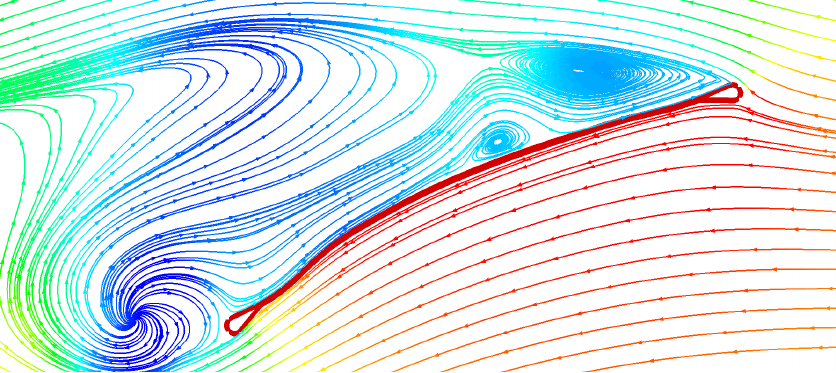}
   	\includegraphics[width=0.24\textwidth]{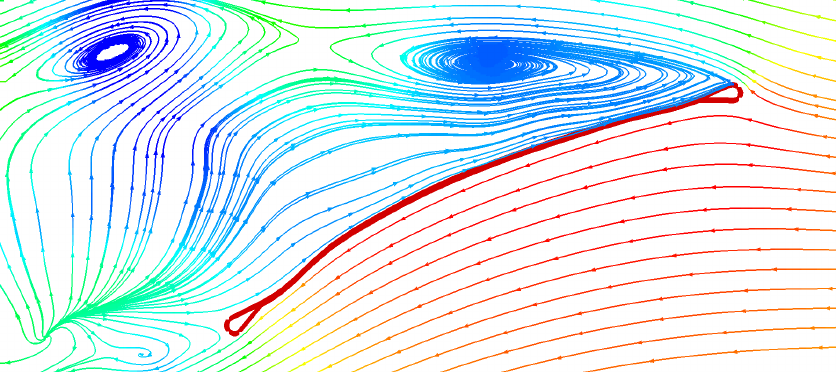}
   	\includegraphics[width=0.24\textwidth]{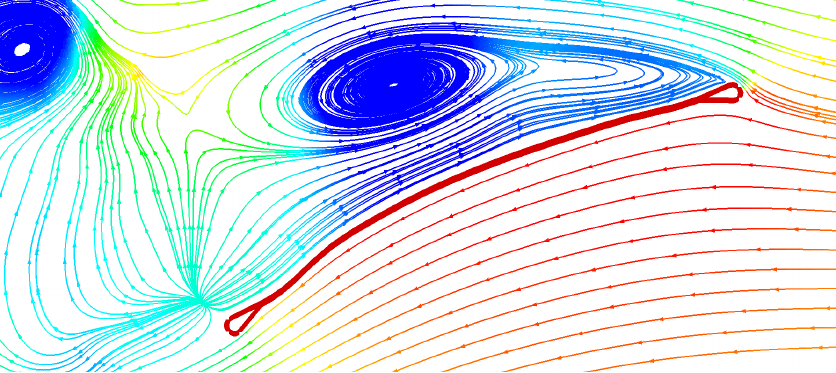}
   	\label{streamline_camber_a25}
   }
    \\
	\includegraphics[width=0.5\textwidth]{./results/v5/cp_leng.pdf}
	\caption{Flow past a 3D rectangular membrane wing: instantaneous streamlines on the mid-span plane colored by pressure coefficient for (a,b,c) a rigid flat wing and (d,e,f) a rigid cambered wing at $\alpha=$ (a,d) 15$^\circ$, (b,e) 20$^\circ$ and (c,f) 25$^\circ$.}
	\label{streamline_com}
\end{figure}

\begin{figure}[H]
	\centering 
	\subfloat[][]{\includegraphics[width=0.6\textwidth]{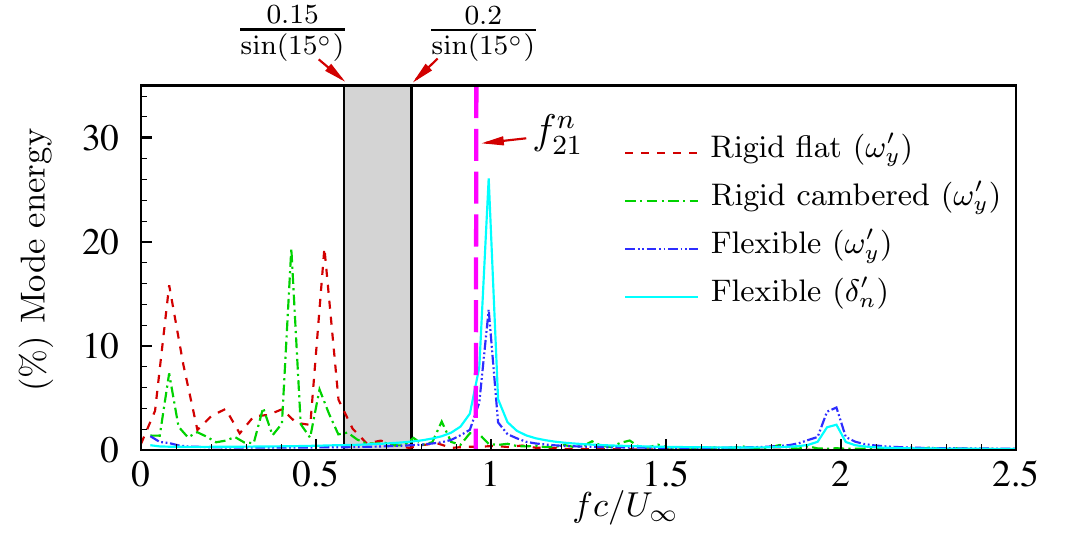}\label{v5_psd_coma}}
	\\
	\subfloat[][]{\includegraphics[width=0.6\textwidth]{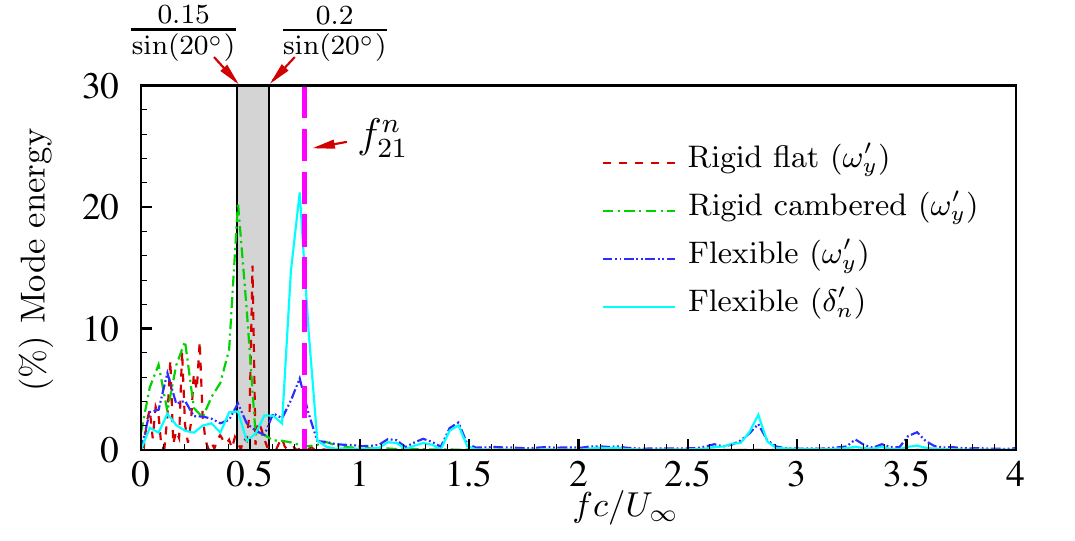}\label{v5_psd_comb}}
	\\
	\subfloat[][]{\includegraphics[width=0.6\textwidth]{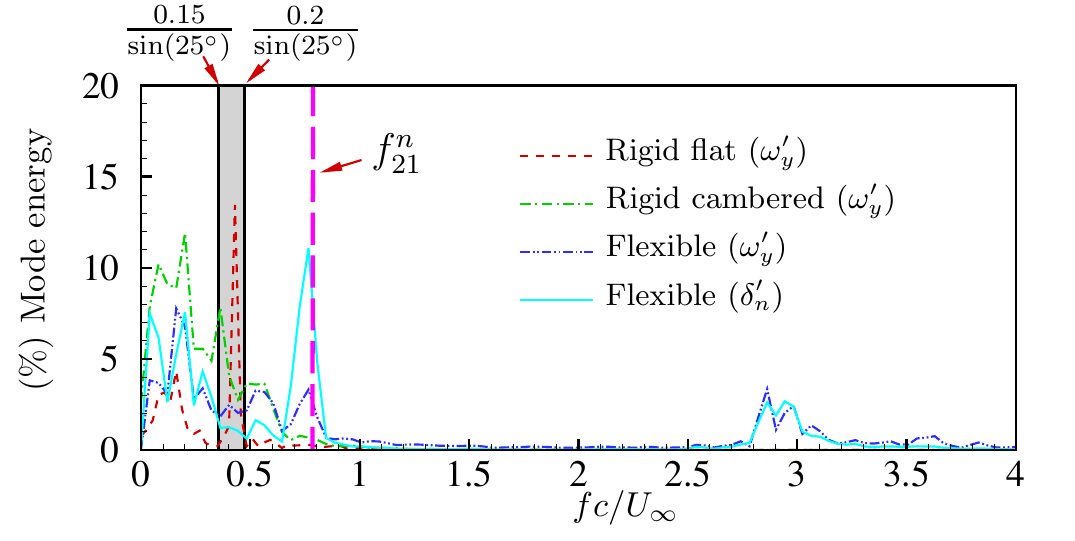}\label{v5_psd_comc}}
	\caption{Comparison of global mode energy spectra of the spatial $Y$-vorticity fluctuation ($\omega_y^{\prime}$) between a rigid flat wing, a rigid cambered wing and a flexible membrane at $\alpha$= (a) 15$^\circ$, (b) 20$^\circ$, and (c) 25$^\circ$. }
	\label{v5_psd_com}
\end{figure}

To further investigate the effect of flexibility on the flow features, we compare the time-averaged pressure coefficient difference on the membrane surface and the turbulent intensity on the mid-span plane for the three types of wings at different angles of attack. It can be observed from \reffigs{rigid_flexible} (d-f) that the wing camber can enlarge the suction area. When the membrane vibration is introduced, the suction area is further extended to the trailing edge. As shown in \reffigs{rigid_flexible} (g-i), compared to the rigid flat wing, the wing camber can suppress the turbulent intensity at $\alpha=15^\circ$, but exhibits smaller influences on the turbulent intensity at higher angles of attack. When the membrane vibration is coupled with the unsteady flow, the high turbulent intensity region gets closer to the membrane surface.

A comparison of the instantaneous streamlines on the mid-span plane for rigid flat wings and rigid cambered wings at three angles of attack is presented in \reffig{streamline_com}. By comparing with the instantaneous streamlines of the flexible membrane shown in \reffig{streamline_aoa}, we find that the vortical structures are significantly changed by coupling with the membrane vibration at $\alpha=15^\circ$. As the angle of attack further increases, the vortical structures are affected by the induced membrane vibration to some extent. However, the rigid cambered wing and the flexible membrane generate some similar vortical structures with large sizes. These produced vortices with similar structures can be regarded as the vortices behind a bluff body, which are associated with the non-periodic responses at higher angles of attack. With the aid of the mode decomposition technique, we further examine the relationship between the membrane aeroelasticity and the bluff body vortex shedding instability.

\begin{figure}[H]
	\centering 
	\includegraphics[width=0.38\textwidth]{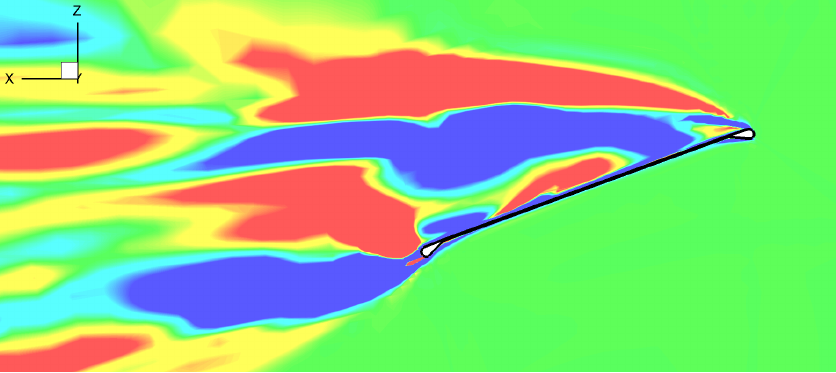}
	\quad
	\includegraphics[width=0.38\textwidth]{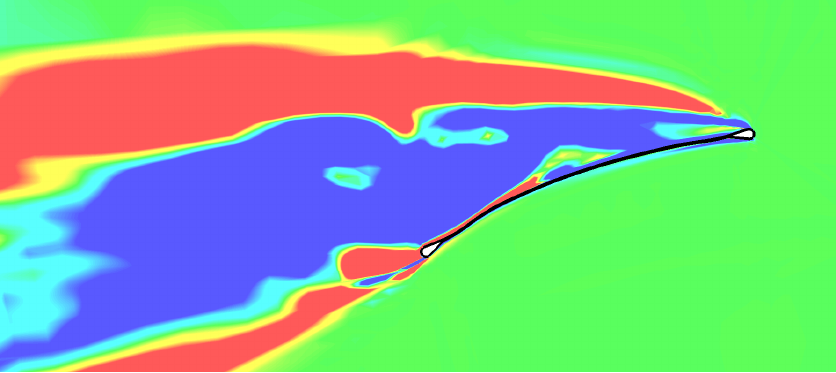}
	\\
	\subfloat[][]{\includegraphics[width=0.38\textwidth]{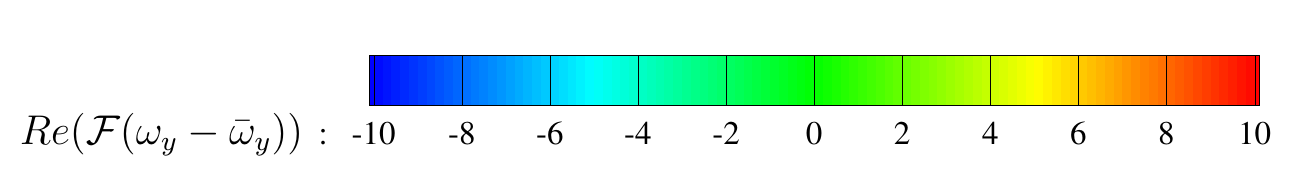}\label{yv_mode_second_a20a}}
	\subfloat[][]{\includegraphics[width=0.38\textwidth]{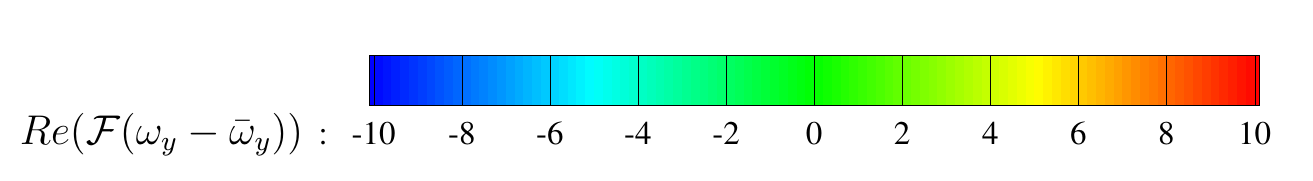}\label{yv_mode_seconda_a20c}}
	\\
	\includegraphics[width=0.38\textwidth]{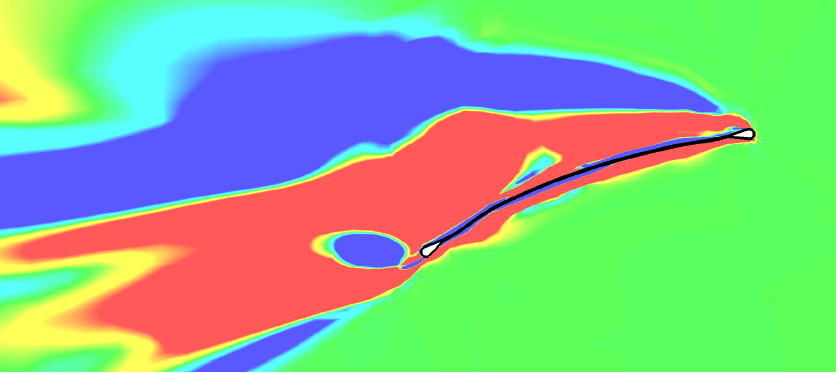}
	\\
	\subfloat[][]{\includegraphics[width=0.38\textwidth]{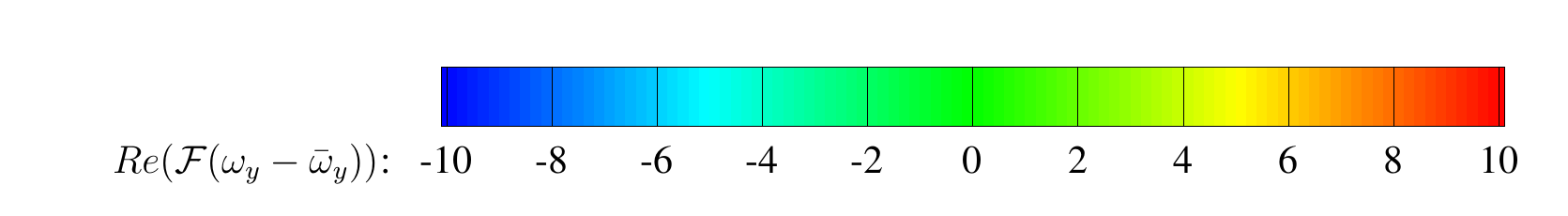}\label{yv_mode_secondac_a20e}}
	\caption{Comparison of the Fourier modes in the $Y-$vorticity fluctuation field associated with the bluff body vortex shedding phenomenon between (a) rigid flat wings, (b) rigid cambered wings and (c) flexible membrane wing at $\alpha=$20$^\circ$. These Fourier modes are selected at the non-dimensional frequency of $f c/ U_{\infty}=$  (a) 0.122, (b) 0.135 and (c) 1.22.}
	\label{yv_mode_second_a20}
\end{figure}

We analyze the Fourier modes and the corresponding mode energy spectra in the fluid domain for the rigid flat wing and the rigid cambered wing. The spanwise $Y$-vorticity is selected to perform the mode decomposition, which reflects the vortex shedding along the chord direction. As Rojratsirikul et al. \cite{rojratsirikul2011flow} reported in their study, the vortex shedding frequency of various finite wings at different angles of attack was observed within a modified Strouhal number range of $f c \sin(\alpha) / U_{\infty} \in [0.15,0.2]$. The modified Strouhal number is scaled by the angle of attack, which reflects the standard bluff body vortex shedding frequency. In \reffig{v5_psd_com}, the summarized Strouhal number range of $f c / U_{\infty} \in [\frac{0.15}{\sin(\alpha)},\frac{0.2}{\sin(\alpha)}]$ indicated by a gray region is added to the plots to explore the connection between the bluff body vortex shedding instability and the membrane aeroelasticity. It can be seen from \reffig{v5_psd_com} that the dominant vortex shedding frequency of a rigid flat wing at different angles of attack is close to or falls into the frequency range of $f c / U_{\infty} \in [\frac{0.15}{\sin(\alpha)},\frac{0.2}{\sin(\alpha)}]$. Except for the dominant frequency detected in the mode energy spectra, some smaller frequency components with lower mode energies are also observed for flow past bluff-body-like wings. As the angle of attack increases, the turbulent vortex structures behind the rigid flat wing become more complex as shown in \reffig{streamline_com}. When the camber effect is taken into account for the rigid cambered wing, we observe that the dominant frequency is reduced slightly than that of the rigid flat wing, but it is still close to the bluff-body-like vortex shedding frequency range.

As shown in \reffig{v5_psd_com}, the mode energy spectra of the flexible membrane wings based on the spanwise $Y$-vorticity and the membrane displacement are also added to the plots. The purpose is to investigate the role of membrane flexibility in membrane aeroelasticity. Based on the aeroelastic mode analysis in \refse{mode_decomposition}, we observe that the flexible membrane vibration locks into a chordwise second and spanwise first mode at the three angles of attack. A natural question to ask is whether the dominant frequency of the coupled system is dependent on the natural frequency of a chordwise second and spanwise first mode. An approximate analytical formula of the nonlinear natural frequency derived in Li et al. \cite{li2020flow} is employed to estimate the natural frequency of the corresponding structural mode for a rectangular membrane immersed in an unsteady flow. This analytical formula is based on a large deflection theory to consider the dynamic strain caused by the membrane vibration. The added mass effect of the coupled system with thin structures is taken into account in the formula. The approximate analytical formula of the nonlinear natural frequency $f_{ij}^n$ corresponding to a chordwise $i$-th and spanwise $j$-the mode is given as
\begin{equation}
f_{ij}^n = \frac{1}{2 \pi} \left( \omega_{ij0} + \frac{3 \beta \kappa \omega_{ij0}}{8}(g_0^2 + h_0^2) \right),
\label{nonlinear_frequency}
\end{equation}
where $\omega_{ij0}$ represents the circular frequency of a rectangular membrane with initial tensions $N_{x0}^s$ and $N_{y0}^s$, which is expressed as
\begin{equation}
\omega_{ij0} = \sqrt{\frac{\pi^2}{\rho^s h + m^{am}} \left[ N_{x0}^s \left(\frac{i}{c}\right)^2 + N_{y0}^s \left(\frac{j}{b}\right)^2 \right]},
\label{mem14}
\end{equation}
where $\rho^s$ is the membrane density and $m^{am}$ denotes the added mass. $c$ and $b$ denote the length of the chord and span. $\beta=\frac{h^2}{bc}$ represents the perturbation parameter in the Poincar$\rm{\acute{e}}$-Lindstedt perturbation method. $g_0$ and $h_0$ denote the initial conditions corresponding to the displacement and velocity of the vibrating membrane. $\kappa$ is a coefficient of the vibration equation, which is given as
\begin{equation}
\begin{split}
&\kappa = \frac{cb}{\omega_{ij0}^2 h} \left[ \frac{3 E^s \pi^4}{32(1- (\nu^s)^2)(\rho^s h + m^{am})} \left( \frac{i^4}{c^4} + \frac{j^4}{b^4} \right) \right. \\
&\left. + \frac{3 E^s \nu^s \pi^4}{16(1- (\nu^s)^2)(\rho^s h + m^{am})} \left( \frac{ij}{cb}  \right)^2 - \frac{E^s \pi^4}{16(1+\nu^s)(\rho^s h + m^{am})} \left( \frac{ij}{cb}  \right)^2 \right],
\label{mem16}
\end{split}
\end{equation}
where $\nu^s$ is the Poisson ratio. The detailed deviation of the approximate analytical formula can be found in Li et al. \cite{li2020flow}. Based on our high-fidelity numerical simulation results, we can determine the relevant parameters to calculate the nonlinear structural natural frequency $f^n_{21}$ corresponding to the chordwise second and spanwise first mode of the coupled system. The estimated structural natural frequency $f^n_{21}$ for different angles of attack is indicated by a purple long dash line in \reffig{v5_psd_com}.  By comparing with the estimated nonlinear natural frequency $f^n_{21}$, it can be seen from \reffig{v5_psd_com} that both the vortex shedding frequency and the membrane vibration frequency are close to the estimated natural frequency $f^n_{21}$. It can be inferred that the vortex shedding frequency locks into the structural natural frequency $f^n_{21}$ to sustain the flow-excited vibration in a chordwise second and spanwise first mode. 

\begin{figure}[H]
	\centering 
	\includegraphics[width=0.38\textwidth]{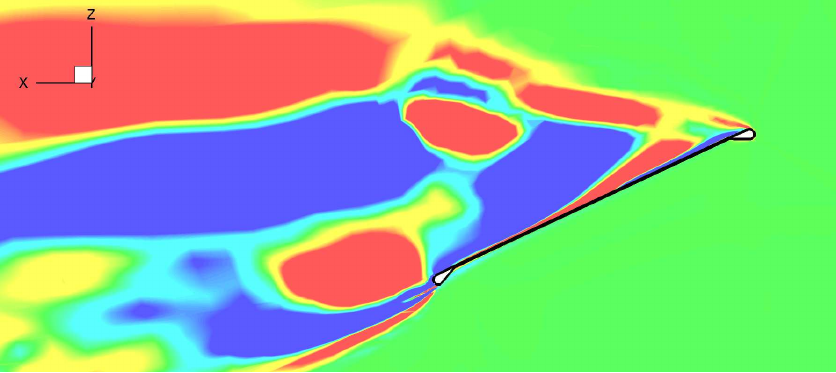}
	\quad
	\includegraphics[width=0.38\textwidth]{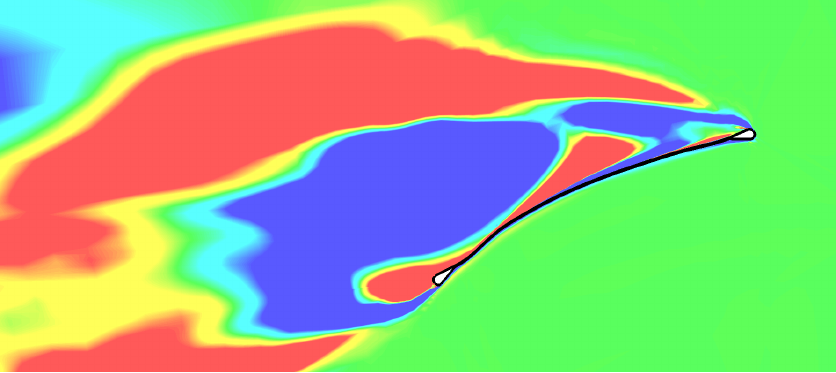}
	\\
	\subfloat[][]{\includegraphics[width=0.38\textwidth]{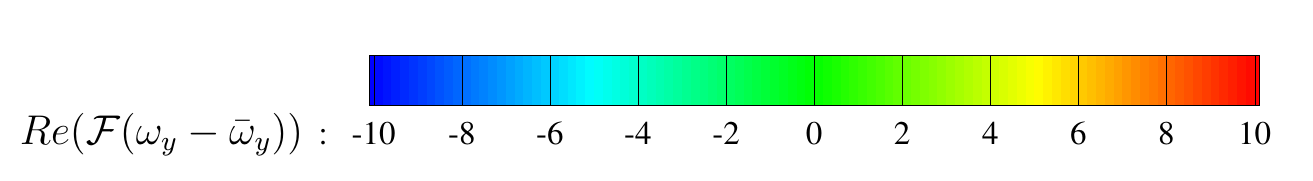}\label{yv_mode_seconda_a25b}}
	\quad
	\subfloat[][]{\includegraphics[width=0.38\textwidth]{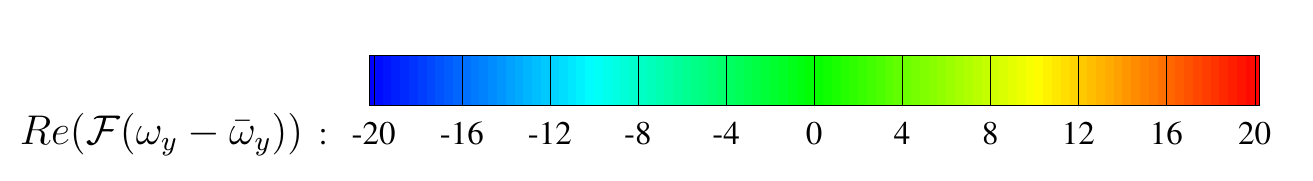}\label{yv_mode_secondac_a25d}}
	\\
	\includegraphics[width=0.38\textwidth]{./results/v5/a25/mode/yv_re_st0_188996_jfs.pdf}
	\\
	\subfloat[][]{\includegraphics[width=0.38\textwidth]{./results/v5/a25/mode/yv_re_st0_188996_leng.pdf}\label{yv_mode_secondace_a25f}}
	\caption{Comparison of the Fourier modes in the $Y-$vorticity fluctuation field associated with the bluff body vortex shedding phenomenon between (a) rigid flat wings, (b) rigid cambered wings and (c) flexible membrane wing at $\alpha=$25$^\circ$. These Fourier modes are selected at the non-dimensional frequency of $f c/ U_{\infty}=$ (a,b,c) 0.2.}
\label{yv_mode_second_a25}
\end{figure}

The frequency corresponding to the bluff-body-like vortex shedding process is not observed in the coupled system at $\alpha=15^\circ$. The vortex structures are regulated to get closer to the membrane surface and shed into the wake in a dominant frequency of $f c / U_{\infty}$=0.99 via the frequency lock-in. As the angle of attack increases, the flexible membrane also responds at several low frequency components to vibrate. These low-frequency components are also observed in the mode energy spectra of the rigid flat wing and the rigid cambered wing, which are associated with the bluff body vortex shedding instability. Except for the dominant vibrational modes caused by the frequency lock-in, we are also interested in the aeroelastic modes at these low frequencies. The Fourier modes in the $Y$-vorticity field corresponding to similar frequencies in the low frequency range at $\alpha=20^\circ$ and $25^\circ$ are compared for the rigid flat wing, the rigid cambered wing and the flexible membrane shown in \reffig{yv_mode_second_a20} and \ref{yv_mode_second_a25}. By investigating the fluid Fourier modes at these low frequencies for the three types of wings, we find that the camber effect and the flow-excited vibration can change the flow features to some extent. The Fourier modes of the $Y$-vorticity field of the rigid cambered wing and the flexible membrane exhibit some similarities in the modal shapes. It can be inferred that the aeroelastic modes corresponding to the low frequency components in the coupled fluid-membrane system are associated with the bluff body vortex shedding instability behind a cambered up wing. The aeroelastic modes caused by the bluff body vortex shedding instability are intertwined with the aeroelastic mode that depends on the frequency lock-in, leading to the non-periodic aeroelastic responses at $\alpha=20^\circ$ and $25^\circ$.

\begin{figure}[H]
	\centering 
	\includegraphics[width=0.7\textwidth]{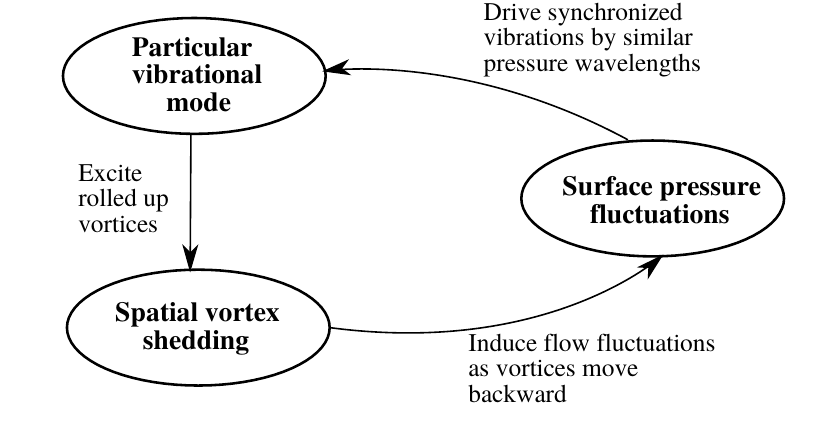}
	\caption{Illustration of a feedback cycle of fluid-membrane coupled mechanism.}
	\label{mechanism}
\end{figure}

The investigation of the decomposed aeroelastic modes suggests a feedback cycle between the vibration mode and the unsteady aerodynamics, as shown in \reffig{mechanism}. The cycle reveals the mode selection mechanism for fluid-membrane interaction problems with the vortex shedding phenomenon. In this coupled system, the vibrational modes excite the separated shear layer to roll up earlier and then form large-scale vortices. As these vortices detach from the membrane surface and are convected downstream, relatively stronger surface pressure fluctuations are induced due to the passing-by vortical structures. Subsequently, the flexible membrane is synchronously driven by the pressure pulsations to excite particular vibrational modes with similar modal shapes. The modal shapes can be observed from the dominant decomposed surface pressure and vibrational modes at the same frequency and the wavenumber-frequency spectra presented in Appendix A. Eventually, the unsteady flow and the membrane vibration enter a strongly coupled state and the frequency synchronization to select the particular aeroelastic modes.

\section{Conclusions}
We presented an aeroelastic mode decomposition framework based on the radial basis function interpolation method and the global Fourier mode decomposition technique. We extracted and identified the Fourier modes of interest both in the fluid and structure fields in a unified manner. The three-dimensional membrane aeroelasticity was simulated by a high-fidelity fluid-structure interaction solver at three angles of attack. The flexible membrane exhibited a periodic aeroelastic response at $\alpha=15^\circ$. The aeroelastic response became non-periodic at higher angles of attack. By comparing the dominant modal shapes observed from the standard deviation analysis and the instantaneous displacement, it was found that the standard deviation was not a reliable indicator to reflect the dominant modes from the coupled system with overlapping modes due to the time-averaged sense. With the aid of the aeroelastic mode decomposition framework, the correlated dominant fluid and structure modes were successfully extracted from the coupled system by detecting the frequency peaks in the mode energy spectra. Based on the mode decomposition analysis, we observed a frequency synchronization between the vortex shedding process and the membrane vibration. The flexible membrane exhibited a similar modal shape with a chordwise second and spanwise first mode at different angles of attack. To explore the role of flexibility in membrane aeroelasticity, we further compared the flow features and the dynamic modes of a rigid flat wing, a rigid cambered wing and their flexible counterpart at three angles of attack. An approximated analytical formula of the nonlinear natural frequency was employed to estimate the natural frequency $f^n_{21}$ corresponding to the chordwise second and spanwise first mode. By comparing with the mode energy spectra of the coupled system, it was found that the vortex shedding frequency locked into the structural natural frequency $f^n_{21}$. Through the comparison of fluid modes corresponding to non-integer low frequency components, the fluid modes of the flexible membrane showed some similarities to those of the rigid cambered wing at $\alpha=20^\circ$ and $25^\circ$. The aeroelastic modes corresponding to the low frequency components can be attributed to the bluff body vortex shedding instability. The non-periodic aeroelastic responses were caused by the interaction between the aeroelastic modes associated with the frequency lock-in and the bluff body vortex shedding instability. Based on the modal analysis, we suggested a feedback cycle between the membrane vibration mode, the vortex shedding process and the surface pressure fluctuations. This feedback cycle revealed that the membrane flexibility acted as a coordinator between the flexible membrane and the unsteady flow to form a frequency lock-in phenomenon to select the dominant mode and sustain the membrane vibration. This mode decomposition method has the potential to be extended to the data analysis of other fluid-structure interaction problems. A combined application with other mode decomposition techniques could offer better physical insight and causal inference for fluid-structure interaction problems.

\section*{Acknowledgements}
The first author wishes to acknowledge supports from the National University of Singapore
and the Ministry of Education, Singapore.  The second author would like to acknowledge the support from the University of British Columbia and the Natural Sciences and Engineering Research Council of Canada (NSERC).

\section*{Declaration of interests}
The authors report no conflict of interest.

\section*{Appendix A: Wavenumber-frequency spectra}
To quantitatively connect the membrane vibration and the pressure pulsation, we analyze the wavelength of each Fourier mode by projecting the mode from the spatial-frequency space to the wavenumber-frequency space. The wavelength corresponding to a specific vibrational frequency can be computed based on the double Fourier transform. The physical signal $y(x_m,t_n)$ sampled at $t_n$ time instant and $x_m$ spatial point is collected to form a time sequence for the double Fourier transform. The equation of double Fourier transform is expressed as
\begin{align}
\mathcal{F}(y(x_m,t_n))(f_k,\kappa_l)=\sum_{n=0}^{N-1} \sum_{m=0}^{M-1}  y(x_m,t_n) e^{-i \frac{2 \pi k}{N}n} e^{-i \frac{2 \pi l}{M}m}
\label{FFT2}
\end{align}
where $\mathcal{F}(y(x_m,t_n))(f_k,\kappa_l)$ is the double Fourier-transformed coefficient. $\kappa_l$ is the wavenumber and $f_k$ represents the frequency. The wavelength can be defined as $\lambda=c/\kappa$ for a length of $c$.

\begin{figure}[H]
	\centering 
	\subfloat[][]{\includegraphics[width=0.4\textwidth]{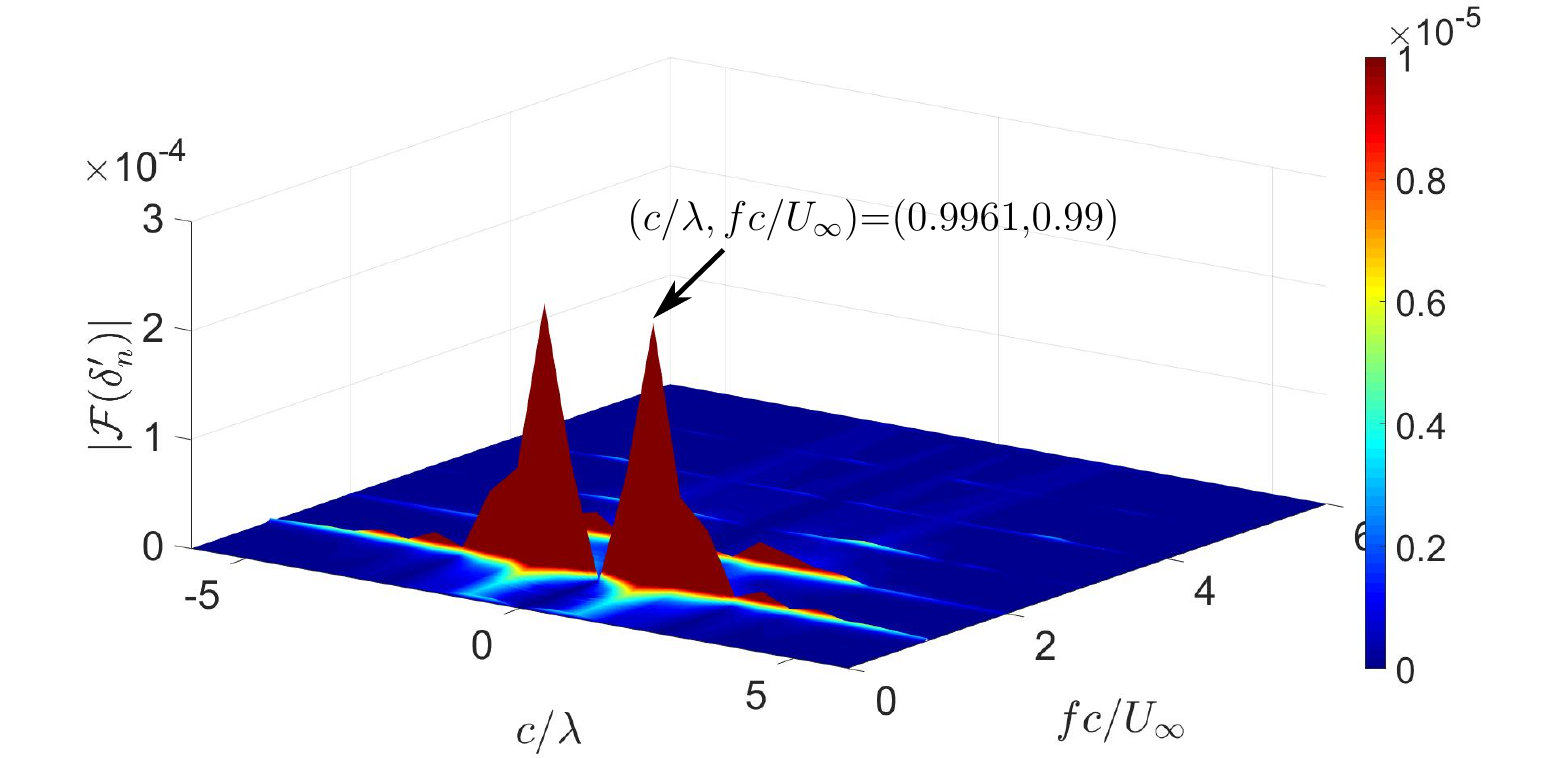}\label{wavefre1}}
	\quad
	\subfloat[][]{\includegraphics[width=0.4\textwidth]{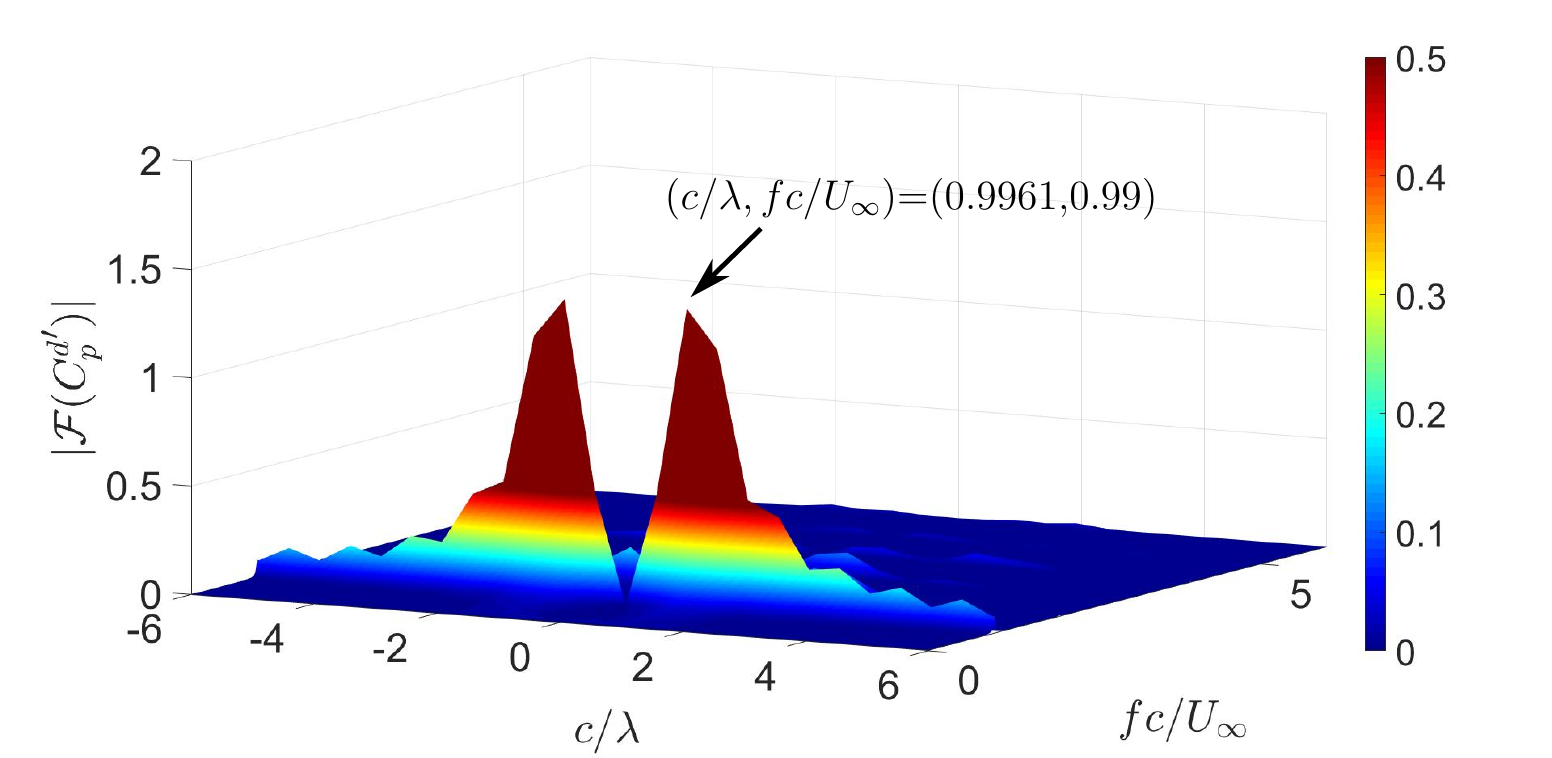}\label{wavefre2}}
	\caption{Wavenumber-frequency spectra of: (a) membrane deflection fluctuations and (b) pressure coefficient difference fluctuations along the flexible membrane at the mid-span location at $\alpha=15^\circ$.}
	\label{wavefre}
\end{figure}

We extract the time-varying membrane vibration and surface pressure signals at 256 equispaced points at the mid-span location along the chord-wise direction of the membrane wing for 1024 non-dimensional time instants at $\alpha=15^\circ$. These data samples are stored in a time-space matrix. The mean values are removed from the full signals and the fluctuations of the analyzed signals are mapped into frequency-wavenumber space through the double Fourier transform. The wavenumber-frequency diagrams of the membrane deflection fluctuations and the pressure difference fluctuations are plotted in \reffig{wavefre}. These 2D diagrams are colored by the amplitude of the double Fourier transformed coefficients of the analyzed signal functions. The $x$-axis is the wavenumber of the unit chord length $c/\lambda$ and the $y$-axis indicates the non-dimensional frequency $fc/U_{\infty}$. The wavenumber at the dominant frequency $fc/U_{\infty}$=0.99 is consistent for the membrane vibration and the pressure coefficient difference.

\section*{References}

\bibliography{referenceBib}

\begin{thebibliography}{10}
\expandafter\ifx\csname url\endcsname\relax
  \def\url#1{\texttt{#1}}\fi
\expandafter\ifx\csname urlprefix\endcsname\relax\def\urlprefix{URL }\fi
\expandafter\ifx\csname href\endcsname\relax
  \def\href#1#2{#2} \def\path#1{#1}\fi

\bibitem{lian2003membrane}
Y.~Lian, W.~Shyy, D.~Viieru, B.~Zhang, Membrane wing aerodynamics for micro air
  vehicles, Progress in Aerospace Sciences 39~(6-7) (2003) 425--465.

\bibitem{shyy2005membrane}
W.~Shyy, P.~Ifju, D.~Viieru, Membrane wing-based micro air vehicles, Applied
  mechanics reviews 58~(4) (2005) 283--301.

\bibitem{platzer2008flapping}
M.~F. Platzer, K.~D. Jones, J.~Young, J.~S.~Lai, Flapping wing aerodynamics:
  progress and challenges, AIAA Journal 46~(9) (2008) 2136--2149.

\bibitem{aldheeb2016review}
M.~A. Aldheeb, W.~Asrar, E.~Sulaeman, A.~A. Omar, A review on aerodynamics of
  non-flapping bird wings, Journal of Aerospace Technology and Management 8~(1)
  (2016) 7--17.

\bibitem{song2008aeromechanics}
A.~Song, X.~Tian, E.~Israeli, R.~Galvao, K.~Bishop, S.~Swartz, K.~Breuer,
  Aeromechanics of membrane wings with implications for animal flight, AIAA
  Journal 46~(8) (2008) 2096--2106.

\bibitem{rojratsirikul2009unsteady}
P.~Rojratsirikul, Z.~Wang, I.~Gursul, Unsteady fluid--structure interactions of
  membrane airfoils at low {Reynolds} numbers, Experiments in Fluids 46~(5)
  (2009) 859.

\bibitem{rojratsirikul2010effect}
P.~Rojratsirikul, Z.~Wang, I.~Gursul, Effect of pre-strain and excess length on
  unsteady fluid--structure interactions of membrane airfoils, Journal of
  Fluids and Structures 26~(3) (2010) 359--376.

\bibitem{rojratsirikul2011flow}
P.~Rojratsirikul, M.~Genc, Z.~Wang, I.~Gursul, Flow-induced vibrations of low
  aspect ratio rectangular membrane wings, Journal of Fluids and Structures
  27~(8) (2011) 1296--1309.

\bibitem{bleischwitz2015aspect}
R.~Bleischwitz, R.~de~Kat, B.~Ganapathisubramani, Aspect-ratio effects on
  aeromechanics of membrane wings at moderate {Reynolds} numbers, AIAA Journal
  53~(3) (2015) 780--788.

\bibitem{sun2016nonlinear}
X.~Sun, J.~Zhang, Nonlinear vibrations of a flexible membrane under periodic
  load, Nonlinear Dynamics 85~(4) (2016) 2467--2486.

\bibitem{sun2017effect}
X.~Sun, J.~Zhang, Effect of the reinforced leading or trailing edge on the
  aerodynamic performance of a perimeter-reinforced membrane wing, Journal of
  Fluids and Structures 68 (2017) 90--112.

\bibitem{sun2017nonlinear}
X.~Sun, X.~Ren, J.~Zhang, Nonlinear dynamic responses of a perimeter-reinforced
  membrane wing in laminar flows, Nonlinear Dynamics 88~(1) (2017) 749--776.

\bibitem{sun2018bifurcations}
X.~Sun, S.~Wang, J.~Zhang, Z.~Ye, Bifurcations of vortex-induced vibrations of
  a fixed membrane wing at {Re} $\leq$ 1000, Nonlinear Dynamics 91~(4) (2018)
  2097--2112.

\bibitem{tregidgo2011fluid}
L.~Tregidgo, Z.~Wang, I.~Gursul, Fluid-structure interactions for a low
  aspect-ratio membrane wing at low {Reynolds} numbers, in: 41st AIAA Fluid
  Dynamics Conference and Exhibit, 2011, p. 3436.

\bibitem{lumley1970stochastic}
J.~L. Lumley, Stochastic tools in turbulence. volume 12. applied mathematics
  and mechanics, Tech. rep., Pennsylvania State University (1970).

\bibitem{sirovich1987turbulence}
L.~Sirovich, Turbulence and the dynamics of coherent structures. i. coherent
  structures, Quarterly of Applied Mathematics 45~(3) (1987) 561--571.

\bibitem{berkooz1993proper}
G.~Berkooz, P.~Holmes, J.~L. Lumley, The proper orthogonal decomposition in the
  analysis of turbulent flows, Annual Review of Fluid Mechanics 25~(1) (1993)
  539--575.

\bibitem{ma2015fourier}
L.~Ma, L.~Feng, C.~Pan, Q.~Gao, J.~Wang, Fourier mode decomposition of {PIV}
  data, Science China Technological Sciences 58~(11) (2015) 1935--1948.

\bibitem{schmid2010dynamic}
P.~J. Schmid, Dynamic mode decomposition of numerical and experimental data,
  Journal of Fluid Mechanics 656 (2010) 5--28.

\bibitem{chen2012variants}
K.~K. Chen, J.~H. Tu, C.~W. Rowley, Variants of dynamic mode decomposition:
  boundary condition, koopman, and fourier analyses, Journal of Nonlinear
  Science 22~(6) (2012) 887--915.

\bibitem{sieber2016spectral}
M.~Sieber, C.~O. Paschereit, K.~Oberleithner, Spectral proper orthogonal
  decomposition, Journal of Fluid Mechanics 792 (2016) 798--828.

\bibitem{towne2018spectral}
A.~Towne, O.~T. Schmidt, T.~Colonius, Spectral proper orthogonal decomposition
  and its relationship to dynamic mode decomposition and resolvent analysis,
  Journal of Fluid Mechanics 847 (2018) 821--867.

\bibitem{seena2011dynamic}
A.~Seena, H.~J. Sung, Dynamic mode decomposition of turbulent cavity flows for
  self-sustained oscillations, International Journal of Heat and Fluid Flow
  32~(6) (2011) 1098--1110.

\bibitem{michelin2008vortex}
S.~Michelin, S.~G. Llewellyn~Smith, B.~J. Glover, Vortex shedding model of a
  flapping flag, Journal of Fluid Mechanics 617~(1) (2008) 1--10.

\bibitem{bozkurttas2009low}
M.~Bozkurttas, R.~Mittal, H.~Dong, G.~Lauder, P.~Madden, Low-dimensional models
  and performance scaling of a highly deformable fish pectoral fin, Journal of
  Fluid Mechanics 631 (2009) 311.

\bibitem{liu2016interaction}
B.~Liu, R.~K. Jaiman, Interaction dynamics of gap flow with vortex-induced
  vibration in side-by-side cylinder arrangement, Physics of Fluids 28~(12)
  (2016) 127103.

\bibitem{miyanawala2019decomposition}
T.~P. Miyanawala, R.~K. Jaiman, Decomposition of wake dynamics in
  fluid--structure interaction via low-dimensional models, Journal of Fluid
  Mechanics 867 (2019) 723--764.

\bibitem{bleischwitz2017fluid}
R.~Bleischwitz, R.~De~Kat, B.~Ganapathisubramani, On the fluid-structure
  interaction of flexible membrane wings for mavs in and out of ground-effect,
  Journal of Fluids and Structures 70 (2017) 214--234.

\bibitem{serrano2018fluid}
S.~Serrano-Galiano, N.~D. Sandham, R.~D. Sandberg, Fluid--structure coupling
  mechanism and its aerodynamic effect on membrane aerofoils, Journal of Fluid
  Mechanics 848 (2018) 1127--1156.

\bibitem{bleischwitz2016aeromechanics}
R.~Bleischwitz, R.~De~Kat, B.~Ganapathisubramani, Aeromechanics of membrane and
  rigid wings in and out of ground-effect at moderate {Reynolds} numbers,
  Journal of Fluids and Structures 62 (2016) 318--331.

\bibitem{tiomkin2019membrane}
S.~Tiomkin, D.~Raveh, On membrane-wing stability in laminar flow, Journal of
  Fluids and Structures 91 (2019) 102694.

\bibitem{goza2018modal}
A.~Goza, T.~Colonius, Modal decomposition of fluid--structure interaction with
  application to flag flapping, Journal of Fluids and Structures 81 (2018)
  728--737.

\bibitem{welch1967use}
P.~Welch, The use of fast fourier transform for the estimation of power
  spectra: a method based on time averaging over short, modified periodograms,
  IEEE Transactions on audio and electroacoustics 15~(2) (1967) 70--73.

\bibitem{jaiman2016stable}
R.~Jaiman, N.~Pillalamarri, M.~Guan, A stable second-order partitioned
  iterative scheme for freely vibrating low-mass bluff bodies in a uniform
  flow, Computer Methods in Applied Mechanics and Engineering 301 (2016)
  187--215.

\bibitem{li2018novel}
G.~Li, Y.~Z. Law, R.~K. Jaiman, A novel 3d variational aeroelastic framework
  for flexible multibody dynamics: Application to bat-like flapping dynamics,
  Computers \& Fluids 180 (2019) 96--116.

\bibitem{li2021high}
G.~Li, G.~Kemp, R.~K. Jaiman, B.~C. Khoo, A high-fidelity numerical study on
  the propulsive performance of pitching flexible plates, Physics of Fluids
  33~(5) (2021) 051901.

\bibitem{li2020computational}
G.~Li, B.~C. Khoo, R.~K. Jaiman, Computational aeroelasticity of flexible
  membrane wings at moderate {Reynolds} numbers, in: AIAA Scitech 2020 Forum,
  2020, p. 2036.

\bibitem{solomon1991psd}
O.~Solomon~Jr, Psd computations using welch's method, NASA STI/Recon Technical
  Report N 92 (1991) 23584.

\bibitem{jwo2021windowing}
D.~J. Jwo, W.~Y. Chang, I.~H. Wu, Windowing techniques, the welch method for
  improvement of power spectrum estimation, Cmc-Computers Materials \& Continua
  67~(3) (2021) 3983--4003.

\bibitem{rojratsirikul2010unsteady}
P.~Rojratsirikul, Z.~Wang, I.~Gursul, Unsteady aerodynamics of low aspect ratio
  membrane wings, in: 48th AIAA Aerospace Sciences Meeting Including the New
  Horizons Forum and Aerospace Exposition, 2010, p. 729.

\bibitem{tregidgo2013unsteady}
L.~Tregidgo, Z.~Wang, I.~Gursul, Unsteady fluid--structure interactions of a
  pitching membrane wing, Aerospace Science and Technology 28~(1) (2013)
  79--90.

\bibitem{li2020flow}
G.~Li, R.~K. Jaiman, B.~C. Khoo, Flow-excited membrane instability at moderate
  {Reynolds} numbers, arXiv preprint arXiv:2011.11422.

\end{thebibliography}

\end{document}